\documentclass[twocolumn]{svjour3}
\usepackage{amsmath,amssymb,graphicx,color,subfigure,wrapfig,epstopdf,diagbox}
\usepackage[titlenumbered,ruled]{algorithm2e}
\usepackage[figuresright]{rotating}
\usepackage{stfloats}
\usepackage{microtype,multicol}
\usepackage{lmodern}
\usepackage{nomencl}

\newcommand{\real}{\mathbb{R}}
\newcommand{\ith}[1]{#1^{\text{th}}}
\newcommand{\vo}[1]{\boldsymbol{#1}}
\newcommand{\Rq}{\vo{R}_{\vo{q}}}

\newcommand{\bcg}{\bar{\vo{b}}_k}
\newcommand{\bcgd}{\dot{\bar{\vo{b}}}_k}

\newcommand{\bv}{\vo{b}_k}
\newcommand{\bvd}{\dot{\vo{b}}_k}

\newcommand{\q}{\vo{q}}
\newcommand{\qd}{\dot{\q}}

\newcommand{\eqnlabel}[1]{\label{eqn:#1}}
\newcommand{\eqn}[1]{(\ref{eqn:#1})}
\renewcommand{\vec}[1]{\boldsymbol{\mathsf{vec}}\left({#1}\right)}

\makenomenclature

\begin{document}
\title{\bf A Lagrangian Method for Constrained Dynamics in Tensegrity Systems with Compressible Bars}
\author{Shao-Chen Hsu$^{1}$ \and
		Vaishnav Tadiparthi$^{2}$ \and
		Raktim Bhattacharya$^{2}$
}

\institute{ Shao-Chen Hsu \at
			\email{addyhsu@gmail.com} \and
			Vaishnav Tadiparthi \at
			\email{vaishnavtv@tamu.edu} \and
			Raktim Bhattacharya \at
			\email{raktim@tamu.edu}
			\at\at
			\begin{tabular}{ll}
			$^1$ & Genesys Aerosystems,\\
			&Mineral Wells, TX, 76067\\
			% &College Station, TX, 77843-3141
			$^2$ & Intelligent Systems Research Laboratory,\\
			&Aerospace Engineering, Texas A\&M University\\
			&College Station, TX, 77843-3141
			\end{tabular}
}

%\author{\begin{tabular}{ccc}Shao-Chen Hsu & Vaishnav Tadiparthi & Raktim Bhattacharya\end{tabular}\\[2mm] \small \textit{ Laboratory for Uncertainty Quantification} \\ \small Aerospace Engineering, Texas A\&M University \\ \small  College Station, TX, 77843-3141.}
\date{}
\maketitle
\begin{abstract}
This paper presents a Lagrangian approach to simulating multibody dynamics in a tensegrity framework with an ability to tackle holonomic constraint violations in an energy-preserving scheme. Governing equations are described using non-minimum coordinates to simplify descriptions of the structure's kinematics.
% leading to an undesirable constraint drift with numerical computation errors.
To minimize constraint drift arising from this redundant system, the direct correction method has been employed in conjunction with a novel energy-correcting scheme that treats the total mechanical energy of the system as a supplementary constraint.
% to nullify any numerical violations occurring in integration.
The formulation has been extended to allow tensegrity structures with compressible bars, allowing for further discussion on potential choices for softer bar materials.
% materials in the context of structural response to force disturbances and mass optimality.
The benchmark example involving a common tensegrity structure demonstrates the superiority of the presented formulation over Simscape Multibody in terms of motion accuracy as well as energy conservation.
The effectiveness of the energy correction scheme is found to be increasing with the extent of deformations in the structure.
% Other tensegrity structures with intricate geometries have been simulated and presented at the end.
% The formulation with compressible bars is used to discuss potential choices for softer bar materials in the context of structural response to force disturbances and mass optimality.

\end{abstract}

% \textcolor{blue}{
\renewcommand{\nomname}{List of Symbols}
\nomenclature{$\vo{n}_i$}{Position of $i^{\rm{th}}$ node}
\nomenclature{$\vo{N}$}{Nodal matrix describing the tensegrity structure}
\nomenclature{$\vo{C}_b$}{Connectivity matrix of bars}
\nomenclature{$\vo{C}_s$}{Connectivity matrix of strings}
\nomenclature{$\vo{L}_{pm}$}{Location matrix of point masses}
\nomenclature{$\vo{B}$}{Bar matrix}
\nomenclature{$\vo{S}$}{String matrix}
\nomenclature{$\vo{P}$}{Point mass matrix}
\nomenclature{$\vo{b}_k$}{$k^{\rm{th}}$ bar}
\nomenclature{$\vo{s}_k$}{$k^{\rm{th}}$ string}
\nomenclature{$\vo{p}_k$}{$k^{\rm{th}}$ point mass}
\nomenclature{$\vo{F}$}{Non-conservative force matrix}
\nomenclature{$l_{\vo{b}_k}$}{Length of $k^{\rm{th}}$ bar}
\nomenclature{$\vo{\omega}_k$}{Angular velocity of $k^{\rm{th}}$ bar}
\nomenclature{$\vo{I}_{\vo{b}_k}$}{Moment of inertia of $k^{\rm{th}}$ bar}
\nomenclature{$\vo{h}_k$}{Angular momentum of $k^{\rm{th}}$ bar}
\nomenclature{$T$}{Total kinetic energy}
\nomenclature{$V_s$}{Potential energy of strings}
\nomenclature{$V_g$}{Potential energy due to gravity}
\nomenclature{$\vo{q}$}{Coordinates in vector form}
\nomenclature{$\vo{R}(\vo{q})$}{Ideal constraints}
\nomenclature{$l_{\vo{s}_k}$}{Natural length of $k^{\rm{th}}$ string}
\nomenclature{$K_{\vo{s}_k}$}{Stiffness of $k^{\rm{th}}$ string}
\nomenclature{$\vo{\lambda}$}{Lagrange multipliers}
\nomenclature{$\sigma_k$}{Force density of $k^{th}$ string}
\nomenclature{$c$}{Damping coefficient}
\nomenclature{$\vo{f}_{d,k}$}{Damper force in $k^{\rm{th}}$ string}
\nomenclature{$E$}{Total energy of the system}
\nomenclature{$W_f$}{Work done by force $f$}
\nomenclature{$K_{\vo{b}_k}$}{Stiffness of $\ith{k}$ bar (compressible)}
\nomenclature{$\Psi_{k}$}{Force density of $\ith{k}$ bar (compressible)}
\nomenclature{$r_{k}$}{Radius of $\ith{k}$ bar (compressible)}
\nomenclature{$\nu_{k}$}{Poisson's ratio of $\ith{k}$ bar material (compressible)}
\printnomenclature

\section{Introduction}
A tensegrity system is an arrangement of axially-loaded elements (no element bends, even though the overall structure bends), that we loosely characterize as a network of bars and cables. The bars take compressive axial loads and the cables handle tensile loads. Since failure due to axial stresses happens at higher loads than at bending, a tensegrity structure has a higher strength-to-weight ratio. Famous architect Buckminster Fuller in the 60's coined the term tensegrity, combining the words tensile and integrity. Since then, tensegrity principles have found applications in diverse domains. Tensegrity systems have been widely adopted in architecture. Donald E. Ingber \cite{ingber1998architecture} explained the behavior of cells by modeling them as tensegrity structures. He further showed that tensegrity structures exist at all detectable scales of the human body. Tensegrity icosahedrons are used to model biologic organisms from viruses to vertebrates, their cells, systems, and subsystems.
Biotensegrity \cite{levin2002tensegrity,scarr2014biotensegrity,dischiavi2018biotensegrity,levin2012biotensegrity} is quite an active area of research.
Beyond architecture and biology, tensegrity principles are gaining popularity in robotics. NASA is considering a new terrestrial robot design based on tensegrity principles \cite{agogino2013super}.
Tensegrity structures, through use of pre-stresses in the bars and cables, can also achieve controlled stiffness in the structure, which makes it attractive in applications such as soft-robotics \cite{rieffel2018adaptive}, robotic locomotion \cite{paul2005gait,shibata2009crawling}, and prosthetics \cite{rifkin2005tensegrity}. In essence, tensegrity principles can be applied in the design of any structure where mass is premium, a high strength-to-weight ratio is critical, and structural stiffness needs to be tailored in both space and time. These include several applications from various engineering sectors such as aerospace (morphing airframes), energy (wind turbine blades, off-shore structures) as well as biomedical engineering (stents, minimally invasive surgical tools) and many more. Clearly, a framework is required that can efficiently model the dynamics of tensegrity structures directly from the topology of bars and cables.

The dynamics of tensegrity systems is governed by multi-body dynamics, given by a set of ordinary differential equations. This paper develops a Lagrangian formulation for deriving these differential equations directly from the given topology of members (bars and strings), and their mass and geometric properties. Three key features of classical tensegrity systems are: a) actuations only occur via cables (though this assumption can be relaxed through the introduction of soft, compressible bars which can actuate through axial deformation), b) bar-to-bar connections are pin joints, and c) the bars do not spin about their respective longitudinal axes. These properties are exploited to simplify the equations of motion. However, the Lagrangian framework presented here is general enough to allow modeling of general multi-body systems with actuated joints.

%The need for software packages that target multibody dynamics (MBD) simulations is becoming increasingly apparent as robotics seems to grow in capability as well as complexity. In this paper, we present a Lagrangian formulation for solving MBD problems within the framework of tensegrity systems. We chose the tensegrity paradigm because it forms a special case of MBD systems in which the disciplines of structures and dynamics are integrated to optimize for mechanical strength and control power simultaneously\cite{sultanThesis}, a notion often overlooked in traditional design approaches\cite{tenseBook}.

The demand for more accurate simulating tools for multi-body dynamics is being challenged quite positively by the open-source community. Physics engines such as Bullet\cite{bulletPhys} and the Open Dynamics Engine (ODE)\cite{odePhys} have become common in robotics applications. NASA's Tensegrity Robotics Toolkit (NTRT)\cite{ntrtsim} is based on the Bullet engine.  They rely on non-minimal coordinate descriptions, while other popular engines, e.g. Simscape Multibody\cite{matlab}, MuJoCo\cite{mujocoPhys}, DART\cite{dartPhys} and Simbody\cite{simbodyPhys} favor using generalized coordinates for describing the kinematics of bodies.
This is because they mostly focus on robotics applications, where the configuration space is naturally reduced in the presence of joints and other constraints\cite{PhysEngine}. However, we have opted to use the Cartesian coordinate system to describe the motion of bodies, most notably, for two reasons. Skelton observed\cite{tenseBook} that in three dimensions, a minimal coordinates approach is prone to singularities developed in the mass matrix, and therefore, the dynamics necessitates an excess coordinates description. Additionally, non-minimal descriptions of vector kinematics allows us to write elegant differential-algebraic equations (DAE), free of trigonometric terms.

To fully express a rigid body motion in Cartesian coordinates, equations describing constraints are written at the acceleration level and augmented to the equations of motion to develop a mass-descriptor form of a set of index-1 DAEs. Since only acceleration level constraints are tackled in the equations, position and velocity level constraints are violated due to errors from numerical integration. Numerous advances have been made in the past few decades addressing this very issue. A prominent method is that of generalized coordinates partitioning
\cite{wehage1982generalized,haug1990generalized} in which, utilizing Gauss-Jordan reduction, independent variables are identified and integrated numerically while dependent variables are preserved through the constraint equations. Baumgarte\cite{baumgarte1972stabilization}, on the other hand, instead of bypassing the problem, introduced two extra terms to the constraint equations so that the violations can be stabilized in the sense of Lyapunov. This method has been studied in different frameworks, such as in adaptive mechanisms \cite{chang1985adaptive}, optimal sense \cite{bae1990stabilization}, and digital control theory \cite{lin2002stabilization}. Stabilization allows for greater computational speed whereas coordinate partitioning is known for its superior error control characteristics, and methods that combine these two techniques
\cite{park1988stabilization,park1986hybrid} to tap into these advantages have been developed as well.

However, parameter selection in the Baumgarte technique is a challenging task \cite{flores2011parametric,ascher1995stabilization}, as systems implemented with the wrong feedback parameters have been found to become unstable.
% \textcolor{blue}{
Coordinate partitioning was also shown to be superior to Baumgarte's method in stabilizing constraint violations during kinematic analyses \cite{neto2003stabilization}.
% }
Therefore, other methods were looked into, the most common being one in which constraint violation is eliminated directly by adding appropriate correction terms to the generalized coordinates after each numerical integration.
Using geometric and energy conservation constraints, Yoon et al. chose corrected positions (constrained through geometry) and velocities (constrained through energy) to be linear in the Jacobian of the constraints\cite{yoon1994geometric}. Yu and Chen developed an algorithm to obtain the corrected terms with the constraints at position and velocity level (both constrained through geometry) by using the Moore-Penrose inverse\cite{yu2000direct}. Citing inconsistency of units and dimensions in generalized coordinates, Blajer added an inverse of the mass matrix to the corrections of \cite{yoon1994geometric} as a weight matrix \cite{blajer2002elimination}. However, Zhang et al. compared the above two formulations in benchmark examples showing that the violation of constraints performed in the same order
\cite{zhang2016constraint}.

Furthermore, compared with the Baumgarte technique, the applied direct correction method performs more efficiently in the context of constraint violations at the position and velocity level\cite{marques2017constraints,flores2013comparison,zhang2016constraint}. However, the extent of inaccuracy in the motion, which can be determined from the violations of the energy constraints is still unclear
\cite{blajer2011methods,blajer2002elimination,yoon1994geometric}.
% \textcolor{blue}{
Therefore, inspired by \cite{marques2017constraints} and
\cite{yu2000direct},
one of the contributions of this paper is to present a novel methodology that attains explicit elimination of not only position and velocity constraints, i.e. holonomic constraints, but also energy variations, i.e. nonholonomic constraints.
The Lagrangian formulation lends itself favorably to an equilibrium analysis of the motion and any corresponding violations in energy conservation. Instead of considering corrected terms of position and velocity separately, we formulate a set of equations linear in these variables with energy constraints and solve the variables simultaneously in the sense of minimal norm.
% }

% \textcolor{blue}{
Additionally, we have extended the formulation to support modeling of compressible bars in a tensegrity structure, i.e, the constraints on bar lengths have been relaxed to allow longitudinal deformation, and in conjunction, a transverse change. This further means that the velocity of a compressible bar would no longer have to be perpendicular to the vector along its length. Accordingly, the kinetic energy has been amended to account for the change in bar lengths and a potential energy term has been added to account for the hitherto insignificant elasticity in the bar. This extension for compressible bars would prove tremendously useful when analysing pneumatic tensegrity systems, like the self-deploying inflatable compression struts introduced in \cite{dessi2019self},
% \textcolor{blue} {
or for examining compliant multistable tensegrity structures like in \cite{sumi2017multistable}.
% }
Further, the axial elasticity of the bars could provide a \textit{passive} actuation mechanism as an alternative to the conventional prestressing and pulling of cables, and therefore warrants a framework that could facilitate the dynamics analysis of such systems.
% }

Recently, Goyal and Skelton \cite{Goyal2019} developed a dynamics formulation for tensegrity systems using Newton and Euler's laws, in which the cables are treated as a finite number of point masses connected with massless strings. Instead of introducing Lagrange multipliers, they combine the bar length constraints with rotational dynamics by implementing the pseudo-inverse technique.
Additionally, a reduced order model has been derived through an intricate use of the singular value decomposition, which avoids other geometrical constraints. Furthermore, a new algorithm has been introduced to tackle bar length correction, which enforces the length of the bars in the structure to be fixed and its velocity vector to be orthogonal to the bar vector.
However, in our paper, we treat all physical limits as constraints and correct the violations simultaneously.

% \textcolor{blue}{
The contributions of this paper are presented as follows. Firstly, a Lagrangian formulation based on Cartesian coordinates is used for deriving DAEs of the governing equations of motion in a tensegrity framework for both rigid and elastic bars. In addition, a general approach for linearization of the equations is determined analytically. Further, a novel technique for improving the accuracy of the simulation is developed to ensure that the errors in states arising from numerical integration are corrected on the position and velocity levels according to both geometric and energy constraints.
Finally, the formulation has been extended to tensegrity systems featuring compressible bars, thereby allowing a deeper analysis into the kinds of materials that could be substituted in place of conventional metallic bars.
% }

The following sections describe the formulation in much greater detail: the nomenclature used in developing the equations, the Lagrangian method for deriving the governing DAEs in the presence of constraints, an elaborate description of the holonomic constraint equations, the direct correction method deployed to ensure that these constraints are not violated at any given time, the proposed energy correction algorithm to nullify energy gain/loss occurring numerically, linearization of the governing equations to facilitate work in control, and finally, the modifications required for tensegrity structures with compressible bars.
% \textcolor{blue}{
A summary has been provided at the end of each of sections 2 and 3 to assist in grasping the salient details of the formulation.
% }
The results for several examples are compared with those from Simscape Multibody (MATLAB's multi-body package)  and presented at the end to discuss the validity of the formulation and the benefits of the approaches proposed in the paper.

\section{Derivation of Tensegrity Dynamics}
\subsection{Nomenclature}
The notations used in the derivation of the tensegrity dynamics are defined as follows, first introduced in \cite{tenseRaman} and \cite{tenseMaziar}.
\begin{enumerate}
\item Let $\vo{n}_i\in\real^{3\times 1}$ be the position of the $i^{\text{th}}$ node.
\item Let $\vo{N}\in\real^{3 \times n}$ be the nodal matrix defined by
$$ \vo{N}:= \begin{bmatrix}\vo{n}_1 & \vo{n}_2 & \cdots & \vo{n}_n\end{bmatrix},$$ where $n$ is the number of nodes in the tensegrity system.
\item Let $\vo{C} \in \real^{m\times n}$ be the connectivity matrix that defines the tensegrity system, where $m$ members are defined by connecting $n$ nodes. Specifically, if the $\ith{k}$ member is defined by connecting nodes $\vo{n}_i$ and $\vo{n}_j$, then
$C(k,i) = -1,\;  C(k,j) = 1,\; \text{ and } C(k,\cdot) = 0$  otherwise.
Moreover, we can partition the $m$ members to bars and strings, resulting in a partitioned connectivity matrix
$$ \vo{C}:=\begin{bmatrix}\vo{C}_b\\\vo{C}_s\end{bmatrix},$$
where $\vo{C}_b\in\real^{n_b\times n}$ defines the $n_b$ bar connections and $\vo{C}_s\in \real^{n_s\times n}$ defines the $n_s$ string connections.

Observing the connectivity matrix $\vo{C}_b$, we derive a matrix $\vo{L}_{p_m}\in \real^{n_{p_m}\times n}$ describing locations of $n_{p_m}$ point masses. These masses are placed at nodes where only strings connect. Specifically, if the $\ith{k}$ point mass is positioned at the node $\vo{n}_i$, then
$$L_{p_m}(k,i) = 1, \text{ and } L_{p_m}(k,\cdot) = 0\text{ otherwise}.$$

\item The bars, strings and point masses are then defined as
\begin{align*}
&\vo{B}:= \vo{N}\vo{C}^T_b\in\real^{3\times n_b},\\
&\vo{S}  := \vo{N}\vo{C}_s^T\in\real^{3\times n_s},\\
&\vo{P} := \vo{N}\vo{L}_{p_m}^T\in\real^{3\times n_{p_m}}.
\end{align*}

The $\ith{k}$ column of $\vo{B}$ represents the $\ith{k}$ bar, denoted by $\bv$. Similarly, the $\ith{k}$ column of $\vo{S}$ represents the $\ith{k}$ string, denoted by $\vo{s}_k$, and the $\ith{k}$ column of $\vo{P}$ represents the $\ith{k}$ point mass, denoted by $\vo{p}_k$. Let $\vo{\theta}_k$, $\vo{\eta}_k$, and $\vo{\phi}_k$ be vectors in $\real^{n_b}$, $\real^{n_s}$, and $\real^{n_{p_m}}$ respectively with the $k^\text{th}$ elements equal to one and the rest zero.
Therefore, we can compactly write
%\begin{equation}\left.
%\begin{aligned}
%\bv &:= \vo{N}\vo{C}_b^T\vo{\theta}_k = \left((\vo{\theta}_k^T\vo{C}_b)\otimes \vo{I}_3\right)\mathbf{vec}(\vo{N}) = \vo{X}_k\vo{q},\\
%\bcg &:= \frac{1}{2} \left((\vo{\theta}_k^T|\vo{C}_b|)\otimes \vo{I}_3\right)\mathbf{vec}(\vo{N}) = \bar{\vo{X}}_k\vo{q},\\
%\vo{s}_k &:= \vo{N}\vo{C}_s^T\vo{\psi}_k = \left((\vo{\psi}_k^T\vo{C}_s)\otimes \vo{I}_3\right)\mathbf{vec}(\vo{N}) = \vo{Y}_k\vo{q},\\
%\vo{p}_k &:= \vo{N}\vo{L}_{p_m}^T\vo{\phi}_k = \left((\vo{\phi}_k^T\vo{L}_{p_m})\otimes \vo{I}_3\right)\mathbf{vec}(\vo{N}) = \vo{P}_k\vo{q},
%\end{aligned}\hspace{5mm} \right\}
%\eqnlabel{bksk}
%\end{equation}
\begin{equation}\left.
\begin{aligned}
\bv &:= \vo{N}\vo{C}_b^T\vo{\theta}_k =  \vo{X}_k\vo{q},\\
\bcg &:= \bar{\vo{X}}_k\vo{q},\\
\vo{s}_k &:= \vo{N}\vo{C}_s^T\vo{\eta}_k = \vo{Y}_k\vo{q},\\
\vo{p}_k &:= \vo{N}\vo{L}_{p_m}^T\vo{\phi}_k = \vo{P}_k\vo{q},
\end{aligned}\hspace{5mm} \right\}
\eqnlabel{bksk}
\end{equation}
where
\begin{align}
\vo{X}_k & :=  \left((\vo{\theta}_k^T\vo{C}_b)\otimes \vo{I}_3\right),\\
\bar{\vo{X}}_k & := \frac{1}{2} \left((\vo{\theta}_k^T|\vo{C}_b|)\otimes \vo{I}_3\right),\\
\vo{Y}_k & := \left((\vo{\eta}_k^T\vo{C}_s)\otimes \vo{I}_3\right),\label{eqn:Yk}\\
\vo{P}_k & := \left((\vo{\phi}_k^T\vo{L}_{p_m})\otimes \vo{I}_3\right),
\end{align}
and $\vo{q}:=\mathbf{vec}(\vo{N})$ represents the Cartesian coordinates and $\otimes$ denotes the Kronecker product.
\item Let $\vo{F}\in\real^{3\times n}$ be the non-conservative force matrix defined by
\begin{align*}
\vo{F}:= \begin{bmatrix}\vo{f}_1 & \vo{f}_2 & \cdots & \vo{f}_n\end{bmatrix},
\end{align*}
where $\vo{f}_i\in \real^{3}$ is the total force acting on the $\ith{i}$ node, and accordingly, the force matrix can be vectorized as $\vo{f}:=\vec{\vo{F}}\in \real^{3n}$. Here we assume a general condition where all the nodes have external forces acting on them. In practice, all nodes may not be loaded. We can set those $f_i$ to zero in the above expression. These external forces can be used to model disturbances and other loads acting on the tensegrity structure.
\end{enumerate}

\subsection{Kinematics}
Consider the motion of $\ith{k}$ bar defined by nodes $\bv :=\vo{n}_{j_k}-\vo{n}_{i_k}$. The center of mass of the bar is given by
\begin{align} \bcg := \frac{\vo{n}_{j_k}+\vo{n}_{i_k}}{2}, \end{align}
and its velocity is given by
\begin{align} \bcgd := \frac{\dot{\vo{N}}_{j_k}+\dot{\vo{N}}_{i_k}}{2}. \end{align}

To determine the angular velocity of the bar we first relate the velocities of $\vo{n}_{j_k}$ and $\vo{n}_{i_k}$ using
\begin{align*}
\dot{\vo{n}}_{j_k} = \dot{\vo{n}}_{i_k} + \vo{\omega}_k \times \bv,
\end{align*}
or
\begin{align*}
\bvd := \dot{\vo{n}}_{j_k} - \dot{\vo{n}}_{i_k} = \vo{\omega}_k \times \bv.
\end{align*}
Taking cross product with $\bv$ on both sides we get
\begin{align*}
\bv\times \bvd = \bv \times (\vo{\omega}_k \times \bv).
\end{align*}
Using the result from triple cross product
\begin{align*}
\vo{a}\times (\vo{b}\times \vo{c}) = \vo{b}(\vo{a} \cdot \vo{c}) - \vo{c}(\vo{a} \cdot \vo{b}),
\end{align*}
we get
\begin{equation}
\bv\times \bvd = \vo{\omega}_k (\bv\cdot \bv) - \bv(\vo{\omega}_k \cdot \bv). \eqnlabel{omega}
\end{equation}

For tensegrity systems, $\vo{\omega}_k \cdot \bv = 0$, i.e. the bar does not spin about its body axis. This is an important difference between tensegrity systems and general multi-body systems.

%In \eqn{omega}, $\vo{\omega}_k \cdot \bv$ need not be zero. That equation simply eliminates the axial component of angular velocity. The expression $\bv\times \bvd$ accounts for the nonaxial angular velocity. If we were to use this approach to derive dynamics for a general multibody system, we would introduce additional variable to include the axial component, which is independent of $\bv$ and $\bvd$.

Therefore, for tensegrity systems, we can write the expression for angular velocity
\begin{align*}
\vo{\omega}_k = \frac{\bv\times \bvd}{\bv^T\bv}.
\end{align*}
Noting that $\bv^T\bv = l_{\vo{b}_k}^2$, where $l_{\vo{b}_k}$ is the length of the bar and is a constant, we can write
\begin{align*}
\vo{\omega}_k = \frac{\bv\times \bvd}{l_{\vo{b}_k}^2}.
\end{align*}

Let the body axis be defined by $(\vo{\hat{b}}_k, \, \vo{\hat{b}}_2,\: \vo{\hat{b}}_3)$.
We can then write the angular velocity in terms of the body axis of the bar as $\vo{\omega}_k := \omega_2 \vo{\hat{b}}_2 + \omega_3 \vo{\hat{b}}_3$  where $\omega_2$, $\omega_3$ are respective components.

Assuming, the bar to be a cylinder with radius $r_k$ and length $l_{\vo{b}_k}$, the moment of inertia of the rod in this body-fixed principal frame is
\begin{align*}
\vo{I}_{\vo{b}_k} := \textbf{diag} \begin{bmatrix} \frac{m_{\vo{b}_k} r^2}{2} & \frac{m_{\vo{b}_k}}{12}(3r_k^2+l_{\vo{b}_k}^2) & \frac{m_{\vo{b}_k}}{12}(3r_k^2+l_{\vo{b}_k}^2) \end{bmatrix}.
\end{align*}
The angular momentum $h_k$ of the bar is therefore
\begin{align*}
	\vo{h}_k &:= \vo{I}_{\vo{b}_k} \begin{bmatrix} 0 & \omega_2 &\omega_3 \end{bmatrix}^T\\
	& =  \frac{m_{\vo{b}_k}}{12}(3r_k^2+l_{\vo{b}_k}^2)( \omega_2 \hat{\vo{b}}_2 \ +\omega_3 \hat{\vo{b}}_3)\\
	& = \frac{m_{\vo{b}_k}}{12}(3r_k^2+l_{\vo{b}_k}^2)\;\vo{\omega}_k,\\
	& = \frac{(3r_k^2+l_{\vo{b}_k}^2)m_{\vo{b}_k}}{12 l_{\vo{b}_k}^2}\,\bv\times \bvd.
\end{align*}
If $\vo{r}_k$ can be ignored, then
$\vo{h}_k \approx \frac{m_{\vo{b}_k}}{12}\,\bv\times \bvd.$
Often, hollow cylinders are used.
In that case, we can substitute the appropriate inertia matrix in the expression for angular momentum.

The inertial position coordinates of $\ith{k}$ point mass are given by $\vo{p}_k  := \vo{n}_{i_k}$
and its velocity given by $\dot{\vo{p}_k}  := \dot{\vo{n}}_{i_k}$.

\subsection{Dynamics Using Lagrangian Approach}
Let $\mathcal{L}:=T-V$ be the Lagrangian, defined over coordinates
$\vo{q}$, with components $q_i$. The equations of motion are then given by
\begin{align*}
\frac{d}{dt}\left(\frac{\partial \mathcal{L}}{\partial \dot{q}_i}\right) - \frac{\partial \mathcal{L}}{\partial q_i} - \vo{\lambda}^T(t) \frac{\partial \vo{R}(\vo{q})}{\partial q_i} = \vo{f}^T \frac{\partial \q}{\partial q_i}
%\hat{\vo{q}}_i,
\end{align*}
where $\vo{R}(\vo{q}): \real^{3n}\mapsto\real^m =\vo{0}$ depict ideal constraints that satisfy the principle of D'Alembert, first stated by Lagrange \cite{lagrange1853mecanique}. On the right, $\vo{f}$ is the non conservative force acting on the system such as externally applied forces, damper forces or disturbances. From the definition of the coordinate $\q$, one can notice that $\frac{\partial \q}{\partial q_i}$ is the $\ith{i}$ column of an identity matrix $\vo{I}_{3n}$.
%where $\vo{R}(\vo{q}): \real^{3n}\mapsto\real^m =\vo{0}$ are the holonomic constraints, $\vo{f}$ is the non conservative force acting on the system and $\hat{\vo{q}}_i$ is the unit vector along coordinate $q_i$. The expression, $\vo{F}\cdot \hat{\vo{q}}_i$ represents the component of $\vo{F}$ along $\hat{\vo{q}}_i$. This term is equivalent to $\frac{\partial (\vo{F}\cdot \vo{q}_i)}{\partial q_i}$.
We can therefore write the equation of motion as
\begin{align*}
\frac{d}{dt}\left(\frac{\partial \mathcal{L}}{\partial \dot{q}_i}\right) - \frac{\partial \mathcal{L}}{\partial q_i} - \vo{\lambda}^T(t) \frac{\partial \vo{R}(\vo{q})}{\partial q_i} = f_i,
\end{align*}
where $f_i$ is the $\ith{i}$ element of $\vo{f}$.

Substituting $\mathcal{L}:=T-V$, we get the equations of motion
\begin{align*}
\frac{d}{dt}\left(\frac{\partial T}{\partial \dot{q}_i}\right) + \frac{\partial}{\partial q_i}\left(V - \vo{\lambda}^T\vo{R}(\vo{q})\right) = f_i,
\end{align*}
for $i=1,\cdots,3n$; or in terms of $\vo{q}$ as
\begin{align}
	\frac{d}{dt}\left(\frac{\partial T}{\partial \dot{\vo{q}}}\right) + \frac{\partial}{\partial \vo{q}}\left(V - \vo{\lambda}^T\vo{R}(\vo{q})\right) = \vo{f}^T.
	\label{eqn:Lagrange}
\end{align}

\subsubsection{Total Kinetic Energy}
Total kinetic energy of the system is
\begin{align*}
T :=& \sum_{k=1}^{n_b}\left(\frac{1}{2}m_{\vo{b}_k}\bcgd \cdot  \bcgd + \frac{1}{2} \vo{h}_k \cdot\vo{\omega}_k\right)\\
&+ \sum_{k=1}^{n_{p_m}}\left(\frac{1}{2}m_{\vo{p}_k} \dot{\vo{p_k}} \cdot \dot{\vo{p_k}}\right)
\end{align*}
The kinetic energy of the $k^\text{th}$ bar is
\begin{align*}
T_{\vo{b}_k} & = \frac{1}{2}\left(m_{\vo{b}_k} \bcgd^T\bcgd + \frac{I_{\vo{b}_k}}{l_k^4}(\bv\times\bvd)\cdot(\bv\times\bvd)\right),
\end{align*}
where $I_{\vo{b}_k}:=\frac{m_{\vo{b}_k}}{12}(3r_k^2+l_{\vo{b}_k}^2)$. Simplifying
\begin{align*}
(\bv\times\bvd)\cdot(\bv\times\bvd) =& (\bv\cdot\bv)(\bvd\cdot\bvd)\\
&- (\bv\cdot\bvd)(\bvd\cdot\bv)\\
=& l_{\vo{b}_k}^2(\bvd\cdot\bvd),
\end{align*}
we get
\begin{align*}
T_{\vo{b}_k} = \frac{1}{2}\left(m_{\vo{b}_k} \bcgd\cdot\bcgd + \frac{I_{\vo{b}_k}}{l_{\vo{b}_k}^2}\bvd\cdot\bvd\right).
\end{align*}

Using \eqn{bksk}, we can write $T_{\vo{b}_k}$ in terms of $\dot{\vo{q}}$ as
\begin{align}
\nonumber T_{\vo{b}_k} & = \frac{1}{2}\dot{\vo{q}}^T\underbrace{\left[m_{\vo{b}_k}\bar{\vo{X}}_k^T\bar{\vo{X}_k}
+ \frac{I_{\vo{b}_k}}{l_{\vo{b}_k}^2}\vo{X}_k^T\vo{X}_k\right]}_{:=\vo{M}_{\vo{b}_k}}\dot{\vo{q}} =
  \frac{1}{2}\dot{\vo{q}}^T\vo{M}_{\vo{b}_k}\dot{\vo{q}}.
%_{\vo{b}_k}\dot{\vo{q}},
\end{align}

The kinetic energy of the $k^\text{th}$ point mass is
$$
T_{\vo{p}_k} =  \frac{1}{2}\left(m_{\vo{p}_k} \dot{\vo{p_k}} \cdot \dot{\vo{p_k}}\right)
$$
Using \eqn{bksk}, we can write $T_{\vo{p}_k}$, in terms of $\dot{\vo{q}}$ as
\begin{align}
	\nonumber T_{\vo{p}_k} & = \frac{1}{2}\dot{\vo{q}}^T\underbrace{\left[m_{\vo{p}_k}{\vo{P}}_k^T{\vo{P}_k}
	\right]}_{:=\vo{M}_{\vo{p}_k}}\dot{\vo{q}} =  \frac{1}{2}\dot{\vo{q}}^T\vo{M}_{\vo{p}_k}\dot{\vo{q}}
\end{align}

\begin{align}
	\implies T &= \nonumber \sum_{k=1}^{n_b} T_{\vo{b}_k} + \sum_{k=1}^{n_{p_m}} T_{\vo{p}_k}  \\
	\nonumber &=  \frac{1}{2}\dot{\vo{q}}^T \underbrace{\left(\sum_{k=1}^{n_b}\vo{M}_{\vo{b}_k}\right)}_{:=\vo{M_b}}\dot{\vo{q}} +
	 \frac{1}{2}\dot{\vo{q}}^T\underbrace{\left(\sum_{k=1}^{n_{p_m}} \vo{M}_{\vo{p}_k}\right)}_{:=\vo{M_p}}\dot{\vo{q}} \\
	\nonumber
	&= \frac{1}{2}\dot{\vo{q}}^T\underbrace{\left(\vo{M_b}+\vo{M_p}\right)}_{:=\vo{M}}\dot{\vo{q}} \\
	&=\frac{1}{2}\dot{\vo{q}}^T\vo{M}\dot{\vo{q}}. \eqnlabel{T}
\end{align}

\subsubsection{Gravity Potential Energy}
Total gravitational potential energy of the system is
\begin{align}
V_g &:= -\sum_{k=1}^{n_b} m_{\vo{b}_k} (\vo{g} \cdot \bcg) - \sum_{k=1}^{n_{p_m}} m_{\vo{p}_k} (\vo{g} \cdot \vo{p}_k)\notag\\
 &= -\underbrace{\vo{g}^T\left(\sum_{k=1}^{n_b}m_{\vo{b}_k}\bar{\vo{X}}_k +
  \sum_{k=1}^{n_{p_m}}m_{\vo{p}_k} \vo{P}_k\right)}_{:= \vo{G}^T}\vo{q} = -\vo{G}^T\vo{q},
\eqnlabel{Vg}
\end{align}
where $\vo{g}:=\begin{bmatrix}0 & 0 & -9.806\end{bmatrix}^T$ is the gravity vector.

\subsubsection{Potential Energy of Strings Modeled as Springs}
We can model the strings as springs. In this case, the spring energy is
\begin{align}
V_s &:= \frac{1}{2} \sum_{k=1}^{n_s} K_k \left(\|\vo{s}_k\| - l_{s_k}\right)^2,
\label{eqn:Vs_original}
\end{align}
adds to the potential energy of the system. In this case, $l_{s_k}$ is the natural length of the spring and $K_k$ is the spring constant. In this formulation, we have to be mindful about $\|\vo{s}_k\| - l_{s_k} \geq 0$, because the strings can only exert tensile force (unidirectional), unlike regular springs.
Force density $\sigma_k$, is defined as
\begin{align}
\sigma_k := K_k\left(1- \frac{l_{s_k}	}{\|\vo{s}_k\|}\right),
\end{align}
which is the control variable.  In the implementation, if the condition $\|\vo{s}_k\| - l_{s_k} \geq 0$ is violated for any string at any point in time, the corresponding force density is set to zero at that instant.

The spring energy in terms of $\sigma_k$ can be written as
\begin{align}
V_s :&= \frac{1}{2} \sum_{k=1}^{n_s}\left(\frac{\sigma_k^2}{K_k}\right) \|\vo{s}_k\|^2\notag\\
& = \frac{1}{2} \sum_{k=1}^{n_s}\left(\frac{\sigma_k^2}{K_k}\right)\vo{s}_k^T\vo{s}_k\notag\\
& = \frac{1}{2} \vo{q}^T\left(\sum_{k=1}^{n_s} \frac{\sigma_k^2}{K_k}\vo{Y}_k^T\vo{Y}_k\right)\vo{q}\notag\\
& = \frac{1}{2} \vo{q}^T\left(\underbrace{\begin{bmatrix}\vo{Y}_1^T\vo{Y}_1 & \cdots & \vo{Y}_{n_s}^T\vo{Y}_{n_s}\end{bmatrix}}_{:=\vo{Y}}\left(\frac{\vo{\sigma}^2}{\vo{K}} \otimes \vo{I}_{3n} \right)\right)\vo{q},\notag\\
& =  \frac{1}{2} \vo{q}^T \vo{Y}\left(\frac{\vo{\sigma}^2}{\vo{K}} \otimes \vo{I}_{3n} \right) \vo{q}
\eqnlabel{Vs}
\end{align}
%We define control variables
%\begin{align}
%	u_k := \left(\frac{\sigma_k^2}{K_k}\right),
%\end{align}
%which results in
%\begin{align}
%\nonumber V_s &:= \frac{1}{2} \sum_{k=1}^{n_s}u_k \|\vo{s}_k\|^2,\\
%\nonumber & = \frac{1}{2} \sum_{k=1}^{n_s} u_k\vo{s}_k^T\vo{s}_k,\\
%\nonumber & = \frac{1}{2} \vo{q}^T\left(\sum_{k=1}^{n_s} u_k\vo{Y}_k^T\vo{Y}_k\right)\vo{q},\\
%\nonumber & = \frac{1}{2} \vo{q}^T\left(\underbrace{\begin{bmatrix}\vo{Y}_1^T\vo{Y}_1 & \cdots & \vo{Y}_{n_s}^T\vo{Y}_{n_s}\end{bmatrix}}_{:=\vo{K}}(\vo{\sigma} \otimes \vo{I}_{3n} )\right)\vo{q},\\
%& =  \frac{1}{2} \vo{q}^T \vo{K}(\vo{\sigma} \otimes \vo{I}_{3n} ) \vo{q}
% \eqnlabel{Vs}
%\end{align}
%where $\vo{\sigma} := \begin{bmatrix}u_1 & \cdots & u_{n_s}\end{bmatrix}^T$.
where $\vo{\sigma}^2 := \begin{bmatrix}\sigma_1^2 & \cdots & \sigma_{n_s}^2\end{bmatrix}^T$, $\vo{K}:=\begin{bmatrix} K_1 &\cdots & K_{n_s} \end{bmatrix}$.

%\subsection{Work by General External Forces at Nodes}
%Work done by external forces $\vo{F}_k\in \real^3$ acting at nodes $\vo{N}_k$
%$$
%Q_{\vo{F}} := \sum_{k=1}^{n} \vo{F}_k \cdot \vo{N}_k.
%$$
%
%Let $\vo{F}:= \begin{bmatrix}\vo{F}_1 & \cdots & \vo{F}_n\end{bmatrix} \in \real^{3\times n}$,
%then $Q_{\vo{F}}$ can be written as
%\begin{equation}
%Q_{\vo{F}} := \mathbf{vec}(\vo{F})^T \mathbf{vec}(\vo{N}) =  \mathbf{vec}(\vo{F})^T\vo{q}. \eqnlabel{QF}
%\end{equation}
%
%
%Here we assume a general condition where all the nodes have external forces acting on them. In practice, all nodes may not be loaded. We can set those $\vo{F}_k$ to zero in the above expression. These external forces can be used to model disturbances and other loads acting on the tensegrity structure.

\subsubsection{Damper force}
We assume a damper force  between two nodes where the string/spring exists and the force is proportional to the changing rate of the string/spring length. Thus the $k^\text{th}$ damper can be modeled as
\begin{align}
\vo{f}_{d,k}&=-c\frac{d\|\vo{s}_k\|}{dt}\frac{\vo{s}_k}{\|\vo{s}_k\|},\\
&=-c\frac{\vo{s}_k^T\dot{\vo{s}}_k}{\|\vo{s}_k\|}\frac{\vo{s}_k}{\|\vo{s}_k\|},\\
&=-c\frac{(\dot{\vo{s}}_k^T\vo{s}_k)\vo{s}_k}{\vo{s}_k^T\vo{s}_k}
\end{align}
where $c$ is the damping coefficient and the direction of the force is always parallel to the string/spring. One should notice that the damper force disappears whenever the string is slack, that is $\vo{f}_{d,k}=0$ if $\|\vo{s}_k\| - l_{s_k} \leq 0$, but the damper force always exists in the spring. To represent the total damper force acting on a node, one can utilize \eqref{eqn:Yk} and obtain $\vo{f}_d=\sum_{k=1}^{n_s}\vo{Y}_k^T\vo{f}_{d,k}$. Considering damper force as one of the members in external force, it can be added to $\vo{f}$ in \eqref{eqn:Lagrange}.

\subsubsection{Equations of Motion}
We are now ready to derive the equations of motion. From \eqn{T}, we have
\begin{align*}
\frac{d}{dt}\left(\frac{\partial T}{\partial \dot{\vo{q}}}\right) = \ddot{\vo{q}}^T\vo{M},
\end{align*}
from \eqn{Vg}, we have
\begin{align*}
\frac{\partial V_g}{\partial \vo{q}} = -\vo{G}^T,
\end{align*}
and finally from \eqn{Vs}, we have
\begin{align}
\frac{\partial V_s}{\partial \vo{q}} &= \frac{\partial}{\partial \vo{q}}\left(\frac{1}{2} \sum_{k=1}^{n_s} \frac{\sigma_k^2}{K_k}\vo{s}_k^T\vo{s}_k\right)\notag\\
&= \frac{1}{2} \sum_{k=1}^{n_s}\left( \frac{\partial}{\partial \vo{q}}\left( \frac{\sigma_k^2}{K_k} \right)\vo{s}_k^T\vo{s}_k+2\frac{{\sigma_k}^2}{K_k}\vo{s}_k^T\frac{\partial \vo{s}_k}{\partial \vo{q}} \right)\notag\\
&= \frac{1}{2} \sum_{k=1}^{n_s}\left( 2l_{s_k}\sigma_k\frac{\vo{s}_k^T}{\|\vo{s}_k\|^3}\vo{Y}_k\|\vo{s}_k\|^2+2\frac{{\sigma_k}^2}{K_k}\vo{s}_k^T\frac{\partial \vo{s}_k}{\partial \vo{q}} \right)\notag\\
&= \sum_{k=1}^{n_s}\left( l_{s_k}\sigma_k\frac{\vo{s}_k^T}{\|\vo{s}_k\|}\vo{Y}_k+\frac{{\sigma_k}^2}{K_k}\vo{s}_k^T\frac{\partial \vo{s}_k}{\partial \vo{q}} \right)\notag\\
&= \sum_{k=1}^{n_s} \sigma_k \left(  l_{s_k}\frac{\vo{s}_k^T}{\|\vo{s}_k\|}+\left(1-\frac{l_{s_k}}{\|\vo{s}_k\|}\right)\vo{s}_k^T \right)\vo{Y}_k\notag\\
&= \sum_{k=1}^{n_s} \sigma_k\vo{q}^T\vo{Y}_k^T\vo{Y}_k\notag\\
&= \vo{q}^T\vo{Y}\left(\vo{\sigma} \otimes \vo{I}_{3n} \right)\eqnlabel{dVs_dq}
\end{align}

%$$
%\frac{\partial V_s}{\partial \vo{q}} = \vo{q}^T\vo{K}(\vo{\sigma} \otimes \vo{I}_{3n} ),
%$$
%and finally from \eqn{QF}, we have
%$$
%\frac{\partial Q_{\vo{F}}}{\partial \vo{q}} = \mathbf{vec}(\vo{F})^T.
%$$

Therefore, the equations of motion are given by
\begin{align*}
\ddot{\vo{q}}^T\vo{M} -\vo{G}^T + \vo{q}^T\vo{Y}(\vo{\sigma} \otimes \vo{I}_{3n} ) - \vo{\lambda}^T\frac{\partial \vo{R}}{\partial \vo{q}} = \vo{f}^T,
\end{align*}
or with transpose
\begin{align}
\vo{M}\ddot{\vo{q}} - \left(\frac{\partial \vo{R}}{\partial \vo{q}}\right)^T\vo{\lambda} =  -
(\vo{\sigma}^T \otimes \vo{I}_{3n})\vo{Y}^T\vo{q} + \vo{G} + \vo{f}.
\eqnlabel{dynamics}
\end{align}

We next look at the constraint equation $\vo{R}(\vo{q}) = \vo{0}$, and compute
\begin{align*}
\frac{d^2\vo{R}(\vo{q})}{dt^2} & = \frac{d}{dt}\left(\frac{d\vo{R}(\vo{q})}{dt}\right),\\
& =  \frac{d}{dt} \left(\frac{\partial \vo{R}}{\partial \vo{q}}\dot{\vo{q}}\right),\\
& = \left(\frac{\partial \vo{R}}{\partial \vo{q}}\right)\ddot{\vo{q}} + \begin{bmatrix} \dot{\vo{q}}^T \left(\frac{\partial^2 R_1}{\partial \vo{q}^2}\right)\dot{\vo{q}} \\
\vdots \\
\dot{\vo{q}}^T \left(\frac{\partial^2 R_m}{\partial \vo{q}^2}\right)\dot{\vo{q}}
\end{bmatrix},
\end{align*}
where $\left(\frac{\partial \vo{R}}{\partial \vo{q}}\right)$ is a Jacobian of $\vo{R}(\vo{q})$ and $\left(\frac{\partial^2 R_i}{\partial \vo{q}^2}\right)$ is the Hessian of $R_i(\vo{q})$.
Therefore, $\frac{d^2\vo{R}(\vo{q})}{dt^2} = 0$ implies
\begin{align}
-\left(\frac{\partial \vo{R}}{\partial \vo{q}}\right)\ddot{\vo{q}} & =   \begin{bmatrix} \dot{\vo{q}}^T \left(\frac{\partial^2 R_1}{\partial \vo{q}^2}\right)\dot{\vo{q}} \\
\vdots \\
\dot{\vo{q}}^T \left(\frac{\partial^2 R_m}{\partial \vo{q}^2}\right)\dot{\vo{q}}
\end{bmatrix}. \eqnlabel{constraints}
\end{align}
Combining \eqn{dynamics} and \eqn{constraints}, we get the final equation
\begin{align}
\begin{bmatrix}
\vo{M} & - \left(\frac{\partial \vo{R}}{\partial \vo{q}}\right)^T \\[2mm]-\left(\frac{\partial \vo{R}}{\partial \vo{q}}\right)  & \vo{0}
\end{bmatrix}&\begin{pmatrix}\ddot{\vo{q}} \\ \vo{\lambda}\end{pmatrix} =\notag\\
&\begin{bmatrix}
 -(\vo{\sigma}^T \otimes \vo{I}_{3n})\vo{Y}^T\vo{q} + \vo{G} + \vo{f}\\
 \dot{\vo{q}}^T \left(\frac{\partial^2 R_1}{\partial \vo{q}^2}\right)\dot{\vo{q}} \\
\vdots \\
\dot{\vo{q}}^T \left(\frac{\partial^2 R_m}{\partial \vo{q}^2}\right)\dot{\vo{q}}
\end{bmatrix}. \eqnlabel{finalEqn}
\end{align}

Defining,
\begin{align*}
\Rq &:= \frac{\partial \vo{R}}{\partial \vo{q}},\\
\vo{\xi}_1 &:=  -(\vo{\sigma}^T \otimes \vo{I}_{3n})\vo{Y}^T\vo{q} + \vo{G} + \vo{f},\\
\vo{\xi}_2 &:= \begin{bmatrix}
\dot{\vo{q}}^T \left(\frac{\partial^2 R_1}{\partial \vo{q}^2}\right)\dot{\vo{q}} \\
\vdots \\
\dot{\vo{q}}^T \left(\frac{\partial^2 R_m}{\partial \vo{q}^2}\right)\dot{\vo{q}}
\end{bmatrix},
\end{align*}
we can analytically express $\ddot{\vo{q}}$ and $\vo{\lambda}$ as
\begin{align}
\ddot{\vo{q}} &= \vo{M}^{-1}\left[\vo{\xi}_1-\Rq^T\left(\Rq\vo{M}^{-1}\Rq^T\right)^{-1}\right.\notag\\
& \qquad\qquad\qquad\qquad\qquad\qquad\qquad\left.\left(\vo{\xi}_2+\Rq\vo{M}^{-1}\vo{\xi}_1\right)\right]\notag\\
&=\vo{\xi}(\q,\dot{\q},\vo{\sigma},\vo{f}),\label{eqn:qdd}\\
\vo{\lambda} & = -\left(\Rq\vo{M}^{-1}\Rq^T\right)^{-1}\left(\vo{\xi}_2+\Rq\vo{M}^{-1}\vo{\xi}_1\right).
\end{align}

In this formulation, numerical difficulties may occur when solving the above equations of motion. Here we assume that the mass matrix $\vo{M}$ is invertible since the kinetic energy is always positive. Small inertia can also cause numerical ill conditioning.
In addition, redundant constraints can also cause singularity in $\Rq\vo{M}^{-1}\Rq^T$.
% \textcolor{blue}{
Such constraints can further lead to non-uniqueness of computed reaction forces and accelerations. These results may be unit-sensitive as well \cite{wojtyra2020utilization}. In such cases, a weighted Moore-Penrose generalized inverse may be used in a meaningful manner to derive the force distribution on the constrained structure \cite{liu2017weighted}.
% }

\subsubsection{Ideal Constraints}
Ideal constraints, as stated earlier, are those that satisfy D'Alembert's principle. In the current derivation, we only consider holonomic constraints, that is, they reduce the dimension of the space of accessible configurations, but do not restrict motion and paths within the reduced dimension \cite{Papastavridis02}. Mathematically, the constraint equations can be expressed as $\vo{R}(\vo{q}) = \vo{0}$, where $\vo{q}$ is a function of time. Commonly constraints will include bar-length constraints that are quadratic in $\vo{q}$, and boundary conditions on $\vo{q}$ that will be linear in $\vo{q}$. Bar length constraints are of the type $\bv^T\bv-l_{\vo{b}_k}^2 = 0$, which in terms of $\vo{q}$ are $\vo{q}^T\vo{X}_k^T\vo{X}_k\vo{q} - l_{\vo{b}_k}^2 = 0$.

Therefore, for these two cases
\begin{align}
& \vo{R}(\vo{q}) := \begin{bmatrix}\vo{A}\vo{q}-\vo{b} \\ \vo{q}^T\vo{X}_1^T\vo{X}_1\vo{q} - l_{b_1}^2\\ \vdots \\ \vo{q}^T\vo{X}_{n_b}^T\vo{X}_{n_b}\vo{q}-l_{b_{n_b}}^2
\end{bmatrix} = 0.
\label{eqn:R}
\end{align}
Therefore,
\begin{align}
 \Rq := \left(\frac{\partial \vo{R}}{\partial \vo{q}}\right) = \begin{bmatrix}
\vo{A} \\  2\vo{q}^T\vo{X}_1^T\vo{X}_1 \\ \vdots \\ 2\vo{q}^T\vo{X}_{n_b}^T\vo{X}_{n_b}
\end{bmatrix},
\label{eqn:Jacobian}
\end{align}
and
\begin{align*}
 \begin{bmatrix} \dot{\vo{q}}^T \left(\frac{\partial^2 R_1}{\partial \vo{q}^2}\right)\dot{\vo{q}} \\
\vdots \\
\dot{\vo{q}}^T \left(\frac{\partial^2 R_m}{\partial \vo{q}^2}\right)\dot{\vo{q}}
\end{bmatrix} = \begin{bmatrix}
\vo{0}\\ 2\dot{\vo{q}}^T \vo{X}_1^T\vo{X}_1\dot{\vo{q}} \\ \vdots \\  2\dot{\vo{q}}^T \vo{X}_{n_b}^T\vo{X}_{n_b}\dot{\vo{q}}
\end{bmatrix}.
\end{align*}

\subsection{DAE Correction}
The holonomic constraints are converted to differential equations by differentiating them twice. This results in constraints on acceleration, which are satisfied exactly. However, the position and velocity constraints get violated due to errors in numerical integration. In addition to the constraints, numerical errors also violate energy conservation. For this reason, inspired by the direct correction approach in \cite{marques2017constraints} and
\cite{yu2000direct}, we use the idea of constraint variations and derive a system of linear equations to correct for errors in numerical integration. To account for energy conservation, we also include variation in the total system energy in the formulation.

Considering the vector of coordinates and its time derivative that need to be corrected for the original constraints,
\begin{align}
	\q^c = \q^u + \delta \q,
	\label{eqn:qc}
\end{align}
where $\q^u$ denotes the uncorrected position, obtained from numerical integration, $\q^c$ the corrected position, and $\delta \q$ is the correction required to satisfy the constraint.
Therefore
\begin{align}
\vo{R}(\q^c) = \vo{R}(\q^u + \delta \q)  = \vo{R}(\q^u) + \Rq\delta \q = \vo{0},
\label{eqn:R_variation}
\end{align}
where $\Rq$ is a Jacobian matrix defined in \eqref{eqn:Jacobian}.

Similarly, the time derivative of the holonomic constraint should satisfy
\begin{align}
\frac{d\vo{R}}{dt}=\frac{\partial \vo{R}}{\partial \q}\dot{\q}=\vo{0}.
\label{eqn:dRdt}
\end{align}

With
\begin{align}
	\dot{\q}^c=\dot{\q}^u+\delta \dot{\q},
		\label{eqn:qcdot}
\end{align}
we get
\begin{align}
\left[\vo{R}_{q}(\q^u)+\frac{\partial\vo{R}_{\q}}{\partial \q}\delta\q\right]\left(\dot{\q}^u+\delta\dot{\q}\right)=\vo{0}, \eqnlabel{constr_deriv}
\end{align}
where $\frac{\partial\vo{R}_{\q}}{\partial \q}$ is a third-order tensor and can be represented in a matrix  form
\begin{align*}
\frac{\partial\vo{R}_{\q}}{\partial \q}\delta\q=\left[ \frac{\partial\vo{R}_{\q}}{\partial q_1}\delta \q \quad \cdots\quad \frac{\partial\vo{R}_{\q}}{\partial q_{3n}}\delta \q \right].
%\frac{\partial\vo{R}_{\q}}{\partial \q}\delta\q=\sum_{i=1}^{3n}\frac{\partial\vo{R}_{\q}}{\partial q_i}\delta q_i.
\end{align*}
Ignoring higher order terms in \eqn{constr_deriv}, we get
\begin{align}
\vo{R}_{\q}\dot{\q}^u+\vo{R}_{\q}\delta\dot{\q}+\left(\frac{\partial\vo{R}_{\q}}{\partial \q}\delta\q\right)\dot{\q}^u=\vo{0},
%\vo{R}_{\q}\dot{\q}^u+\vo{R}_{\q}\delta\dot{\q}+\sum_{i=1}^{3n}\frac{\partial\Rq}{\partial q_i}\delta q_i\dot{\q}^u=0,
\label{eqn:dRdt2}
\end{align}
where the third term can be reformulated as
\begin{align*}
\left(\frac{\partial\vo{R}_{\q}}{\partial \q}\delta\q\right)\dot{\q}^u&=\left[ \frac{\partial\vo{R}_{\q}}{\partial q_1}\delta \q \quad \cdots\quad \frac{\partial\vo{R}_{\q}}{\partial q_{3n}}\delta \q \right]\dot{\q}^u,\\
&= \sum_{i=1}^{3n}\frac{\partial\Rq}{\partial q_i}\dot{q}^u_i\delta\q,\\
&=\left(\sum_{i=1}^{3n}\frac{\partial\Rq}{\partial q_i}\dot{\q}^u\vo{\alpha}_i\right)\delta \q,\\
&=\vo{Q}\delta \q,
\end{align*}
where $\vo{Q} := \left(\sum_{i=1}^{3n}\frac{\partial\Rq}{\partial q_i}\dot{\q}^u\vo{\alpha}_i\right)$, and $\vo{\alpha}_i$ is the $\ith{i}$ column of the identity matrix $\vo{I}_{3n}$.

Then \eqref{eqn:dRdt2} becomes
\begin{align}
\Rq\dot{\q}^u+\Rq\delta\dot{\q}+\vo{Q}\delta \q=\vo{0}.
\label{eqn:dRdt_variation}
\end{align}

Combining \eqref{eqn:R_variation} and \eqref{eqn:dRdt_variation}, we obtain the following system of linear equations
\begin{align}
\begin{bmatrix}
\Rq & \vo{0}\\
\vo{Q} & \Rq
\end{bmatrix}
\begin{bmatrix}
\delta \q\\ \delta \dot{\q}
\end{bmatrix}
=
\begin{bmatrix}
-\vo{R}\\-\Rq\dot{\q}^u
\end{bmatrix}.
\label{eqn:correction_direct}
\end{align}

Since the matrix in \eqref{eqn:correction_direct} has fewer rows than columns ($2m < 6n$), it doesn't have full column rank and there exist infinite solutions. In \cite{marques2017constraints}, the author applied Moore-Penrose inverse to minimize the  $2$-norm of the solution. However, the corrections in $\delta \q$ and $\delta \dot{\q}$ from such a formulation, modifies the potential and kinetic energy of the system and violates the conservation of mechanical energy. In this work, we extend the work in \cite{marques2017constraints}, by explicitly constraining the energy change, due to $\delta \q$ and $\delta \dot{\q}$, to be zero. This will result in an additional linear equation in $\delta \q$ and $\delta \dot{\q}$. The derivation of that constraint equation is as follows. We consider a general formulation, where the work done by external forces are accounted for.

The total energy of the system is defined as
\begin{align}
E(\q,\dot{\q}) := T(\qd) + V_s(\q) + V_g(\q),
\eqnlabel{totEnergy}
\end{align}
and energy conservation states that the total energy at any time $t$ is the sum of the total energy at initial time and the work done by non conservative forces, i.e.
\begin{align}
E(\q,\dot{\q}) = E(\q_0,\dot{\q}_0) + \int_{\mathit{C}}\vo{f}\cdot d\q,
\eqnlabel{conservation}
\end{align}
where $(\q_0,\dot{\q}_0)$ is the initial condition, $\vo{f}$ is the external force, which can be either state or time dependent, and the integration is done over path $C$ connecting $\q_0$ to $\q$.

Let the work done by force $\vo{f}$ be $W_{\vo{f}}$, i.e.
\begin{align}
W_{\vo{f}}&=\int_{\mathit{C}}\vo{f}\cdot d\q\notag ,\\
&=\int_{t_0}^{t}\vo{f}^T\frac{d\q(\tau)}{d\tau}d\tau.
\eqnlabel{WF}
\end{align}

Here we treat $W_{\vo{f}}$ as an additional state variable, and augment the state-dynamics in \eqn{qdd}, with
\begin{align}
\dot{W_{\vo{f}}} = \vo{f}^T\dot{\q(t)}. \eqnlabel{workState}
\end{align}
Integration of \eqn{workState} results in the time evolution of $W_{\vo{f}}$, which will also incur errors due to numerical integration, and hence must be corrected like $\q$, and $\dot{\q}$. Similar to the correction for  $\q$ and  $\dot{\q}$, we consider
\begin{align}
W_{\vo{f}}^c=W_{\vo{f}}^u+\delta W_{\vo{f}}.
\label{eqn:WF_variation}
\end{align}

Due to numerical errors in integration of dynamics and \eqn{WF}, \eqn{conservation} will not be satisfied. Therefore, the corrections $\delta\q$, $\delta\qd$, and $\delta W_{\vo{f}}$ must be such that \eqn{conservation} is satisfied with the corrected quantities $\q^c$, $\qd^c$, and $W_{\vo{f}^c}$, i.e,
\begin{align}
E(\q^c,\qd^c) = E(\q_0,\dot{\q}_0) + W_{\vo{f}}^c. \eqnlabel{energyBal}
\end{align}

%From conservation of energy, we have
%\begin{align}
%E(\q_0,\dot{\q}_0) = T(\dot{\q}) + V_g(\q) + V_s(\q).
%\end{align}

Substituting $\q^c$, $\dot{\q}^c$, in $T(\qd^c)$, $V_g(\q^c)$, $V_s(\q^c)$, and retaining linear terms only, we get

\begin{align}
T(\qd^c) = T(\qd^u+\delta\qd) &\approx T(\qd^u) + \left.\frac{\partial T}{\partial \qd}\right|_{\qd^u}\delta\qd  \label{eqn:T_variation} \\
V_g(\q^c) = V_g(\q^u + \delta \q) &\approx  V_g(\q^u) + \left.\frac{\partial V_g}{\partial \q}\right|_{\q^u}\delta\q, \eqnlabel{Vg_variation}\\
V_s(\q^c) = V_s(\q^u+\delta\q) & \approx V_s(\q^u) + \left.\frac{\partial V_s }{\partial \q}\right|_{\q^u}\delta\q.\end{align}

Therefore, \eqn{energyBal} becomes
\begin{align}
&\begin{bmatrix}
\left(\left.\frac{\partial V_g}{\partial \q}\right|_{\q^u} + \left.\frac{\partial V_s}{\partial \q}\right|_{\q^u}\right) &\left.\frac{\partial T}{\partial \qd}\right|_{\qd^u} & -1
\end{bmatrix}\begin{pmatrix}\delta\q\\\delta \qd\\\delta W_{\vo{f}}
\end{pmatrix} \notag\\
&= E(\q_0,\qd_0) - E(\q^u,\qd^u) + W^u_{\vo{f}}. \eqnlabel{energyCorrection}
\end{align}

%Since the string potential energy is nonlinear in $\q$, we linearize $\|\vo{s}_k(\q)\|$ about $\q=\q^u$, as

Combining \eqn{correction_direct} and \eqn{energyCorrection}, we arrive at a final linear system of equations
\begin{align}
&\underbrace{\begin{bmatrix}
\Rq & \vo{0} & 0\\
\vo{Q} & \Rq & 0\\
\left(\left.\frac{\partial V_g}{\partial \q}\right|_{\q^u} + \left.\frac{\partial V_s}{\partial \q}\right|_{\q^u}\right) &\left.\frac{\partial T}{\partial \qd}\right|_{\qd^u} & -1
\end{bmatrix}}_{:=\vo{A}_c}
\begin{bmatrix}
\delta \q\\ \delta \dot{\q}\\\delta W_{\vo{f}}
\end{bmatrix}\notag\\
&=
\underbrace{\begin{bmatrix}
-\vo{R}\\-\Rq\dot{\q}^u\\E(\q_0,\qd_0) - E(\q^u,\qd^u) + W^u_{\vo{f}}
\end{bmatrix}}_{:=\vo{b}_c},
\label{eqn:correction_final}
\end{align}
and the minimum norm corrections $\delta \q$, $\delta \dot{\q}$, and $\delta W_{\vo{f}}$ are determined using pseudoinverse of $\vo{A}_c$, i.e.
\begin{align}
\begin{bmatrix}
\delta \q\\ \delta \dot{\q}\\\delta W_{\vo{f}}
\end{bmatrix}
=
\vo{A}_c^T(\vo{A}_c\vo{A}_c^T)^{-1}\vo{b}_c.
\label{eqn:correction_solution}
\end{align}
These corrections are done after every integration in each time step once the 2-norm of the constraints violations or the energy violation is greater than a given threshold $\gamma$. Further, since the constraints and energy equation are approximated through linearization, the solution to \eqref{eqn:correction_final} does not fully satisfy the nonlinear equations \eqref{eqn:R} and \eqref{eqn:conservation} depending on the error coming from the integration. So, an iteration method presented in Algorithm \ref{alg:correction} is used to ensure the performance of the corrections.

\begin{algorithm}[h]
\SetKwInOut{Input}{input}
\SetKwInOut{Output}{output}

\Input{uncorrected terms $\q^u$, $\dot{\q}^u$ and $W_f^u$}
\Output{corrected terms $\q^c$, $\dot{\q}^c$ and $W_f^c$}
\While{$\|\vo{R}\|_2 > \gamma$ \textbf{or} $E-E_0-W_f>\gamma$}{
implement \eqref{eqn:correction_solution}\;
update $\q^c$, $\dot{\q}^c$ and $W_f^c$ by \eqref{eqn:qc},\eqref{eqn:qcdot}, \eqref{eqn:WF_variation}\;
update $\|\vo{R}\|_2 $, $E$ and $W_f$ with corrected term\;
}
\caption{An iteration method to minimize the constraints and energy violations.}
\label{alg:correction}
\end{algorithm}

\subsection{Linearization}
\label{subsec:linear}
The nonlinear dynamics of the system can be written as
\begin{align}
	\ddot{\q} = \vo{\xi}(\q,\dot{\q},\vo{\sigma},\vo{f}),
\end{align}
where $\q,\dot{\q}$ are the states, $\vo{\sigma}$ is control, and $\vo{f}$ is external force (or disturbance).
Let $\q_0, \dot{\q}_0, \vo{\sigma}_0$ and $\vo{f}_0$ be the state, control, and external loading about which linearization is desired. The linear system is therefore
\begin{align}
\delta \ddot{\q}=& \frac{\partial \vo{\xi} }{ \partial \q}|_{\q_0,\dot{\q}_0,\vo{\sigma}_0,\vo{f}_0}\delta \q + \frac{\partial \vo{\xi}}{\partial \dot{\q}}|_{\q_0,\dot{\q}_0,\vo{\sigma}_0,\vo{f}_0}\delta \dot{\q}\notag\\
 &+ \frac{\partial \vo{\xi}}{\partial \vo{\sigma}}|_{\q_0,\dot{\q}_0,\vo{\sigma}_0,\vo{f}_0}\delta \vo{\sigma} + \frac{\partial \vo{\xi}}{\partial \vo{f}}|_{\q_0,\dot{\q}_0,\vo{\sigma}_0,\vo{f}_0}\delta \vo{f}. \label{eqn:linDyn}
\end{align}

From \eqref{eqn:qdd} the first term in \eqref{eqn:linDyn} is represented by
\begin{align*}
\frac{\partial\vo{\xi}}{\partial\q}= \vo{M}^{-1}\left[ \frac{\partial \vo{\xi}_1}{\partial \q}-\frac{\partial \Rq^T}{\partial \q}\tilde{\vo{M}}\tilde{\vo{\xi}}\right.\left.-~\Rq^T\right.&\frac{\partial \tilde{\vo{M}}}{\partial\q}\tilde{\vo{\xi}}\\
&\left. -~\Rq^T\tilde{\vo{M}}\frac{\partial \tilde{\vo{\xi}}}{\partial\q} \right],
\end{align*}
where
\begin{align*}
&\tilde{\vo{M}}=(\Rq \vo{M}^{-1}\Rq^T)^{-1}\\[1mm]
&\tilde{\vo{\xi}}=(\vo{\xi}_2+\Rq \vo{M}^{-1}\vo{\xi}_1)\\[1mm]
&\frac{\partial \Rq^T}{\partial \q}\tilde{\vo{M}}\tilde{\vo{\xi}}=\left[ \frac{\partial \Rq^T}{\partial q_1}\tilde{\vo{M}}\tilde{\vo{\xi}} \quad \cdots \quad \frac{\partial \Rq^T}{\partial q_{3n}}\tilde{\vo{M}}\tilde{\vo{\xi}} \right],\\[1mm]
&\Rq^T\frac{\partial \tilde{\vo{M}}}{\partial\q}\tilde{\vo{\xi}}=\left[ \Rq^T\frac{\partial}{\partial q_{1}}\tilde{\vo{M}}\tilde{\vo{\xi}}\quad \cdots \quad \Rq^T\frac{\partial}{\partial q_{3n}}\tilde{\vo{M}}\tilde{\vo{\xi}} \right],\\[1mm]
&\Rq^T\tilde{\vo{M}}\frac{\partial \tilde{\vo{\xi}}}{\partial\q}=\left[ \Rq^T\tilde{\vo{M}}\frac{\partial}{\partial q_1}\tilde{\vo{\xi}} \quad \cdots \quad \Rq^T\tilde{\vo{M}}\frac{\partial}{\partial q_{3n}}\tilde{\vo{\xi}} \right],\\[1mm]
&\frac{\partial\tilde{\vo{M}}}{\partial q_i} = -~\tilde{\vo{M}}\frac{\partial\left(\Rq \vo{M}^{-1}\Rq^T\right)}{\partial q_i}\tilde{\vo{M}}.
\end{align*}

Since only $\vo{\xi}_2$ is dependent on $\dot{\q}$, the second term in \eqref{eqn:linDyn} is
\begin{align*}
\frac{\partial \vo{\xi}}{\partial\dot{\q}} &= - \vo{M}^{-1}\Rq^T\tilde{\vo{M}}\frac{\partial \vo{\xi}_2}{\partial \dot{\q}}.
\end{align*}

Finally, $\vo{\xi}_1$ depends on the control $\vo{\sigma}$ and external force $\vo{f}$, the third term and fourth term in \eqref{eqn:linDyn} are
\begin{align*}
\frac{\partial \vo{\xi}}{\partial \vo{\sigma}}=\vo{M}^{-1}\left[ \frac{\partial\vo{\xi}_1}{\partial \vo{\sigma}}-\Rq^T\tilde{\vo{M}}\Rq\vo{M}^{-1} \frac{\partial\vo{\xi}_1}{\partial \vo{\sigma}} \right]
%\frac{\partial \vo{f}}{\partial \vo{\sigma}}&=\frac{\partial}{\partial \vo{\sigma}}\left[ -(\vo{\sigma}^T \otimes \vo{I}_{3n})\vo{Y}^T\vo{q} + \vo{G} + \vec{\vo{f}} \right]\\
%&=\frac{\partial}{\partial \vo{\sigma}}\left[ -\vec{\vo{Zu}}+\vo{G}+\vec{\vo{f}} \right]\\
%&=-\vo{Q},
\end{align*}
and
\begin{align*}
\frac{\partial \vo{\xi}}{\partial \vo{f}}=\vo{M}^{-1}\left[\vo{I}-\Rq^T\tilde{\vo{M}}\Rq\vo{M}^{-1}\right].
\end{align*}
%since $\vo{\xi}_2$ is independent of $\vo{f}$ and $\frac{\partial \vo{k_1}}{\partial \vo{f}} = \vo{I}$.

To determine $\frac{\partial\vo{\xi}_1}{\partial \vo{\sigma}}$, using the Kronecker identity
\begin{align*}
(\vo{B}^T\otimes\vo{A})\vec{\vo{C}} = \vec{\vo{A}\vo{C}\vo{B}}
\end{align*}
we write
\begin{align*}
(\vo{\sigma}^T \otimes \vo{I}_{3n})\vo{Y}^T\q &=
(\vo{\sigma}^T \otimes \vo{I}_{3n})\vec{\hat{\vo{Y}}} \\
&= \vec{\hat{\vo{Y}}\vo{\sigma}} \\
&= \hat{\vo{Y}}\vo{\sigma},
\end{align*}
where $\hat{\vo{Y}}\in \real^{3n\times n_u}$ such that $\vec{\hat{\vo{Y}}} = \vo{Y}^T\q$, which is a rearrangement of the elements of column vector $\vo{Y}^T\q$ into a matrix of dimension $3n\times n_u$. Therefore,

\begin{align}
\frac{\partial\vo{\xi}_1}{\partial \vo{\sigma}} &= \frac{\partial}{\partial \vo{\sigma}}(-(\vo{\sigma}^T \otimes \vo{I}_{3n})\vo{Y}^T\vo{q}
) = -\frac{\partial}{\partial \vo{\sigma}}\left(\hat{\vo{Y}}\vo{\sigma}\right)  =  -\hat{\vo{Y}}.
\label{eqn:linSigma}
\end{align}

Defining,
\begin{align}
\vo{A} &:=
\begin{bmatrix}
\vo{0}&\vo{I}\\
\frac{\partial \vo{\xi}}{\partial \q}|_{\q_0,\dot{\q}_0,\vo{\sigma}_0,\vo{f}_0}&
\frac{\partial \vo{\xi}}{\partial \dot{q}}|_{\q_0,\dot{\q}_0,\vo{\sigma}_0,\vo{f}_0}
\end{bmatrix},\\
\vo{B_{\vo{\sigma}}} & :=\begin{bmatrix}
\vo{0}\\
\frac{\partial \vo{\xi}}{ \partial \vo{\sigma}}|_{\q_0,\dot{\q}_0,\vo{\sigma}_0,\vo{f}_0}
\end{bmatrix},\\
\vo{B_f} &:=
\begin{bmatrix}
\vo{0}\\
\frac{\partial \vo{\xi}}{\partial \vo{f}}|_{\q_0,\dot{\q}_0,\vo{\sigma}_0,\vo{f}_0}
\end{bmatrix},
\end{align}
we can write the linear system as
\begin{align}
\dot{\vo{x}} = \vo{A} \vo{x} + \vo{B_\sigma}\delta\vo{\sigma} + \vo{B_f}\delta\vo{f}, \eqnlabel{linSys}
\end{align}
where
\begin{align*}
\vo{x} :=\begin{bmatrix}\delta \vo{\q} \\ \delta\dot{\vo{q}}\end{bmatrix}.
\end{align*}

The linear system in \eqn{linSys} is not a minimum realization, due to the algebraic constraints on $\delta\vo{q}$. Controlling tensegrity systems using modern control theory, requires minimum realization. The system can be transformed to a minimum realization using standard techniques \cite{laub1980computation,moore1981principal,laub1987computation,gawronski1990model}.

% \textcolor{blue}{
\subsection{Summary}
Section 2 presented a detailed description of the derivations required to arrive at the final set of equations governing the dynamics of our multibody tensegrity system in the Lagrangian framework.
A gist of the approach is presented as follows.
\begin{enumerate}
\item \textbf{Input}: For the tensegrity system under analysis, provide material, geometric, and connectivity properties that would fully describe the structure under stasis.
	% Material properties include ones such as density and the Young's Modulus. Mass and stiffness would be calculated accordingly.
	% Geometric properties include bars and strings' radii and lengths.
	% Connectivity matrices describe network locations of the bars and strings, and further information regarding connectivity would include which nodes would be inertially fixed throughout.
\item \textbf{Structure Generation}: Construct a tensegrity structure with the following matrices: $\vo{N}$, $\vo{C}$, $\vo{X}$, $\vo{Y}$, $\vo{P}$ which describe the nodal configuration and the appropriately defined connectivity matrices for bars, strings, and the point masses.
\item \textbf{Lagrangian Dynamics}: Compute the following quantities in order: total kinetic energy, total potential energy, constraint equations, and non-conservative forces.
\begin{align*}
	% \textrm{Total Kinetic Energy:} \\
	{T} &= \frac{1}{2}\qd^T \vo{M} \qd \\
	% \textrm{Potential Energy:} \\
	V &= V_g + V_s \\
	&= -\vo{G}^T \q + \frac{1}{2} \q^T  \vo{Y} (\frac{\vo{\sigma^2}}{\vo{K}} \otimes \vo{I}_{3n}) \q \\
	% \textrm{Constraints:} \\
	\vo{R}(\q) &= \begin{bmatrix}\vo{A}\vo{q}-\vo{b} \\ \vo{q}^T\vo{X}_1^T\vo{X}_1\vo{q} - l_{b_1}^2\\ \vdots \\ \vo{q}^T\vo{X}_{n_b}^T\vo{X}_{n_b}\vo{q}-l_{b_{n_b}}^2
	\end{bmatrix} = 0. \\
	% \textrm{Non-conservative forces:} \\
	\vo{f} &= \vo{f}_{\textrm{ext}} + \vo{f}_d \\
\end{align*}
\item \textbf{Equations of Motion}: The first equation describes the governing equations of motion while the second one describes the additional constraint equations.
\begin{align*}
	\vo{M}\ddot{\vo{q}} - \left(\frac{\partial \vo{R}}{\partial \vo{q}}\right)^T\vo{\lambda} =  -
	(\vo{\sigma}^T \otimes \vo{I}_{3n})\vo{Y}^T\vo{q} + \vo{G} + \vo{f}.
\end{align*}
\begin{align*}
	-\left(\frac{\partial \vo{R}}{\partial \vo{q}}\right)\ddot{\vo{q}} =   \begin{bmatrix} \dot{\vo{q}}^T \left(\frac{\partial^2 R_1}{\partial \vo{q}^2}\right)\dot{\vo{q}} \\
	\vdots \\
	\dot{\vo{q}}^T \left(\frac{\partial^2 R_m}{\partial \vo{q}^2}\right)\dot{\vo{q}}
\end{bmatrix}
\end{align*}
\item \textbf{Correction}:
Correcting for position,
\begin{align*}
	\vo{R}(\q^c) = \vo{R}(\q^u) + \Rq\delta \q = \vo{0},
\end{align*}
Correcting for velocity,
\begin{align*}
	\vo{R}_{\q}\dot{\q}^u+\vo{R}_{\q}\delta\dot{\q}+\left(\frac{\partial\vo{R}_{\q}}{\partial \q}\delta\q\right)\dot{\q}^u=\vo{0},
\end{align*}
Correcting for energy,
\begin{align*}
	E(\q^c,\qd^c) = E(\q_0,\dot{\q}_0) + W_{\vo{f}}^c
\end{align*}
\item \textbf{Linearization}:
The nonlinear dynamics given by:
\begin{align*}
	\ddot{\q} = \vo{\xi}(\q,\dot{\q},\vo{\sigma},\vo{f}),
\end{align*}
can be linearized into:
\begin{align*}
	\dot{\vo{x}} = \vo{A} \vo{x} + \vo{B_\sigma}\delta\vo{\sigma} + \vo{B_f}\delta\vo{f}
\end{align*}
where
\begin{align*}
	\vo{x} := \begin{bmatrix}
		\delta \q \\ \delta \qd
\end{bmatrix}
\end{align*}
\end{enumerate}
% }

\section{Tensegrity Systems with Compressible Bars}
\subsection{Dynamics}
Since bars are assumed to be rigid, the holonomic constraint $\vo{R}$ is a must to ensure that the bar lengths stay fixed during simulation. However, in reality, the bar is not perfectly rigid but compressible with a large Young's modulus, which allows us to release the constraints and treat the bar as an elastic body. In this section, we allow for bars to deform longitudinally and, in accordance with the Poisson effect, transversely as well, i.e., $\dot{r}_k=-\nu_k r_k\frac{\dot{l}_{\vo{b}_k}}{l_{\vo{b}_k}}$, where $r_k$ and $l_{\vo{b}_k}$ are the radius and length of $\ith{k}$ bar respectively, and $\nu_k$ is Poisson's ratio. It must be stated that this elasticity does not allow for bending of the bar in any case.

One should note that $\dot{l}_{\vo{b}_k}$ is no longer zero due to the flexibility of the bar. Specifically, $\dot{l}_{\vo{b}_k}=\frac{\vo{b}_k^T\dot{\vo{b}}_k}{l_{\vo{b}_k}}$, which implies that the velocity vector of the bar is not perpendicular to the bar vector and yields that
\begin{align*}
	\bvd = \dot{l}_{\vo{b}_k} \hat{\vo{b}}_k + \vo{\omega}_k \times \bv
\end{align*}
where $\hat{\vo{b}}_k$ is the unit vector along the axis of the bar represented by $\bv$.
Taking cross product on both sides with $\bv$
\begin{align*}
	 \bv \times \bvd &= \vo{0} + \bv \times (\vo{\omega}_k \times \bv) \\
	 &= \vo{\omega}_k (\bv\cdot \bv) - \bv(\vo{\omega}_k \cdot \bv)
\end{align*}
The assumption for tensegrity systems with regards to the bar not spinning about the body axis holds, i.e., $(\vo{\omega}_k \cdot \bv) = \vo{0}$, and hence,
\begin{align*}
	\vo{\omega}_k = \frac{\bv \times \bvd}{l_{\bv}^2}
\end{align*}
which is the same expression for angular velocity as that for the rigid bar case. However, when computing the rotational kinetic energy of the bar, we observe
\begin{align*}
	T_{\vo{b}_k} & = \frac{1}{2}\left(m_{\vo{b}_k} \bcgd^T\bcgd + \frac{I_{\vo{b}_k}}{l_k^4}(\bv\times\bvd)\cdot(\bv\times\bvd)\right),
\end{align*}
where now,
\begin{align*}
(\bv\times\bvd)\cdot(\bv\times\bvd) = l_{\vo{b}_k}^2(\bvd\cdot\bvd) - (l_{\vo{b}_k}\dot{l}_{\vo{b}_k})^2.
\end{align*}
The kinetic energy of the $\ith{k}$ bar therefore becomes
\begin{align*}
T_{\vo{b}_k} = \frac{1}{2}\dot{\vo{q}}^T\vo{M}_{\vo{b}_k}\dot{\vo{q}}-\frac{1}{2}\left(\frac{I_{\vo{b}_k}}{l_{\vo{b}_k}^2}\dot{l}_{\vo{b}_k}^2\right),
\end{align*}
and the total kinetic energy is
\begin{align*}
T = \frac{1}{2}\dot{\vo{q}}^T\vo{M}\dot{\vo{q}}-T_f
%{1}{2}m_f,
\end{align*}
where $T_f=\frac{1}{2}\sum_{k=1}^{n_b}\frac{I_{\vo{b}_k}}{l_{\vo{b}_k}^2}\dot{l}_{\vo{b}_k}^2$. On comparing the differences with  \eqref{eqn:T}, we note that $\vo{M}$  is no longer a constant matrix and there exists an extra term $T_f$, so
\begin{align*}
\frac{\partial T}{\partial \dot{\q}}&=\dot{\q}^T\vo{M}-\frac{\partial T_f}{\partial \dot{\q}}\\
&= \dot{\q}^T\vo{M}-\frac{1}{2}\sum_{k=1}^{n_b}\frac{I_{\vo{b}_k}}{l_{\vo{b}_k}^2}(2\dot{l}_{\vo{b}_k}\frac{\partial \dot{l}_{\vo{b}_k}}{\partial\dot{\q}})\\
&= \dot{\q}^T\vo{M}-\sum_{k=1}^{n_b}\frac{I_{\vo{b}_k}}{l_{\vo{b}_k}^2}\frac{\q^T\vo{X}_k^T\vo{X}_k}{l_{\vo{b}_k}}\dot{l}_{\vo{b}_k}\\
&= \dot{\q}^T\vo{M}-\q^T\vo{M}_f,
\end{align*}
where $\vo{M}_f= \sum_{k=1}^{n_b} \frac{I_{\vo{b}_k}\vo{X}_k^T\vo{X}_k\dot{l}_{\vo{b}_k}}{l_{\vo{b}_k}^3}$, and
\begin{align}
\frac{d}{dt}\left(\frac{\partial T}{\partial \dot{\q}}\right)=\ddot{\q}^T\vo{M}+\dot{\q}^T\dot{\vo{M}}-\dot{\q}^T\vo{M}_f-\q^T\dot{\vo{M}}_f,
\label{eqn:dTdqd}
\end{align}
where
\begin{align*}
\dot{\vo{M}}&=\sum_{k=1}^{n_b}\left[\frac{d}{dt}\left( \frac{I_{\vo{b}_k}}{l_{\vo{b}_k}^2} \right)\vo{X}_k^T\vo{X}_k\right]\\
&= \sum_{k=1}^{n_b}\left[ -2\left( \frac{I_{\vo{b}_k}}{l_{\vo{b}_k}^3} \right)\vo{X}_k^T\vo{X}_k\dot{l}_{\vo{b}_k}+\frac{dI_{\vo{b}_k}}{dt}\frac{1}{l_{\vo{b}_k}^2}\vo{X}_k^T\vo{X}_k \right],
\end{align*}
and
\begin{align*}
\dot{\vo{M}}_f=\sum_{k=1}^{n_b}\left[ \frac{dI_{\vo{b}_k}}{dt}\frac{\vo{X}_k^T\vo{X}_k\dot{l}_{\vo{b}_k}}{l_{\vo{b}_k}^3}
-3\right.&\left(\frac{I_{\vo{b}_k}\vo{X}_k^T\vo{X}_k}{l_{\vo{b}_k}^4} \right)\dot{l}_{\vo{b}_k}^2 \\
&\qquad\left.+
 \ddot{l}_{\vo{b}_k}\frac{I_{\vo{b}_k}\vo{X}_k^T\vo{X}_k}{l_{\vo{b}_k}^3} \right],
\end{align*}
with
\begin{align*}
\ddot{l}_{\vo{b}_k}&=\frac{d}{dt}\left[ \frac{\q^T\vo{X}_k^T\vo{X}_k\dot{\q}}{l_{\vo{b}_k}} \right]\\
&=\frac{\dot{\q}^T\vo{X}_k^T\vo{X}_k\dot{\q}}{l_{\vo{b}_k}}+\frac{\ddot{\q}^T\vo{X}_k^T\vo{X}_k\q}{l_{\vo{b}_k}}-\frac{\q^T\vo{X}_k^T\vo{X}_k\dot{\q}}{l_{\vo{b}_k}^2}\dot{l}_{\vo{b}_k},
\end{align*}
and
\begin{align*}
\frac{dI_{\vo{b}_k}}{dt}=\frac{m_{\vo{b}_k}}{12}(6r_k\dot{r}_k+2l_{\vo{b}_k}\dot{l}_{\vo{b}_k}).
\end{align*}
Rearranging \eqref{eqn:dTdqd} where the terms associated with $\ddot{\q}$ are combined together, we arrive
\begin{align}
\frac{d}{dt}\left(\frac{\partial T}{\partial \dot{\q}}\right)=\ddot{\q}^T\vo{M}_{\ddot{\q}}+\dot{\q}^T\vo{M}_{\dot{\q}}+\q^T\vo{M}_{\q},
\end{align}
where
\begin{align*}
&\vo{M}_{\ddot{\q}}=\vo{M}-\sum_{k=1}^{n_b}
\frac{\vo{X}_k^T\vo{X}_k\q}{l_{\vo{b}_k}}\frac{\q^T I_{\vo{b}_k}\vo{X}_k^T\vo{X}_k}{l_{\vo{b}_k}^3},\\
&\vo{M}_{\dot{\q}}=\dot{\vo{M}}-\vo{M}_f,\\
&\vo{M}_{\q}=-\sum_{k=1}^{n_b}\left[ \frac{dI_{\vo{b}_k}}{dt}\frac{\vo{X}_k^T\vo{X}_k\dot{l}_{\vo{b}_k}}{l_{\vo{b}_k}^3}
-3\left( \frac{I_{\vo{b}_k}\vo{X}_k^T\vo{X}_k}{l_{\vo{b}_k}^4} \right)\dot{l}_{\vo{b}_k}^2 \right.\\
&\qquad\left.+ \left( \frac{\dot{\q}^T\vo{X}_k^T\vo{X}_k\dot{\q}}{l_{\vo{b}_k}}-\frac{\q^T\vo{X}_k^T\vo{X}_k\dot{\q}}{l_{\vo{b}_k}^2}\dot{l}_{\vo{b}_k} \right)\frac{I_{\vo{b}_k}\vo{X}_k^T\vo{X}_k}{l_{\vo{b}_k}^3} \right].
\end{align*}

Considering the elasticity of the bars, a potential energy term will also be required. Similar to the expression for potential energy in strings, we can model this term as
\begin{align*}
V_b&=\frac{1}{2}\sum_{k=1}^{n_b}K_{\vo{b}_k}\left( \|\vo{b}_k\|-l_{\vo{b}_k}(0) \right)^2,
\end{align*}
where $K_{\vo{b}_k}$ is the stiffness of the $\ith{k}$ bar which is usually a large value, and $l_{\vo{b}_k}(0)$ is the rest length of the $\ith{k}$ bar.
For solid bars under tension or compression,
\begin{align*}
	K_{\vo{b}_k} = \frac{A_{\vo{b}_k} E_{\vo{b}_k}}{l_{\vo{b}_k}(0)}
\end{align*}
where $A_{\vo{b}_k}$ is the area of the bar under consideration and $E_{\vo{b}_k}$ is its Young's modulus.
Then,
\begin{align}
\frac{\partial V_b}{\partial \q}&=\sum_{k=1}^{n_b}K_{\vo{b}_k}\left[ \q^T\vo{X}_k^T\vo{X}_k-\frac{\q^T\vo{X}_k^T\vo{X}_k}{l_{\vo{b}_k}}l_{\vo{b}_k}(0) \right]\notag\\
&=\q^T\sum_{k=1}^{n_b}K_{\vo{b}_k}\left[ \vo{X}_k^T\vo{X}_k-\frac{\vo{X}_k^T\vo{X}_k}{l_{\vo{b}_k}}l_{\vo{b}_k}(0) \right]\notag\\
&=\q^T \vo{X} (\vo{\Psi} \otimes \vo{I}_{3n})
% }
\label{eqn:dVbdq}
\end{align}
where $\vo{\Psi} \in \real^{n_b}$, $\Psi_k = K_{\vo{b}_k} \left[1 -  \cfrac{l_{\vo{b}_k}(0)}{l_{\vo{b}_k}}\right]$, and \\
$\vo{X} = [\vo{X}_1^T \vo{X}_1 \cdots \vo{X}_{n_b}^T \vo{X}_{n_b}]$.

Substituting \eqref{eqn:dTdqd} and \eqref{eqn:dVbdq} into \eqref{eqn:Lagrange}, the equations of motion are given by
\begin{align}
\vo{M}_{\ddot{\q}}\ddot{\q} - \vo{R}_{\vo{q}}^T\vo{\lambda} =& -\left[\vo{M}_{\q}^T +
(\vo{\Psi}^T \otimes \vo{I}_{3n})\vo{X}^T \right. \notag\\
 &\left. +(\vo{\sigma}^T \otimes \vo{I}_{3n})\vo{Y}^T\right]\q \notag\\
&-\vo{M}_{\dot{\q}}^T\dot{\q} + \vo{G} + \vo{f} + \vo{f}_d:=\vo{\xi}_3,
\end{align}
where the holonomic constraints $\vo{R}$ do not contain bar length constraints anymore, but a few more terms are added to the governing equation to capture the elasticity of the bars.
% The negative sign for the term $(\vo{\Psi}^T \otimes \vo{I}_{3n})\vo{X}^T$ indicates compression, as $\vo{\Psi} < 0$.
% \comment{
% We then arrive at the final expression of $\ddot{\q}$ as
% \begin{align}
% \ddot{\q}&=\vo{M}_{\ddot{\q}}^{-1}\left[\vo{\xi}_3-\Rq^T\tilde{\vo{M}}_{\ddot{\q}}\left(\vo{\xi}_2+\Rq\vo{M}_{\ddot{\q}}^{-1}\vo{\xi}_3\right)\right]\notag\\
% &=\vo{\zeta}(\q,\dot{\q},\vo{\sigma},\vo{\Psi},\vo{f}),
% \label{eqn:qdd_flex}
% \end{align}
% where $\tilde{\vo{M}}_{\ddot{\q}}=\left(\Rq\vo{M}_{\ddot{\q}}^{-1}\Rq^T\right)^{-1}$.
% $\vo{M}_{\ddot{\q}}$ is always invertible because the second term of $\vo{M}_{\ddot{\q}}$ is also positive definite.}
% Finally, the linearization of \eqref{eqn:qdd_flex} can be done via the approach described in section \ref{subsec:linear}.

\subsection{Impact on Structural Stiffness}
Under equilibrium, the equations are:
\begin{align}
- \vo{R}_{\vo{q}}^T\vo{\lambda} =& -\left[ (\vo{\Psi}^T \otimes \vo{I}_{3n})\vo{X}^T
% \right. \notag \\
% & \left.
 + (\vo{\sigma}^T \otimes \vo{I}_{3n})\vo{Y}^T\right]\q \notag \\
& + \vo{G} + \vo{f} - \frac{\partial \vo{M}_{\qd}}{\partial \q}\qd - \vo{M}_{\q}^T \q
\end{align}

Note that compressibility allows for actuation of the tensegrity structure through pre-stressing of bars.
Suppose, in a regular tensegrity structure with rigid bars, we wished to replace the tension-bearing cables in the structure with compression-bearing bars.
\begin{align*}
	\vo{q}^T \vo{Y} (\vo{\sigma} \otimes \vo{I}_{3n}) %&= - \q^T\sum_{k=1}^{n_b}K_{\vo{b}_k}\left[
	 %\vo{X}_k^T\vo{X}_k-\frac{\vo{X}_k^T\vo{X}_k l_{\vo{b}_k}(0)}{l_{\vo{b}_k}} \right] \\
	% &= - \q^T\sum_{k=1}^{n_b} \vo{X}_k^T\vo{X}_k K_{\vo{b}_k} \left[1 -  \frac{l_{\vo{b}_k}(0)}{l_{\vo{b}_k}}\right]\\
	&= -\q^T \vo{X} (\vo{\Psi} \otimes \vo{I}_{3n})
\end{align*}
where the negative sign indicates a change from tension to compression.
% where $\vo{\Psi} \in \real^{n_b}$, $\Psi_k = K_{\vo{b}_k} \left[1 -  \cfrac{l_{\vo{b}_k}(0)}{l_{\vo{b}_k}}\right]$, and \\
% $\vo{X} = [\vo{X}_1^T \vo{X}_1 \cdots \vo{X}_{n_b}^T \vo{X}_{n_b}]$.
% }

% \textcolor{blue}{
The stiffness of the tensegrity structure comes from the strings as well as the compressible bars present in the structure.
Stiffness is:
\begin{align*}
	\vo{K}_{sys} &= \frac{\partial}{\partial \q} \left( \left. \frac{\partial V_s}{\partial \vo{q}}\right|_{\vo{q} = \vo{q}_{eq}} + \left. \frac{\partial V_b}{\partial \vo{q}}\right|_{\vo{q} = \vo{q}_{eq}} \right)^T \\
	&= \frac{\partial}{\partial \q} \left( \sum_{k=1}^{n_s} \sigma_k \q^T \vo{Y}_k^T \vo{Y}_k + \sum_{j=1}^{n_b} \Psi_j \q^T \vo{X}_j^T \vo{X}_j  \right)^T
	\\
	&= \sum_{k=1}^{n_s} \left( \sigma_k  \vo{Y}_k^T \vo{Y}_k + \frac{\partial \sigma_k}{\partial \q} \vo{Y}_k^T \vo{Y}_k \q \right)
	\\
	& + \sum_{j=1}^{n_b} \left( \Psi_j  \vo{X}_j^T \vo{X}_j + \frac{\partial \Psi_j}{\partial \q} \vo{X}_j^T \vo{X}_j \q \right)
	\\
	&= \sum_{k=1}^{n_s} \left( \sigma_k  \vo{Y}_k^T \vo{Y}_k +  \frac{K_{\vo{s}_k} l_{{0\vo{s}}_k}}{||\vo{s}_k ||^3} \vo{Y}_k^T \vo{Y}_k \q \q^T \vo{Y}_k^T \vo{Y}_k \right)
	\\
	&+ \sum_{j=1}^{n_b} \left( \Psi_j  \vo{X}_j^T \vo{X}_j +  \frac{K_{\vo{b}_j} l_{{0\vo{b}}_j}}{||\vo{b}_j ||^3} \vo{X}_j^T \vo{X}_j \q \q^T \vo{X}_j^T \vo{X}_j \right)
	% \\
	% &= \frac{\partial}{\partial \q}((\vo{\sigma}^T \otimes \vo{I}_{3n})\vo{Y}^T \q + (\vo{\Psi}^T \otimes \vo{I}_{3n}) \vo{X}^T \q) \\
	% &=  (\vo{\sigma}^T \otimes \vo{I}_{3n})\vo{Y}^T  + (\vo{\Psi}^T \otimes \vo{I}_{3n}) \vo{X}^T\\
	% &= \sum_{k=1}^{n_s} {\sigma}_k \vo{Y}_k^T \vo{Y}_k + \sum_{k=1}^{n_b} {\Psi}_k \vo{X}_k^T \vo{X}_k
\end{align*}
% Given the nature of $\vo{Y}_k$ and $\vo{X}_k$, $\vo{Y}_k^T \vo{Y}_k$ and $\vo{X}_k^T \vo{X}_k$ are always positive semi-definite $\forall k$. Since $\sigma_k \geq 0$ and $\psi_k \leq 0$ $\forall {} k$, $\vo{K}_{sys}$ will be positive semi-definite as well.
% \begin{align*}
% 	\frac{\partial \vo{K}_{sys}}{\partial \vo{\sigma}} = \vo{\hat{Y}}
% \end{align*}
% where, as discussed before in the linearization section, $\vo{\hat{Y}} \in \real^{3n \times n_s}$, such that $\vec{\vo{\hat{Y}}} = \vo{Y}^T \vo{q}$.
% Similarly,
% \begin{align*}
% 	\frac{\partial \vo{K}_{sys}}{\partial \vo{\Psi}} = \vo{\hat{X}}
% \end{align*}
% where, $\vo{\hat{X}} \in \real^{3n \times n_b}$, such that $\vec{\vo{\hat{X}}} = \vo{X}^T \vo{q}$.
$l_{{0\vo{s}}_k}$ and $l_{{0\vo{b}}_j}$ are the natural lengths of the $k^{th}$ string and $j^{th}$ bar respectively.
Clearly, there is a linear dependence of the stiffness matrix on the individual stiffnesses of the bars and the cables present in the structure.
This allows for flexibility in choosing the material of the bars as well. The stiffer the bar material (steel or aluminium and the like), lower will be the deflection, and vice-versa.

% \textcolor{blue}{
\subsection{Linearization of Augmented Lagrangian System}
Consider the set of equations:
\begin{align*}
	\vo{M}_{\ddot{\q}}(\q)\ddot{\q} - \vo{R}_{\vo{q}}^T \vo{\lambda} &=\vo{\xi}_3 (\q,\dot{\q},\vo{\sigma},\vo{\Psi},\vo{f}),  \\
	-\Rq \ddot{\q} &= \vo{\xi}_2
\end{align*}
Note that in the absence of bar length constraints, $\vo{R}(\q) := [\vo{A}\q - \vo{b}] = 0$ and $\vo{\xi}_2 = 0$. Therefore,
$\Rq := \left(\frac{\partial \vo{R}}{\partial \vo{q}}\right) =\vo{A} \textrm{ (a constant)}$.
% which yields
% \begin{align*}
% 	\frac{\partial \Rq^T}{\partial \q} = \vo{0} \quad \textrm{and} \quad \vo{\xi}_2 = \vo{0}.
% \end{align*}
This means that the above set of dynamics equations reduces to:
\begin{align*}
	\vo{M}_{\ddot{\q}}(\q)\ddot{\q} - \vo{R}_{\vo{q}}^T \vo{\lambda} &=\vo{\xi}_3 (\q,\dot{\q},\vo{\sigma},\vo{\Psi},\vo{f}),  \\
	-\Rq \ddot{\q} &= \vo{0}
\end{align*}
Linearizing this set of equations at $[\q_0, \dot{\q}_0], [\vo{\sigma}_0$, $\vo{\Psi}_0]$, and $\vo{f}_0$, i.e., the state, control, and external loading that satisfy equilibrium conditions,
\begin{align*}
	&\left(\frac{\partial \vo{M}_{\ddot{\q}}}{\partial \q} \ddot{\q} \right) \delta \q  +
	 \vo{M}_{\ddot{\q}} \delta \ddot{\q} - \Rq^T \delta \vo{\lambda} = \\
	 & \frac{\partial \vo{\xi}_3 }{ \partial \q}|_{\q_0,\dot{\q}_0,\vo{\sigma}_0,\vo{\Psi}_0,\vo{f}_0}\delta \q +
	 \frac{\partial \vo{\xi}_3}{\partial \dot{\q}}|_{\q_0,\dot{\q}_0,\vo{\sigma}_0,\vo{\Psi}_0,\vo{f}_0}\delta \dot{\q} \\
	  &+ \frac{\partial \vo{\xi}_3}{\partial \vo{\sigma}}|_{\q_0,\dot{\q}_0,\vo{\sigma}_0,\vo{\Psi}_0,\vo{f}_0}\delta \vo{\sigma} +
	   \frac{\partial \vo{\xi}_3}{\partial \vo{\Psi}}|_{\q_0,\dot{\q}_0,\vo{\sigma}_0,\vo{\Psi}_0,\vo{f}_0}\delta \vo{\Psi} + \\
	  & \frac{\partial \vo{\xi}_3} {\partial \vo{f}}|_{\q_0,\dot{\q}_0,\vo{\sigma}_0,\vo{\Psi}_0,\vo{f}_0}\delta \vo{f}. \\
	& \Rq \delta \ddot{\q}		 = \vo{0}
\end{align*}
Since at equilibrium, $\ddot{\q} = 0$, the first term in the first equation above can be eliminated.
% Here,
% \begin{align*}
% 	\frac{\partial \vo{M}_{\ddot{\q}}}{\partial \q} \ddot{\q} =
% 	\left [\frac{\partial \vo{M}_{\ddot{\q}}}{\partial q_1} \ddot{\q} \cdots \frac{\partial \vo{M}_{\ddot{\q}}}{\partial q_{3n}} \ddot{\q} \right]
% \end{align*}
Further,
\begin{align*}
	\frac{\partial \vo{\xi}_3}{\partial \q} &= - \left[ \vo{M}_{\q}^T + (\vo{\Psi}^T \otimes \vo{I}_{3n})\vo{X}^T + (\vo{\sigma}^T \otimes \vo{I}_{3n})\vo{Y}^T  \right] \\
	& \quad - \frac{\partial \vo{M}_{\dot{\q}}^T}{\partial \q} \dot{\q} + \frac{\partial \vo{f}_d}{\partial \q},\\
 	\frac{\partial \vo{\xi}_3}{\partial \dot{\q}} &=  - \frac{\partial \vo{M}_{\q}^T}{\partial \dot{\q}} \q - \vo{M}_{\dot{\q}}^T -
	\frac{\partial \vo{M}_{\dot{\q}}^T}{\partial \dot{\q}} {\dot{\q}}
	+ \frac{\partial \vo{f}_d}{\partial \qd}
\end{align*}
where
% \textcolor{red}{
\begin{align*}
	\frac{\partial \vo{M}_{\dot{\q}}^T}{\partial \q} \qd &= \frac{\partial \dot{\vo{M}}^T}{\partial \q} \qd - \frac{\partial \vo{M}_f^T}{\partial \q} \qd\\
	&= \sum_{k=1}^{n_b} \left[ -2 \frac{I_{\vo{b}_k} \vo{X}_k^T \vo{X}_k \qd}{{l}_{\vo{b}_k}^3} \frac{\partial \dot{l}_{\vo{b}_k}}{\partial \q} \right.\\
	& -2 \frac{\dot{l}_{\vo{b}_k} \vo{X}_k^T \vo{X}_k \qd}{{l}_{\vo{b}_k}^3} \frac{\partial I_{\vo{b}_k}}{\partial \q}\\
	& -6 \frac{I_{\vo{b}_k} \dot{l}_{\vo{b}_k} \vo{X}_k^T \vo{X}_k \qd}{{l}_{\vo{b}_k}^4} \frac{\partial {l}_{\vo{b}_k}}{\partial \q}
 	+ \frac{  \vo{X}_k^T \vo{X}_k \qd}{{l}_{\vo{b}_k}^2} \frac{\partial \cfrac{dI_{\vo{b}_k}}{dt}}{\partial \q} \\
	& \left. -2 \frac{dI_{\vo{b}_k}}{dt} \frac{  \vo{X}_k^T \vo{X}_k \qd}{{l}_{\vo{b}_k}^3} \frac{\partial {l}_{\vo{b}_k}}{\partial \q} \right] \\
	& - \sum_{k=1}^{n_b} \left[ \frac{I_{\vo{b}_k} \vo{X}_k^T \vo{X}_k \qd}{{l}_{\vo{b}_k}^3} \frac{\partial \dot{l}_{\vo{b}_k}}{\partial \q} \right.\\
	& \left. -3 \frac{I_{\vo{b}_k} \dot{l}_{\vo{b}_k} \vo{X}_k^T \vo{X}_k \qd}{{l}_{\vo{b}_k}^4} \frac{\partial {l}_{\vo{b}_k}}{\partial \q}
	+ \frac{\dot{l}_{\vo{b}_k} \vo{X}_k^T \vo{X}_k \qd}{{l}_{\vo{b}_k}^3} \frac{\partial I_{\vo{b}_k}}{\partial \q} \right] \\
	&= \sum_{k=1}^{n_b} \vo{X}_k^T \vo{X}_k \qd \left[ -3 \frac{I_{\vo{b}_k}}{{l}_{\vo{b}_k}^3} \frac{\partial \dot{l}_{\vo{b}_k}}{\partial \q}
	 - 3 \frac{I_{\vo{b}_k} \dot{l}_{\vo{b}_k}}{{l}_{\vo{b}_k}^4} \frac{\partial {l}_{\vo{b}_k}}{\partial \q} \right.\\
 	& \left. + \frac{ 1}{{l}_{\vo{b}_k}^2} \frac{\partial \cfrac{dI_{\vo{b}_k}}{dt}}{\partial \q}
	 -2 \frac{dI_{\vo{b}_k}}{dt} \frac{1}{{l}_{\vo{b}_k}^3} \frac{\partial {l}_{\vo{b}_k}}{\partial \q} - 3
  \frac{\dot{l}_{\vo{b}_k}}{{l}_{\vo{b}_k}^3} \frac{\partial I_{\vo{b}_k}}{\partial \q}  \right]
\end{align*}
Since $ \vo{f}_{d,k} = -c \dot{l}_{\vo{s}_k} \cfrac{\vo{s}_k}{{l}_{\vo{s}_k}}$ is the damping force present in the $k^{th}$ string, $\frac{\partial \vo{f}_{d}}{\partial \q}  = \sum_{k=1}^{n_s} \vo{Y}_k^T \frac{\partial \vo{f}_{d,k}}{\partial \q}$ where
\begin{align*}
	\frac{\partial \vo{f}_{d,k}}{\partial \q} &= -c \left( \frac{\vo{Y}_k \q}{{l}_{\vo{s}_k}} \frac{\partial \dot{l}_{\vo{s}_k}}{\partial \q}
	 +  \frac{\dot{l}_{\vo{s}_k} \vo{Y}_k}{{l}_{\vo{s}_k}}
	 -  \frac{\dot{l}_{\vo{s}_k} \vo{Y}_k \q}{{l}_{\vo{s}_k}^2} \frac{\partial {l}_{\vo{s}_k}}{\partial \q}
	 \right)
\end{align*}
Here,
\begin{align*}
	\frac{\partial {l}_{\vo{b}_k}}{\partial \q} &= \frac{\q^T \vo{X}_k^T \vo{X}_k}{ {l}_{\vo{b}_k}} \\
	\frac{\partial \dot{l}_{\vo{b}_k}}{\partial \q} &= \frac{\qd^T \vo{X}_k^T \vo{X}_k}{ {l}_{\vo{b}_k}}
	 - \frac{\qd^T \vo{X}_k^T \vo{X}_k \q}{ {l}_{\vo{b}_k}^2} \frac{\partial {l}_{\vo{b}_k}}{\partial \q}
	 \\
  \frac{\partial I_{\vo{b}_k}}{\partial \q} = &\frac{\partial \left (\cfrac{m_{\vo{b}_k}}{12} \left(3 r_k^2 + l_{\vo{b}_k}^2 \right) \right)}{\partial \q}
	\\
	&= \frac{{m_{\vo{b}_k} {l}_{\vo{b}_k}}}{6} \frac{\partial {l}_{\vo{b}_k}}{\partial \q}
\\
	\frac{\frac{dI_{\vo{b}_k}}{dt}}{\partial \q} &= \frac{\partial \left[\frac{m_{\vo{b}_k}}{12}\left(6 r_k \dot{r}_k + 2 \dot{l}_{\vo{b}_k}{l}_{\vo{b}_k} \right) \right]}{\partial \q}
	\\
	 &= \frac{m_{\vo{b}_k}}{12} \left(6 r_k \frac{\partial \dot{r}_k}{\partial \q} +
	2 \dot{l}_{\vo{b}_k} \frac{\partial {l}_{\vo{b}_k}}{\partial \q}
	+ 2 {l}_{\vo{b}_k} \frac{\partial \dot{l}_{\vo{b}_k}}{\partial \q} \right)
\end{align*}
\begin{align*}
	% \textrm{where} \\
	\frac{\partial \dot{r}_k}{\partial \q} &= \frac{\partial \left(- \cfrac{{\nu}_k r_k \dot{l}_{\vo{b}_k}}
	{{l}_{\vo{b}_k}} \right)}
	{\partial \q}\\
	&= - \frac{{\nu}_k r_k}{{l}_{\vo{b}_k}} \frac{\partial \dot{l}_{\vo{b}_k}}{\partial \q} +
	 \frac{{\nu}_k r_k \dot{l}_{\vo{b}_k}}{{l}_{\vo{b}_k}^2} \frac{\partial {l}_{\vo{b}_k}}{\partial \q}
	 \\
	 \frac{\partial {l}_{\vo{s}_k}}{\partial \q} &= \frac{\q^T \vo{Y}_k^T \vo{Y}_k}{ {l}_{\vo{s}_k}} \\
 	\frac{\partial \dot{l}_{\vo{s}_k}}{\partial \q} &= \frac{\qd^T \vo{Y}_k^T \vo{Y}_k}{ {l}_{\vo{s}_k}}
 	 - \frac{\qd^T \vo{Y}_k^T \vo{Y}_k \q}{ {l}_{\vo{s}_k}^2} \frac{\partial {l}_{\vo{s}_k}}{\partial \q}
\end{align*}
\begin{align*}
	\frac{\partial \vo{M}_{\q}^T}{\partial \dot{\q}} \q = - &\sum_{k=1}^{n_b} \left[  \frac{1}{l_{\vo{b}_k}^3} \vo{X}_k^T \vo{X}_k \q \dot{l}_{\vo{b}_k}
	\frac{\partial \frac{dI_{\vo{b}_k}}{dt}}{\partial \qd} \right.
	\\
	&\left. + \frac{dI_{\vo{b}_k}}{dt} \frac{\vo{X}_k^T \vo{X}_k \q}{l_{\vo{b}_k}^3} \frac{\partial \dot{l}_{\vo{b}_k}}{\partial \qd} \right. \\
	& \left. -6 \frac{I_{\vo{b}_k} \vo{X}_k^T \vo{X}_k \q \dot{l}_{\vo{b}_k}}{l_{\vo{b}_k}^4} \frac{\partial \dot{l}_{\vo{b}_k}}{\partial \qd}  \right. \\
	&+	\left(\frac{I_{\vo{b}_k} \vo{X}_k^T \vo{X}_k}{l_{\vo{b}_k}^3} \right)
	\left(\frac{2 \vo{X}_k^T \vo{X}_k \q}{{l}_{\vo{b}_k}} \qd^T  \right. \\
	&\left. \left. - \frac{\q \q^T \vo{X}_k^T \vo{X}_k \dot{l}_{\vo{b}_k}}{{l}_{\vo{b}_k}^2} -\frac{\q \q^T \vo{X}_k^T \vo{X}_k \qd}{{l}_{\vo{b}_k}^2} \frac{\partial \dot{l}_{\vo{b}_k}}{\partial \qd} \right) \right]
\end{align*}
\begin{align*}
	\frac{\partial \vo{M}_{\dot{\q}}^T}{\partial \dot{\q}} \qd &= \frac{\partial \dot{\vo{M}}^T}{\partial \qd} \qd -\frac{\partial \vo{M}_f^T}{\partial \qd} \qd \\
	&= \sum_{k=1}^{n_b}\left[ -2\left( \frac{I_{\vo{b}_k}}{l_{\vo{b}_k}^3} \right)\vo{X}_k^T\vo{X}_k \qd \frac{\partial \dot{l}_{\vo{b}_k}}{\partial \qd} \right. \\
	& \quad+
	\left. \frac{1}{l_{\vo{b}_k}^2}\vo{X}_k^T\vo{X}_k \qd \frac{\partial \frac{dI_{\vo{b}_k}}{dt}}{\partial \qd} \right]
	\\
  &- \sum_{k=1}^{n_b} \frac{I_{\vo{b}_k}\vo{X}_k^T\vo{X}_k \qd}{l_{\vo{b}_k}^3}\frac{\partial \dot{l}_{\vo{b}_k}}{\partial \qd} \\
 &= \sum_{k=1}^{n_b}\left[ -3 \left( \frac{I_{\vo{b}_k}}{l_{\vo{b}_k}^3} \right)\vo{X}_k^T\vo{X}_k \qd \frac{\partial \dot{l}_{\vo{b}_k}}{\partial \qd} \right. \\
 & \left. \quad  + \frac{1}{l_{\vo{b}_k}^2}\vo{X}_k^T\vo{X}_k \qd \frac{\partial \frac{dI_{\vo{b}_k}}{dt}}{\partial \qd} \right]
\end{align*}
As before, for every $k^{th}$ string,
\begin{align*}
		 \frac{\partial \vo{f}_{d,k}}{\partial \qd} &= -c \frac{\vo{Y}_k \q}{{l}_{\vo{s}_k}} \frac{\partial \dot{l}_{\vo{s}_k}}{\partial \qd} \quad \textrm{hence,}
		 \quad \frac{\partial \vo{f}_{d}}{\partial \qd} = \sum_{k=1}^{n_s} \vo{Y}_k^T \frac{\partial \vo{f}_{d,k}}{\partial \qd}
\end{align*}
Here,
\begin{align*}
\frac{\partial \dot{l}_{\vo{b}_k}}{\partial \qd}&=\frac{\q^T\vo{X}_k^T\vo{X}_k}{l_{\vo{b}_k}}
\end{align*}
\begin{align*}
\frac{\partial \frac{dI_{\vo{b}_k}}{dt}}{\partial \qd}&=\frac{\partial}{\partial \qd}\left( \frac{m_{\vo{b}_k}}{12}(6r_k\dot{r}_k+2l_{\vo{b}_k}\dot{l}_{\vo{b}_k}) \right)\\
&=\left[ \frac{m_{\vo{b}_k}}{12}(-\frac{6r_k^2 \nu_k}{l_{\vo{b}_k}}+2l_{\vo{b}_k}) \right]\frac{\partial \dot{l}_{\vo{b}_k}}{\partial \qd}\\
&=\left[ \frac{m_{\vo{b}_k}}{12}(-\frac{6r_k^2 \nu_k}{l_{\vo{b}_k}^2}+2) \right]\q^T\vo{X}_k^T\vo{X}_k
\\
\frac{\partial \dot{l}_{\vo{s}_k}}{\partial \qd}&=\frac{\q^T\vo{Y}_k^T\vo{Y}_k}{l_{\vo{s}_k}}
\end{align*}
$\cfrac{\partial \vo{\xi}_3 }{ \partial \vo{\sigma}}$ can be computed as described in the procedure in \eqref{eqn:linSigma}.
Therefore,
\begin{align*}
	\frac{\partial \vo{\xi}_3 }{ \partial \vo{\sigma}} = -\vo{\hat{Y}}
\end{align*}
where $\hat{\vo{Y}}\in \real^{3n\times n_s}$ such that $\vec{\hat{\vo{Y}}} = \vo{Y}^T\q$, which is a rearrangement of the elements of column vector $\vo{Y}^T\q$ into a matrix of dimension $3n\times n_s$.
Similarly,
% \begin{align*}
% 	\vo{X}_{\q}^T = (\vo{\Psi}^T \otimes \vo{I}_{3n}) \vo{X}^T
% \end{align*}
% and therefore,
\begin{align*}
	\frac{\partial \vo{\xi}_3 }{ \partial \vo{\Psi}} = \vo{\hat{X}}
\end{align*}
where $\hat{\vo{X}}\in \real^{3n\times n_b}$ such that $\vec{\hat{\vo{X}}} = \vo{X}^T\q$, which is a rearrangement of the elements of column vector $\vo{X}^T\q$ into a matrix of dimension $3n\times n_b$.
Finally,
\begin{align*}
	\frac{\partial \vo{\xi}_3}{\partial \vo{f}} = \vo{I}
\end{align*}
Let
\begin{align*}
	\vo{\xi}_4 = \frac{\partial \vo{\xi}_3}{\partial \q} \delta \q + \frac{\partial \vo{\xi}_3}{\partial \dot{\q}} \delta \dot{\q} +
	\frac{\partial \vo{\xi}_3}{\partial \vo{\sigma}} \delta \vo{\sigma}
	+ \frac{\partial \vo{\xi}_3}{\partial \vo{\Psi}} \delta \vo{\Psi}
	+\frac{\partial \vo{\xi}_3}{\partial \vo{f}} \delta \vo{f}
\end{align*}
Then, since at equilibrium, $\ddot{\q} = 0$, equations are reduced to:
\begin{align*}
	\begin{bmatrix}
		\vo{M}_{\ddot{\q}} & -\Rq^T \\
		-\Rq & \vo{0}
	\end{bmatrix}
	\begin{bmatrix}
		\delta \ddot{\q} \\
		\delta \vo{\lambda}
	\end{bmatrix} =
	\begin{bmatrix}
		\vo{\xi}_4 \\
		\vo{0}
	\end{bmatrix}
\end{align*}
The matrix on the left is invertible, irrespective of configuration.
This means,
\begin{align}
	\begin{bmatrix}
		\delta \ddot{\q} \\
		\delta \vo{\lambda}
	\end{bmatrix} =
	\begin{bmatrix}
		\vo{M}_{\ddot{\q}} & -\Rq^T \\
		-\Rq & \vo{0}
	\end{bmatrix}^{-1}
	\begin{bmatrix}
		\vo{\xi}_4 \\
		\vo{0}
	\end{bmatrix}
	\label{eqn:linSysCompInv}
\end{align}
Let
\begin{align*}
		\vo{M}_{\vo{\alpha}} &= \begin{bmatrix}
			\vo{M}_{\ddot{\q}} & -\Rq^T \\
			-\Rq & \vo{0}
		\end{bmatrix}^{-1}, \quad \textrm{and} \\
		\vo{M}_{\vo{\beta}} &=\vo{M}_{\vo{\alpha}}(1:3n,1:3n),
\end{align*}
i.e., $\vo{M}_{\vo{\beta}}$ is derived from the first $3n$ rows and columns of $\vo{M}_{\vo{\alpha}}$.
Reducing \eqref{eqn:linSysCompInv}, we get:
\begin{align*}
	\delta \ddot{\q} = \vo{M}_{\vo{\beta}} \vo{\xi}_4
\end{align*}
Defining,
\begin{align*}
\vo{A} &:=
\begin{bmatrix}
\vo{0}&\vo{I}\\
  \vo{M}_{\vo{\beta}} \cfrac{\partial \vo{\xi}_3}{\partial \q}|_{\q_0,\dot{\q}_0,\vo{\sigma}_0,  \vo{\Psi}_0,\vo{f}_0}&
	\vo{M}_{\vo{\beta}} \cfrac{\partial  \vo{\xi}_3}{\partial \dot{\q}}|_{\q_0,\dot{\q}_0,\vo{\sigma}_0, \vo{\Psi}_0,\vo{f}_0}
\end{bmatrix}\\
\vo{B_{\vo{u}}} & :=  \begin{bmatrix}
\vo{0} & \vo{0}\\
 \vo{M}_{\vo{\beta}} \cfrac{\partial \vo{\xi}_3}{ \partial \vo{\sigma}}|_{\q_0,\dot{\q}_0,\vo{\sigma}_0, \vo{\Psi}_0,\vo{f}_0}&
   \vo{M}_{\vo{\beta}} \cfrac{\partial  \vo{\xi}_3}{ \partial \vo{\Psi}}|_{\q_0,\dot{\q}_0,\vo{\sigma}_0, \vo{\Psi}_0,\vo{f}_0}
\end{bmatrix}\\
\vo{B_f} &:=
\begin{bmatrix}
\vo{0}\\
\vo{M}_{\vo{\beta}} \cfrac{\partial \vo{\xi}_3}{\partial \vo{f}}|_{\q_0,\dot{\q}_0,\vo{\sigma}_0, \vo{\Psi}_0, \vo{f}_0}
\end{bmatrix}
\end{align*}
we can write the linear system as
\begin{align}
\dot{\vo{x}} = \vo{A} \vo{x} + \vo{B_u}\delta\vo{u} + \vo{B_f}\delta\vo{f}, \eqnlabel{linSysCompInvRed}
\end{align}
where
$\vo{x} :=\begin{bmatrix}\delta \vo{\q} \\ \delta\dot{\vo{q}}\end{bmatrix}$
and $\vo{u} :=\begin{bmatrix}\delta \vo{\sigma} \\ \delta{\vo{\Psi}}\end{bmatrix}$.\\
Again, note that \eqref{eqn:linSysCompInvRed} is not a minimal realization of the tensegrity system due to the presence of the constraints acting on $\delta \q$, despite the relaxation of bar length constraints. Linear constraints may still be acting on the system, thereby rendering it non-minimal.

\subsection{Summary}
Section 3 presented a detailed description of the derivations that had to be amended in the general multibody system approach to account for compressibility in bars.
A gist of the approach is presented as follows.
\begin{enumerate}
\item \textbf{Energies}: Additional terms will be introduced into the kinetic and potential energies to account for flexibility in bars.
\begin{align*}
	{T} &= \frac{1}{2}\qd^T \vo{M} \qd  - T_f\\
	V_s &= \frac{1}{2} \q^T  \vo{Y} (\frac{\vo{\sigma^2}}{\vo{K}_s} \otimes \vo{I}_{3n}) \q + \frac{1}{2} \q^T  \vo{Y} (\frac{\vo{\psi^2}}{\vo{K}_b} \otimes \vo{I}_{3n}) \q
\end{align*}
Here, $T_f$ is the kinetic energy injected to account for longitudinal deformation in the bar, i.e., $\dot{l}_{\vo{b}_k}$ is no longer zero.
In the absence of nonlinear bar length constraints, for a generic system,
\begin{align*}
	\vo{R}(\q) := \begin{bmatrix}
		\vo{A}\q - \vo{b}
\end{bmatrix}
\end{align*}
\item \textbf{Equations of Motion}:
The governing equations of motion have now been amended to:
\begin{align*}
\vo{M}_{\ddot{\q}}\ddot{\q} - \vo{R}_{\vo{q}}^T\vo{\lambda} =& -\left[\vo{M}_{\q}^T +
(\vo{\Psi}^T \otimes \vo{I}_{3n})\vo{X}^T \right. \notag\\
 &\left. +(\vo{\sigma}^T \otimes \vo{I}_{3n})\vo{Y}^T\right]\q \notag\\
&-\vo{M}_{\dot{\q}}^T\dot{\q} + \vo{G} + \vo{f} + \vo{f}_d%:=\vo{\xi}_3,
\end{align*}
% The holonomic constraints in $\vo{R}$ do not contain bar length constraints anymore.
Linearization follows the same approach described as before, but the introduction of elasticity necessitates re-evaluation of state and control matrices.
\end{enumerate}
% }

\section{Example}
\renewcommand{\thefootnote}{\fnsymbol{footnote}}
In this section, we model a simple tensegrity structure in 2 dimensions as an example to demonstrate the accuracy of the constraints and motion trajectories using the proposed approach. We also compare our results with those obtained using a commercial tool, i.e. Simscape \cite{matlab}. In addition, a robotic arm and a ball based on tensegrity structures as two examples are presented to show the efficiency of the method applied to models with higher complexity. In particular, corrected numerical integration is utilized by Matlab to these different models. The equation of motion \eqref{eqn:finalEqn} is integrated based on the Dormand-Prince method \cite{dormand1980family,shampine1997matlab} with relative and absolute tolerances of $10^{-10}$, both constraint correction and energy correction turned on, and the given threshold $\gamma=10^{-10}$.

\begin{figure}[!ht]
\centering
\subfigure[A 2-bar structure in 2 dimensions.]{\includegraphics[width=0.32\textwidth]{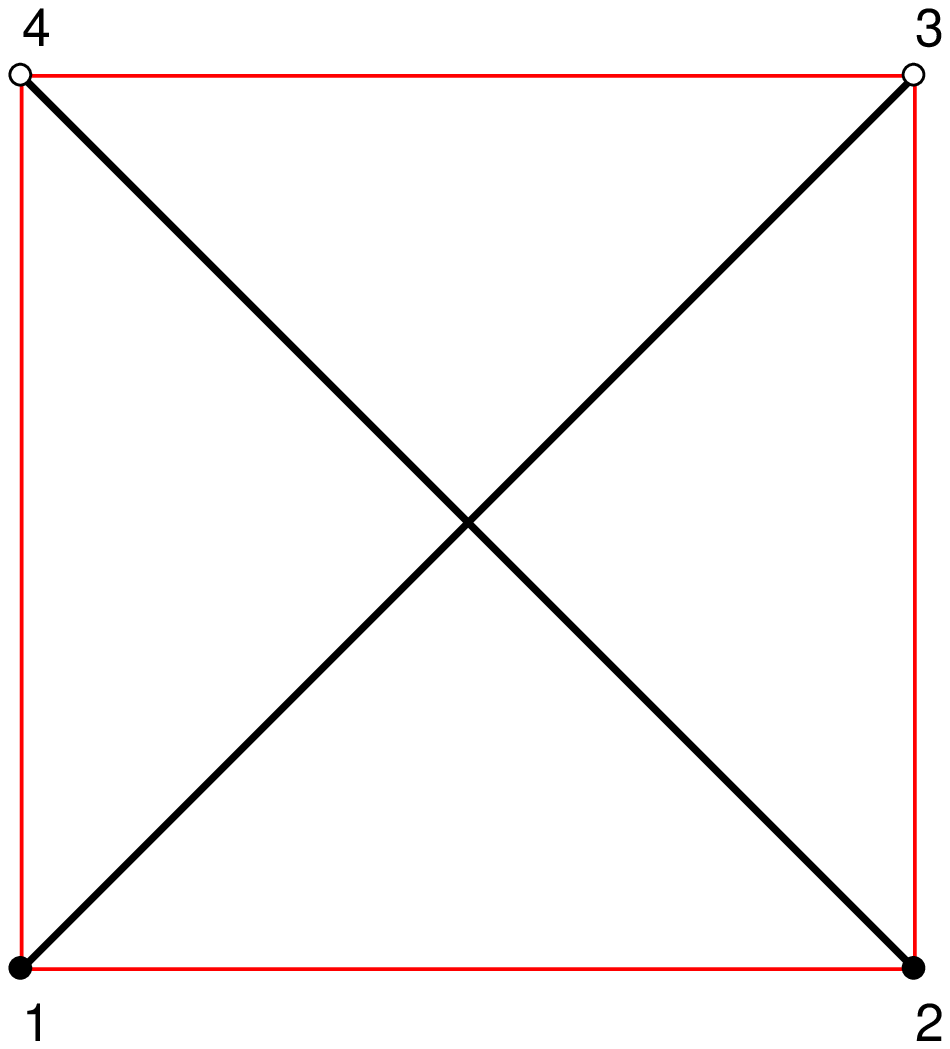}\label{fig:Tbar}}
\subfigure[A robotic arm in 2 dimensions.]{\includegraphics[width=0.32\textwidth]{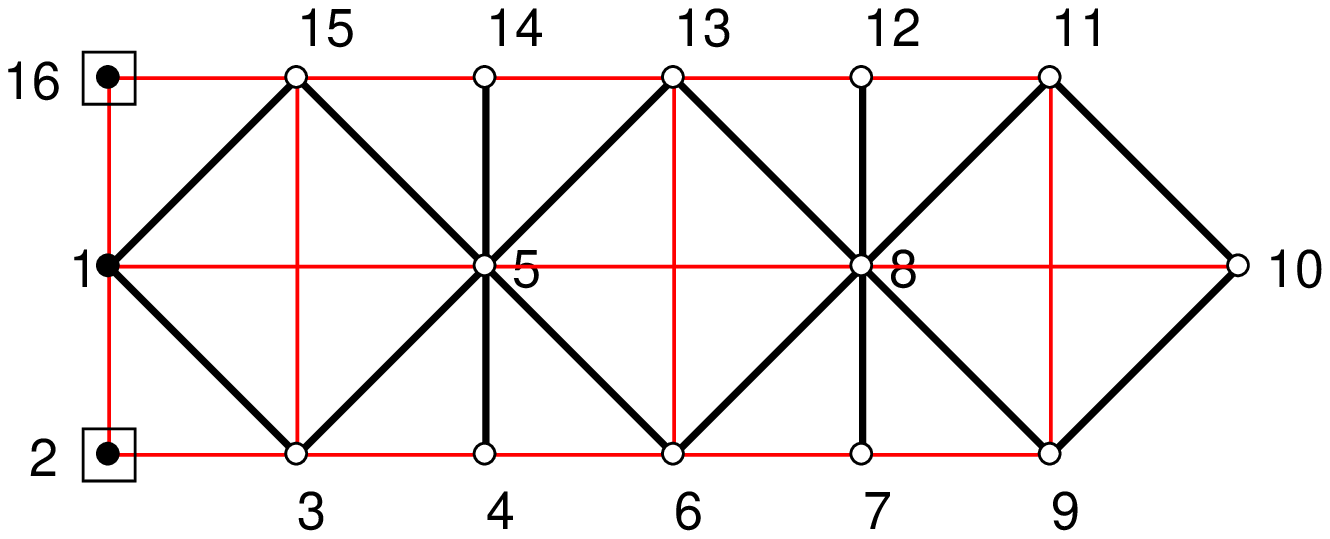}\label{fig:Arm}}
\subfigure[A ball in 3 dimensions.]{\includegraphics[width=0.32\textwidth]{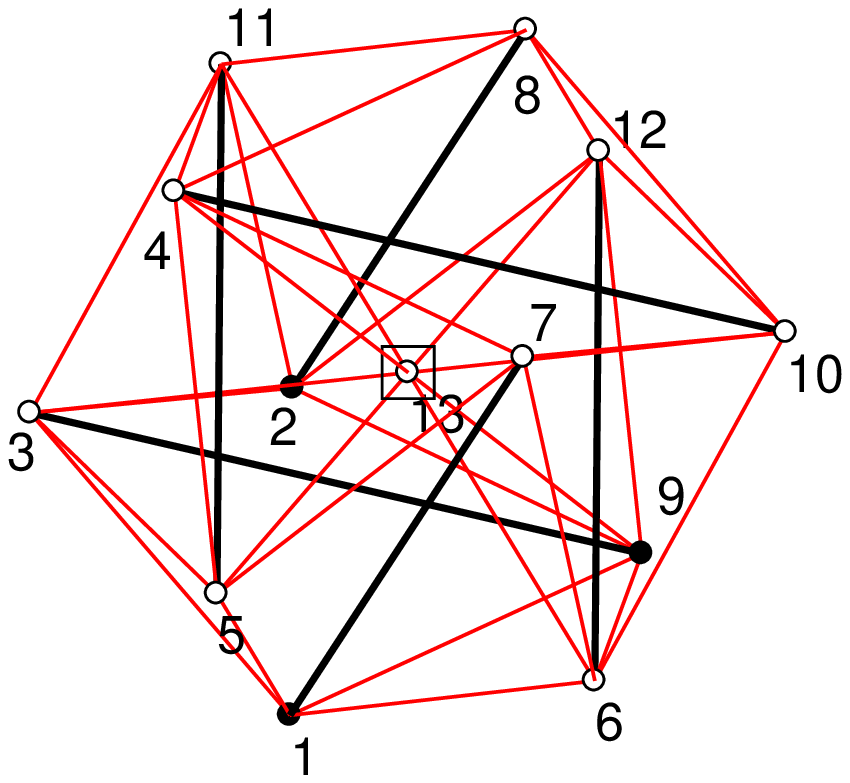}\label{fig:ball}}
\caption{Structures of the examples where red lines indicate strings, black lines indicate rigid bars, squares indicate point masses, black dots are fixed nodes, white dots are free nodes, and numbers are node notations.}
\end{figure}

First of all, a 2D tensegrity structure built using 2 bars and 4 springs shown in Figure \ref{fig:Tbar} is used, where the nodes at the bottom are fixed and the left and right springs are pre-stressed at $90\%$ of the rest length of the springs. Secondly, a robotic arm built from 3 sets of squares is shown in Figure \ref{fig:Arm}, where strings made of nylon are prestressed so that the structure is in equilibrium under gravity. An external force of a time-dependent sinusoidal function is applied vertically to the tip of the arm. Thirdly, a 3D ball with a payload is shown in Figure \ref{fig:ball}, where 6 bars and 32 strings are used. Here too, strings are prestressed so that the structure can be in equilibrium under gravity. A time-dependent external force of a sinusoidal function is given to  the top 3 nodes in different directions, i.e. along x,y,z axes respectively (in order of numbering).

\begin{table*}[t]
\centering
\begin{tabular}[center]{|l||*{3}{r|}}\hline
	\diagbox{Attribute}{Model}
	&\makebox[3em]{T-Bar}&\makebox[3em]{Arm}&\makebox[3em]{Ball}\\
	\hline \hline
	Gravity ($m/s^2$) & 0 & -9.806 & -9.806\\
\hline
	Bars: Length ($m$) & 5  & 1 & 1 \\
	Bars: Radius ($m$) & 0.05 & 0.01 & 0.01\\
	Bars: Density ($kg/m^3$)& 500 & 1300 & 1300\\
\hline
	Springs: Stiffness ($N/m$)& 100 & - & - \\
	Springs: Rest Length Percentage (\% of initial) & 90 (vertical only) & - & -\\
\hline
	Strings: Young's Modulus ($GPa$)& 2 & 2 & 2 \\
	Strings: Radius ($m$)& 0.001 & 0.001 & 0.001\\
%	Strings: Rest Length (\% of initial) & - & 80 & 90 \\
 \hline
	External Force ($N$)& 0 & $300\times \sin(t)$ & $300\times \sin(t)$ \\
\hline
\end{tabular} \\
\caption{User-defined properties of the 3 models}
\label{tab:prop}
\end{table*}

% \begin{table}
% \centering
% \begin{tabular}[center]{|l | p{5 cm}|}
% \hline
% 	Attribute & Value \\
% \hline
% 	Gravity & -9.806 $m/s^2$\\
% \hline
% 	Bars: Length & 1 $m$ \\
% 	Bars: Mass & 1 $kg$ \\
% \hline
% 	Strings: Young's Modulus & 2E9 $N/m^2$ \\
% 	Strings: Radius & 0.001 $m$ \\
% 	Strings: Rest Length & 0.8 (all strings) \\
% \hline
% 	External Force & $300\times sin(t)$ $N$ (applied vertically at tip) \\
% \hline
% \end{tabular}
% \caption{User-defined properties of the Arm}
% \label{tab:prop_Arm}
% \end{table}
%
% \begin{table}
% \centering
% \begin{tabular}{|l | p{5 cm}|}
% \hline
% 	Attribute & Value \\
% \hline
% 	Gravity & -9.806 $m/s^2$\\
% \hline
% 	Bars: Length & 1 $m$ \\
% 	Bars: Radius & 0.01 $m$ \\
% 	Bars: Density & 1300 $kg/m^3$ \\
% \hline
% 	Strings: Young's Modulus & 2E9 $N/m^2$ \\
% 	Strings: Radius & 0.001 $m$ \\
% 	Strings: Rest Length & 0.9 (all strings) \\
% \hline
% 	External Force & $300\times sin(t)$ $N$ (applied in x,y,z directions respectively on top 3 nodes in order)\\
% \hline
% \end{tabular} \\
% \caption{User-defined properties of the Ball}
% \label{tab:prop_Ball}
% \end{table}

\begin{figure*}[!ht]
\centering
\subfigure[Minimum Realization vs Proposed Method]
{\includegraphics[width=0.48\textwidth]
{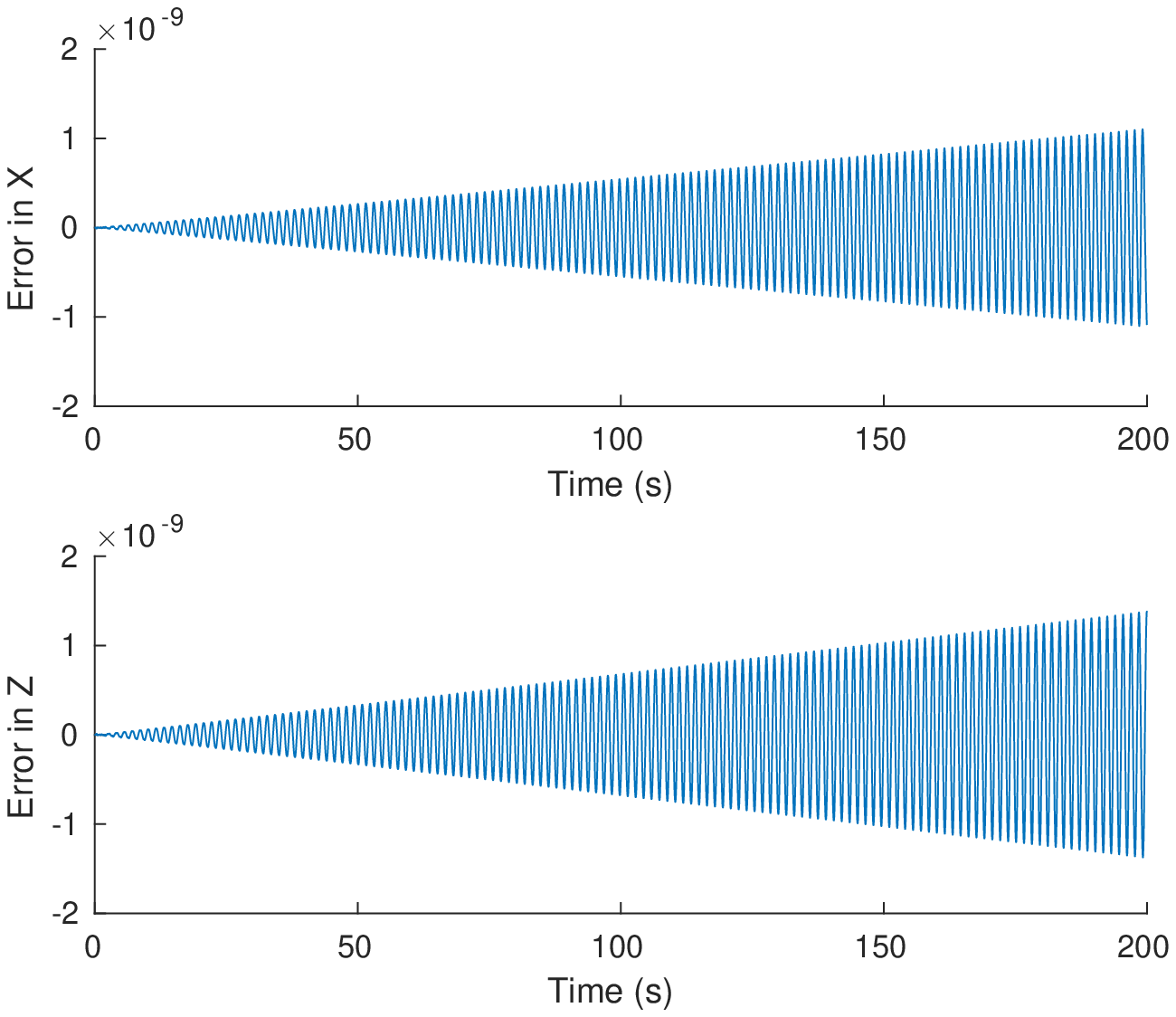}
\label{fig:motion_TbarL}
}\hfill
% \subfigure[Node 4.]
% {\includegraphics[width=0.48\textwidth]{figs/TBAR_MotionError_Node4LMin}\label{fig:motion_TbarL2}}\hfill
% \caption{Difference in motion of node 3 between minimum realization and proposed method in the example of T-bar.}
% \end{figure}
%
% \begin{figure}[!ht]
% \centering
\subfigure[Minimum Realization vs Simscape Multibody]
{\includegraphics[width=0.48\textwidth]
{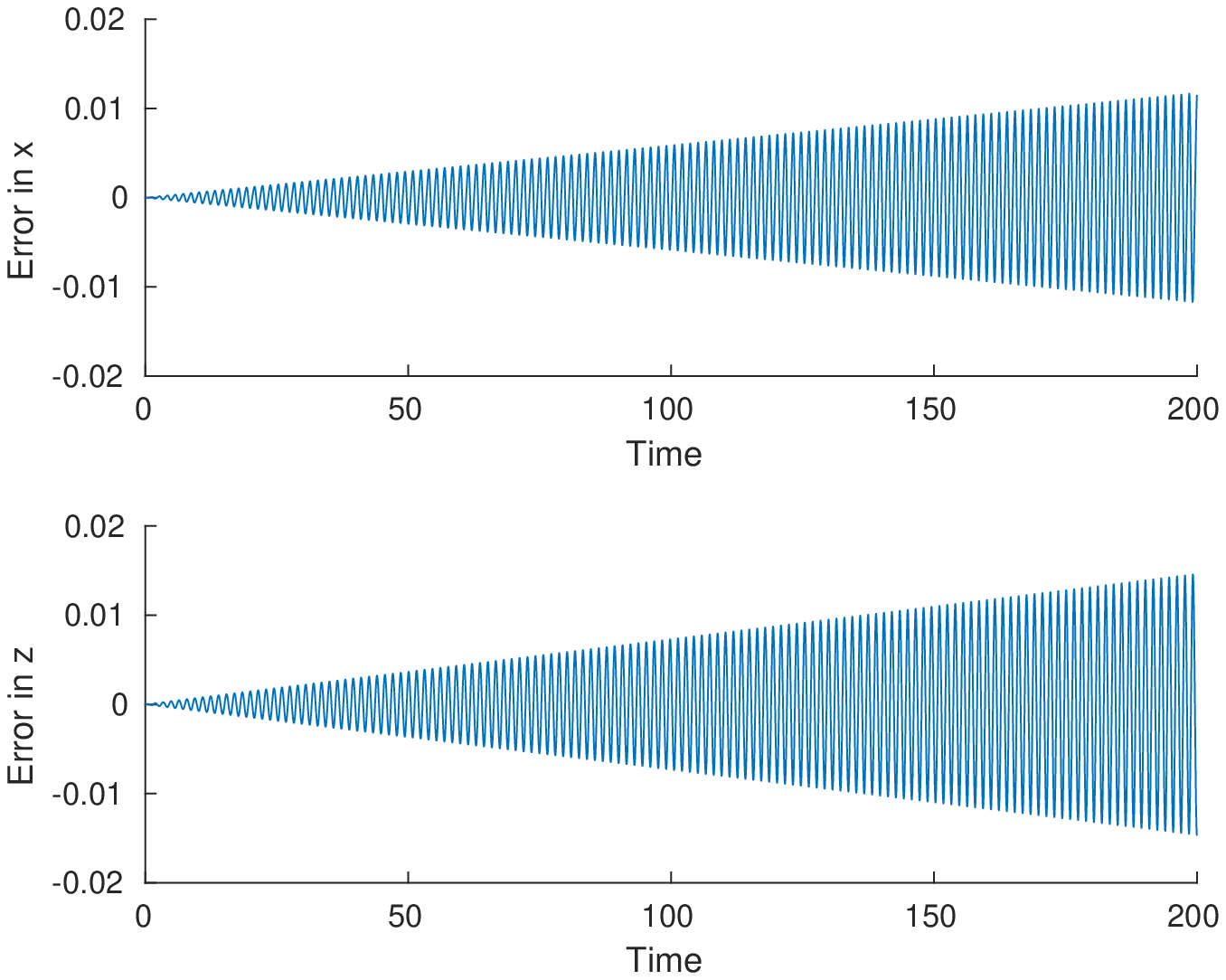}
\label{fig:motion_TbarS}}\hfill
% \subfigure[Node 4.]
% {\includegraphics[width=0.48\textwidth]{figs/TBAR_MotionError_Node4SMin}\label{fig:motion_TbarS2}}\hfill
\caption{Difference in motion of node 3 between minimum realization, the proposed method and Simscape Multibody in the example of the 2-bar structure shown in Figure \ref{fig:Tbar}.}
\end{figure*}

\begin{figure*}[!ht]
	\centering
\subfigure[Bar 1 length error]
{\includegraphics[width=0.48\textwidth]
{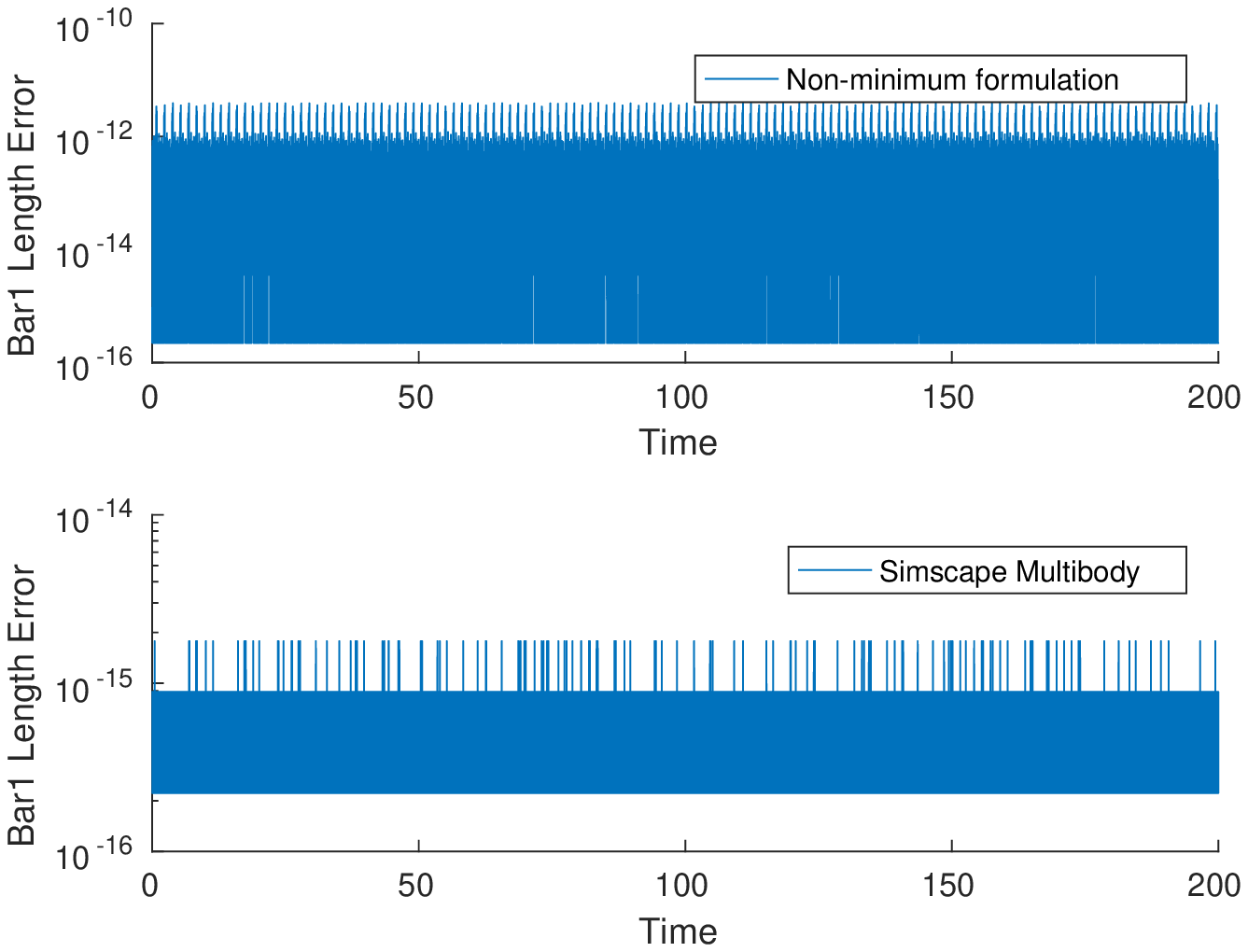}
\label{fig:Bar1LengthErrorLSim}}\hfill
% \caption{Bar Length constraint violation observed for Bar 1.}
% \end{figure}
%
% \begin{figure}[!ht]
% \centering
\subfigure[Bar 2 length error]
{\includegraphics[width=0.48\textwidth]
{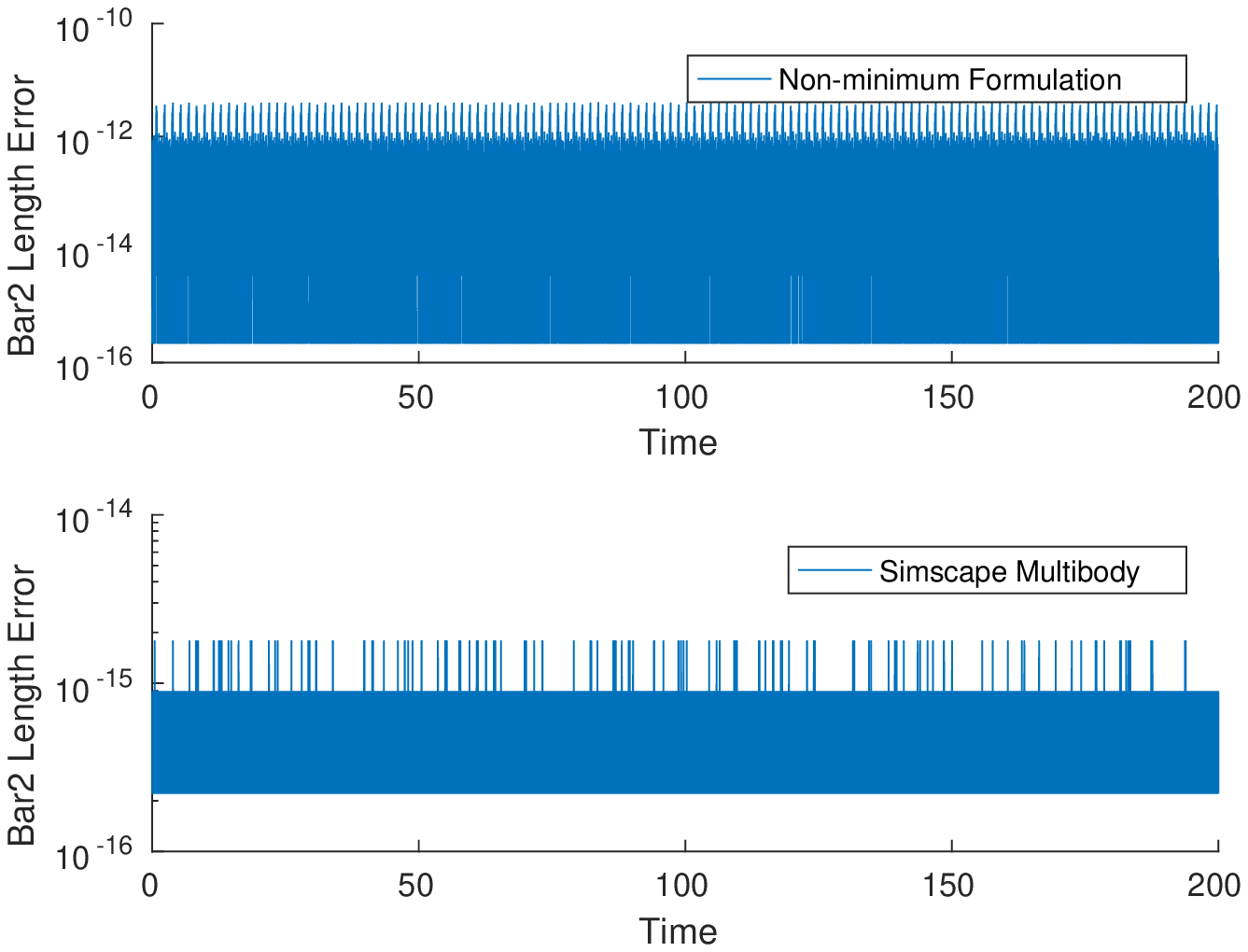}
\label{fig:Bar2LengthErrorLSim}}\hfill
\caption{Bar Length constraint violations observed in the example of the 2-bar structure described in Figure \ref{fig:Tbar}.}
\end{figure*}

\begin{figure*}[!ht]
\centering
% \subfigure[Energy violation of \eqref{eqn:conservation}.]
{\includegraphics[width=0.4\textwidth]
{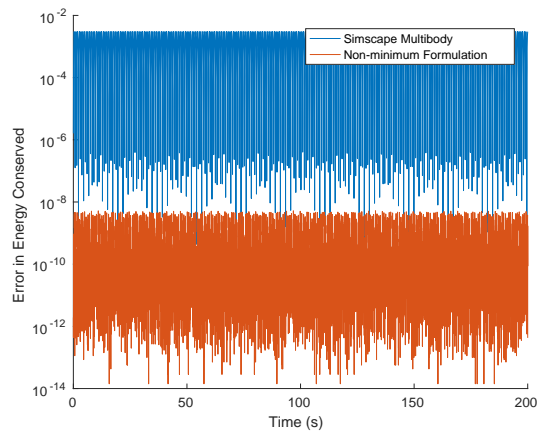}}\hfill
\caption{Energy violation \eqref{eqn:conservation} observed in the example of the 2-bar structure shown in Figure \ref{fig:Tbar}.}
\label{fig:EnergyErrorLSim}
\end{figure*}

Figure \ref{fig:motion_TbarL} shows the discrepancies between the motion trajectories obtained using the proposed approach and the minimum realization, where we consider the latter as the benchmark since the equations of motion are derived using generalized coordinates that preserve the geometric constraints and the relative and absolute tolerance of numerical integration is $2.2\times 10^{-14}$ (This is the minimum value of relative tolerance that can be chosen in Matlab.)
% \comment{( footnote causes me error so I use parenthesis here.)}
%\footnote{This is the minimum value of relative tolerance that can be chosen in Matlab.}
and $10^{-14}$ respectively, while Figure \ref{fig:motion_TbarS} shows the differences between Simscape and minimum realization, where Simscape is with the same numerical method and tolerance as minimum realization. Comparing the figures, we observe that our proposed method produces the motion $10^7$ times closer to the benchmark than Simscape, which indicates a significant improvement in accuracy.
Figures \ref{fig:Bar1LengthErrorLSim}, \ref{fig:Bar2LengthErrorLSim} and \ref{fig:EnergyErrorLSim} present the magnitude of constraint violations in bar length and total energy, which shows that the proposed method of constraint correction reduces the violation of the energy to around $10^{-11}$ and keeps the bar length constraint violations at about $10^{-12}$ simultaneously. One can observe that the bar length violations in the simulation produced using Simcape are of a smaller magnitude. This is because it utilizes generalized coordinates, thereby automatically satisfying the bar length constraints. The differences in motion seem to be increasing in time, but in fact, simulating for very long durations would show them to be within bounds, as can also be said by observing the total energy variation.
Since the energy is always stable from Figure \ref{fig:EnergyErrorLSim}, the motion must be stable and therefore, the differences are all bounded. The attached video \textit{TEST\_TBAR.mp4} demonstrates the motions of the 2-bar structure with 3 approaches in real time.

% \textcolor{blue}{
We also compared the time required to run a 10 second simulation in each of the approaches being compared here.
Simscape Multibody required $\sim$ 1.1 seconds on average to run a T-bar simulation, while the proposed approach required $\sim$ 2.4 seconds with correction built in.
We expect to see a bigger difference when simulating a model with higher complexity, as with more bars present in the structure, more computational effort would be required to prevent constraint violation.
The computational advantage of Simscape could also be due to several code optimizations, and an underlying mixture of programming engines including C and Fortran that gives it the additional speed up.
These are implementation details that can be accounted in our formulation as well.
Our code currently does not have any speed optimization, and we will address it in our future work.
Nonetheless, our framework allows for easier modeling of tensegrity structures, which becomes prohibitively complex in Simscape.
% }

% \begin{table*}[!ht]
% \centering
% \begin{tabular}[center]{|l||*{2}{r|}}\hline
% 	\diagbox{Configuration}{Approach}
% 	&\makebox[7em]{Proposed Method}&\makebox[7em]{Simscape}\\
% 	\hline \hline
% 	D-bar  & 23.645 s. & 4.0721 s.\\
% \hline
% 	T-bar  & 2.43 s. & 1.0947 s. \\
% 	\hline
% \end{tabular} \\
% \caption{Time required to run same simulation in proposed method vs. Simscape Multibody}
% \label{tab:compTimeSimLag}
% \end{table*}

\begin{figure*}[!ht]
% \centering
\subfigure[Motion Error of Node 3.]
{\includegraphics[width=0.48\textwidth]
{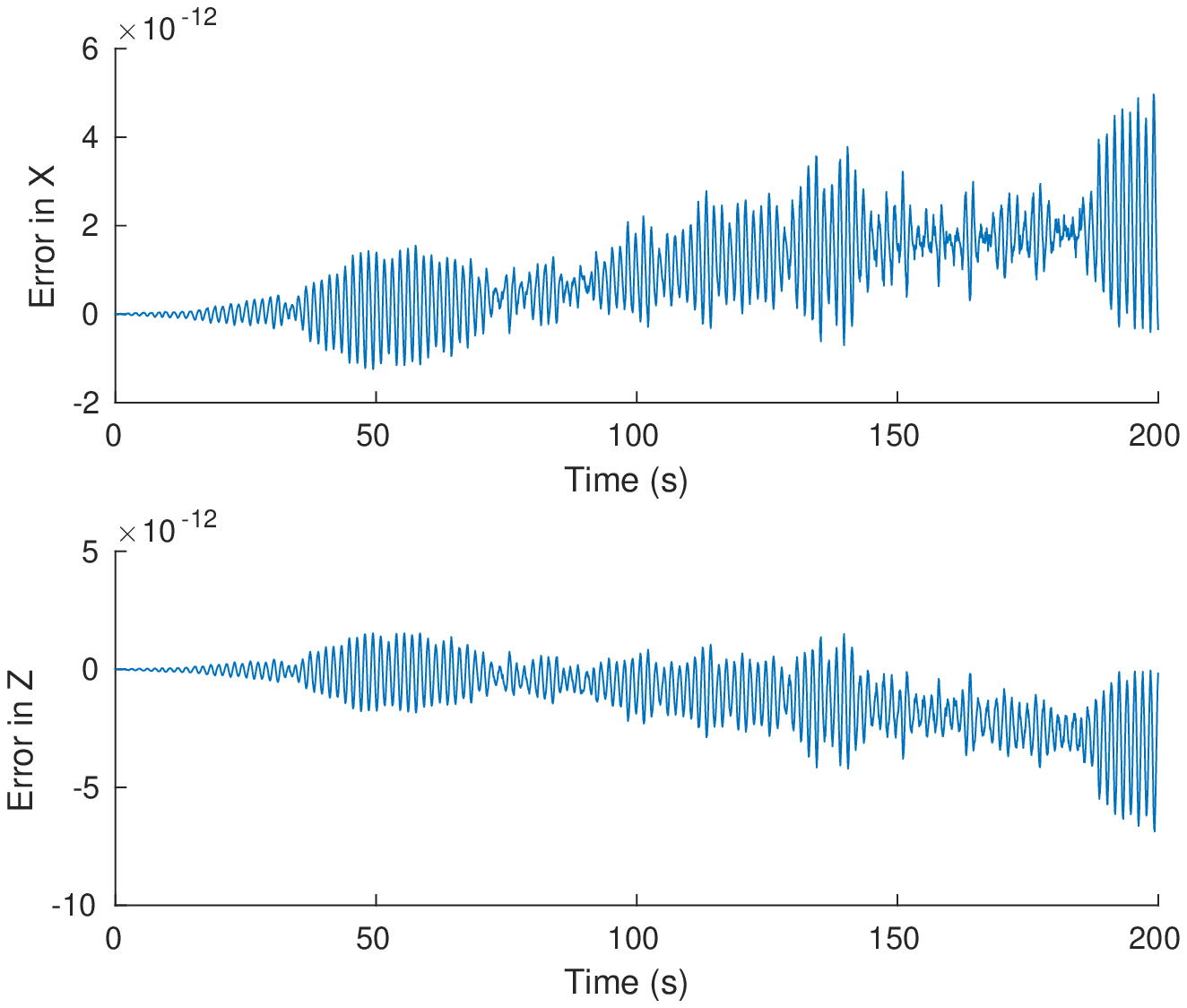}
\label{fig:MotionTol14_Tbar}}\hfill
% \caption{Difference in motion of node 3 between minimum realization and proposed method simulated at $10^{-14}$ tolerance.}
% \end{figure}
% \begin{figure}[!ht]
% \centering
\subfigure[Energy violation \eqref{eqn:conservation}.]
{\includegraphics[width=0.48\textwidth]
{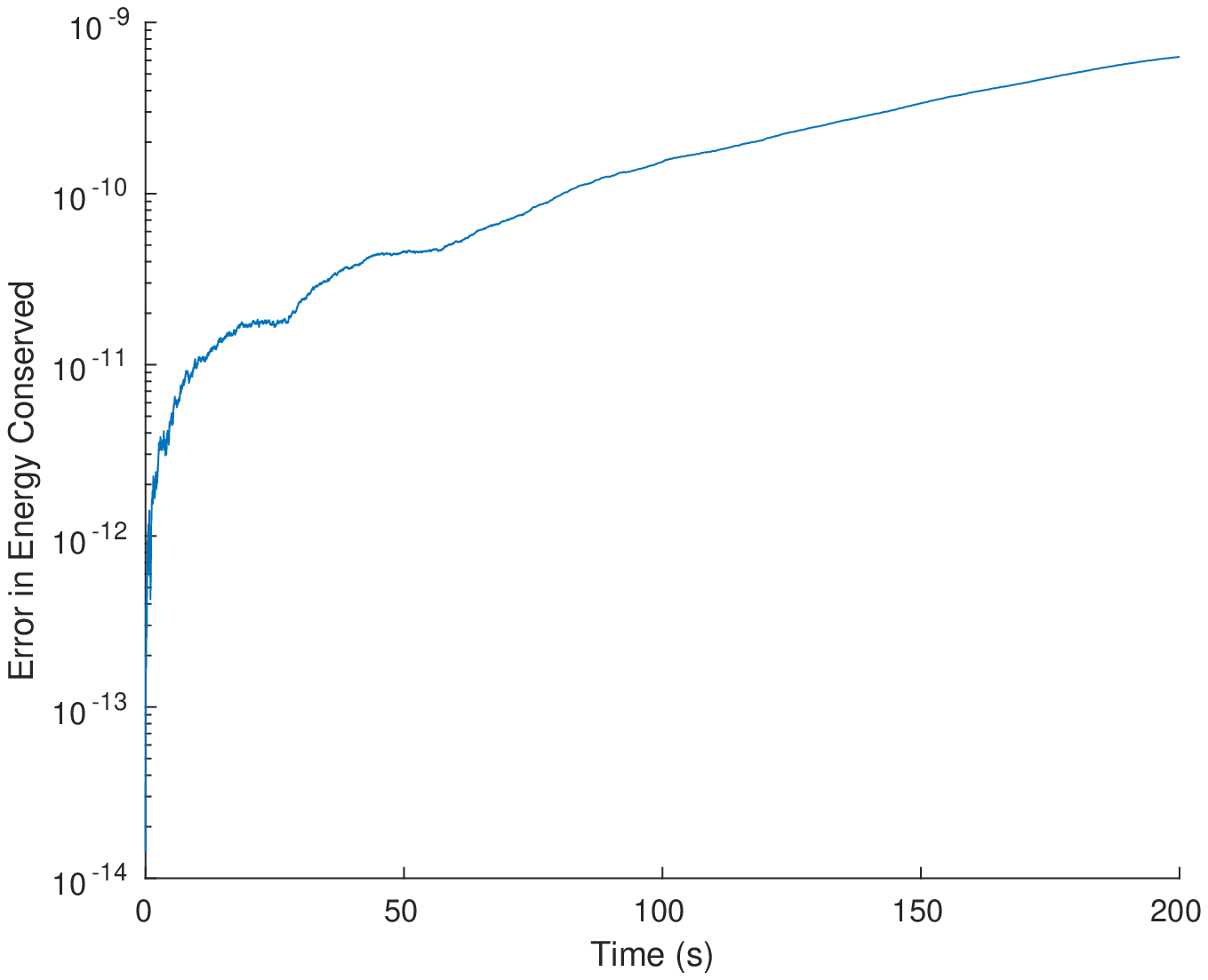}
\label{fig:EnergyErrorTol14_Tbar}}\hfill
% \caption{Energy violation of \eqref{eqn:conservation} at $10^{-14}$ tolerance.}
% \end{figure}

% \begin{figure}[!ht]
% 	\centering
\subfigure[Bar 1 length error]
{\includegraphics[width=0.48\textwidth]
{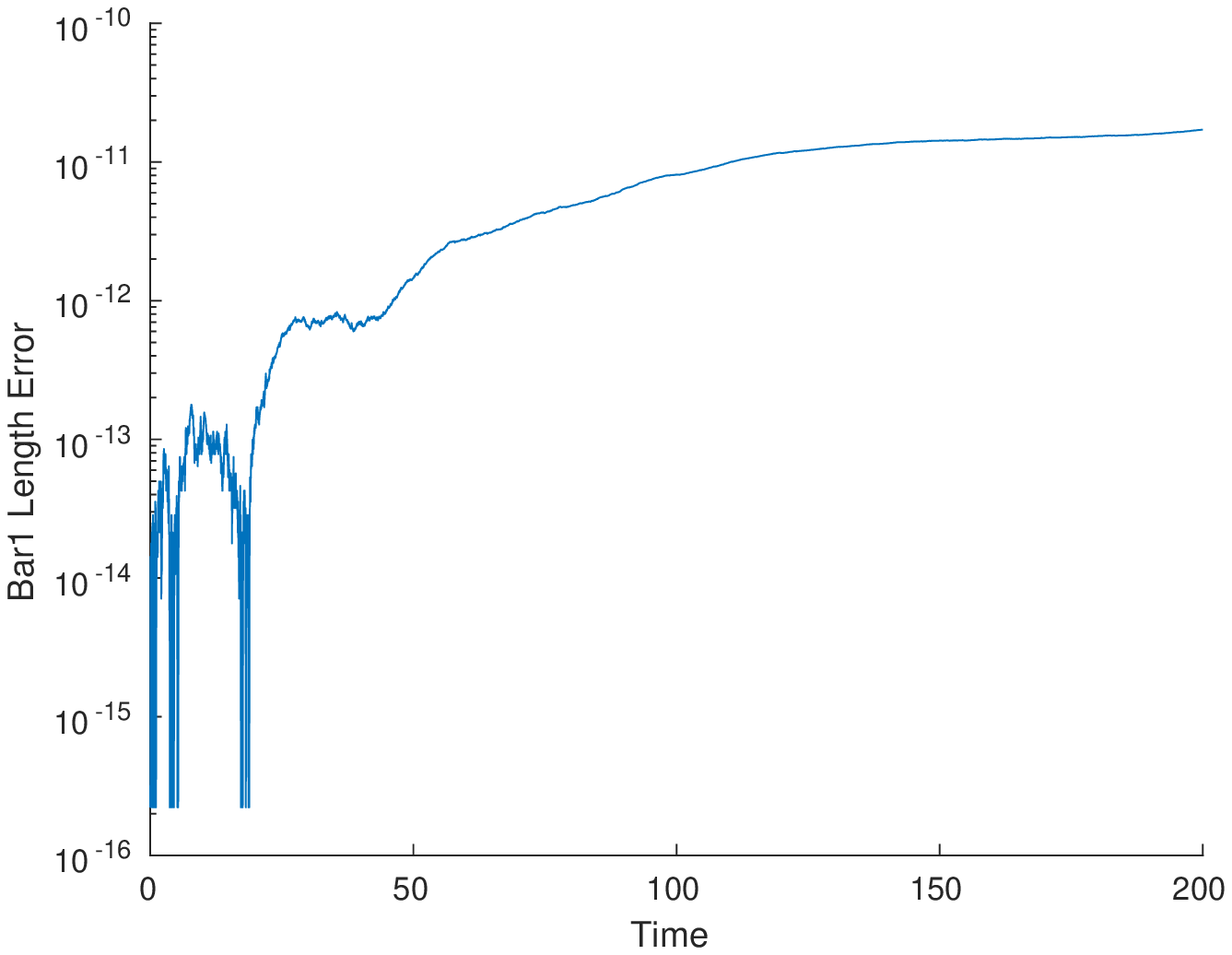}
\label{fig:Bar1LengthErrorTol14_Tbar}}\hfill
% \caption{Bar Length constraint violation observed for Bar 1 at $10^{-14}$ tolerance.}
% \end{figure}
%
% \begin{figure}[!ht]
% \centering
\subfigure[Bar 2 length error]
{\includegraphics[width=0.48\textwidth]
{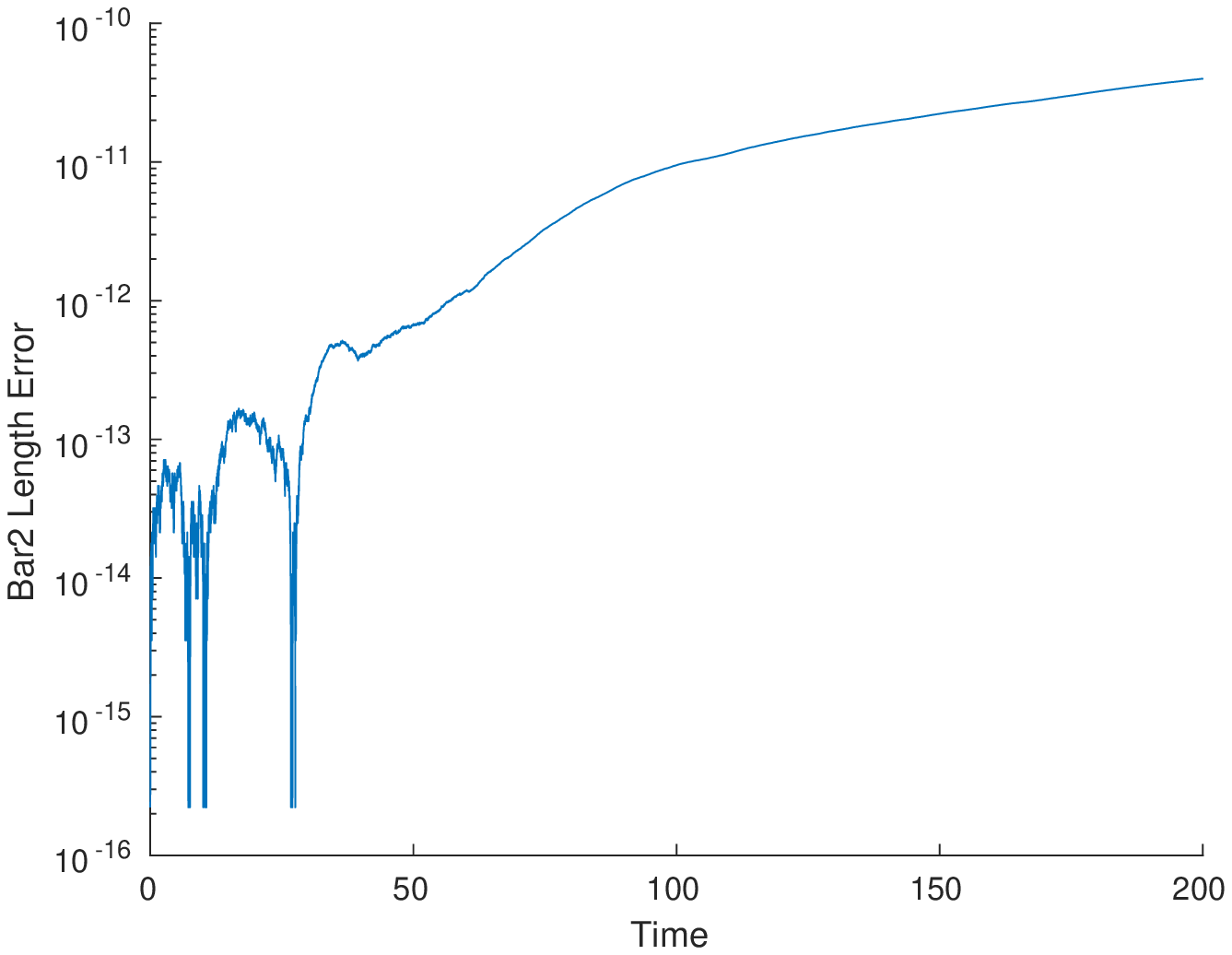}
\label{fig:Bar2LengthErrorTol14_Tbar}}\hfill
% \caption{Bar Length constraint violation observed for Bar 2 at $10^{-14}$ tolerance.}
\caption{Motion error, constraint and energy violations if simulated at $10^{-14}$ tolerance without any correction in the example of the 2-bar structure shown in Figure \ref{fig:Tbar}.}
\end{figure*}

\begin{figure*}[!ht]
\centering
% \subfigure[Energy violation of \eqref{eqn:conservation}.]
{\includegraphics[width=0.5\textwidth]
{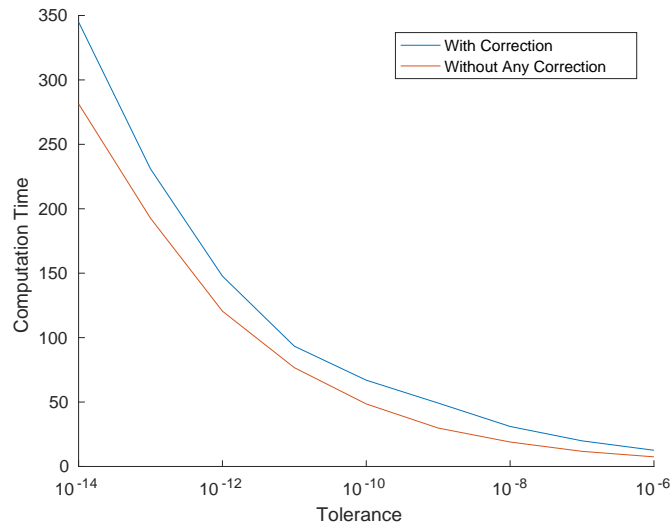}}\hfill
\caption{Computation Times for the 2-bar structure example, simulated at different tolerances. }
\label{fig:CompTime_Tbar}
\end{figure*}

It is important to note here that, when simulating tensegrity systems with non-minimum coordinates at machine-level precision tolerance settings (at the expense of speed), the user might be tempted to do away with correction altogether. However, the solution, while still being accurate in its motion trajectory for short time lengths, drifts away from the constraint space and consequently, tends to become inaccurate if simulated for long durations. As Yoon\cite{YoonThesis} points out, it is a necessary condition for accurate simulation that both geometric and energy constraints be satisfied during integration.
Hence, it would be advisable to keep the correction algorithm turned on at all times.
Figure \ref{fig:MotionTol14_Tbar} shows how accurately the non-minimum formulation performs without the need for correction at the tolerance settings of $10^{-14}$, and the constraint and energy violations as presented in Figure \ref{fig:EnergyErrorTol14_Tbar}, \ref{fig:Bar1LengthErrorTol14_Tbar},
and \ref{fig:Bar2LengthErrorTol14_Tbar} present the extent of the associated constraint drift.
Figure \ref{fig:CompTime_Tbar} shows the plot for computation times for the T-bar example, simulated at different tolerance settings ranging from $10^{-6}$ to $10^{-14}$. This goes to show that if computation speeds are a higher priority than accuracy, it would be much more prudent to perform simulations at lower tolerance settings like $10^{-9}$ or $10^{-10}$ with correction turned on than to do it at $10^{-14}$ without any correction at all.

\begin{figure*}[!ht]
\centering
\subfigure[Motion Error of Node 3.]
{\includegraphics[width=0.48\textwidth]
{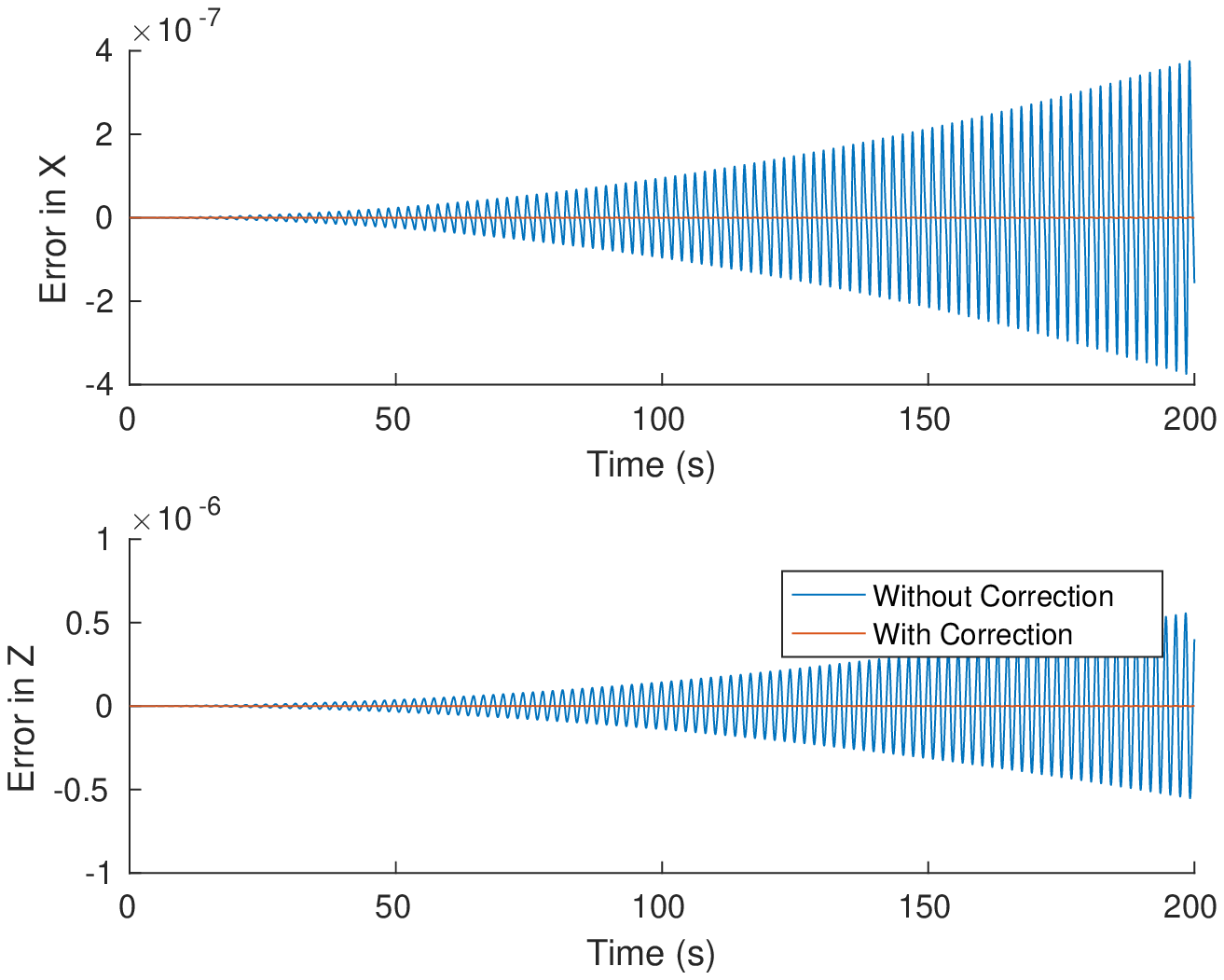}
\label{fig:MotionErrTol10EnergyCorrection_Tbar}}\hfill
% \caption{Difference in motion of node 3 in the T-bar example, observed with and without energy correction at $10^{-10}$ relative and absolute tolerance.}
% \end{figure}
%
% \begin{figure}[!ht]
% \centering
\subfigure[Energy violation of \eqref{eqn:conservation}.]
{\includegraphics[width=0.48\textwidth]
{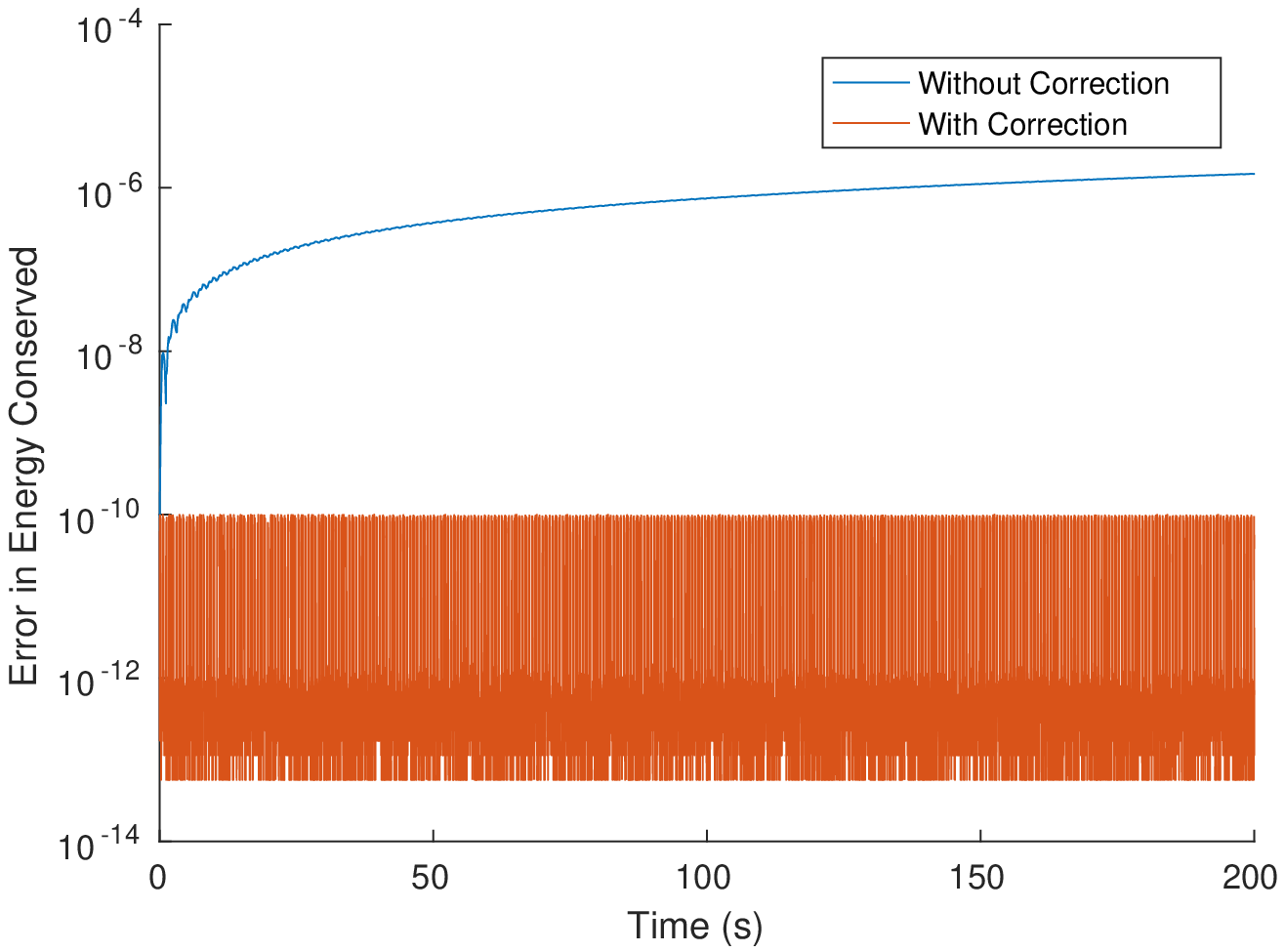}
\label{fig:EnergyErrTol10EnergyCorrection_Tbar}}\hfill
% \caption{Bar Length constraint violation observed for Bar 2 in the T-bar example, observed with and without energy correction at $10^{-10}$ relative and absolute tolerance.}
% \end{figure}
%
% \begin{figure}[!ht]
% \centering
\subfigure[Bar 1 Length Error.]
{\includegraphics[width=0.48\textwidth]
{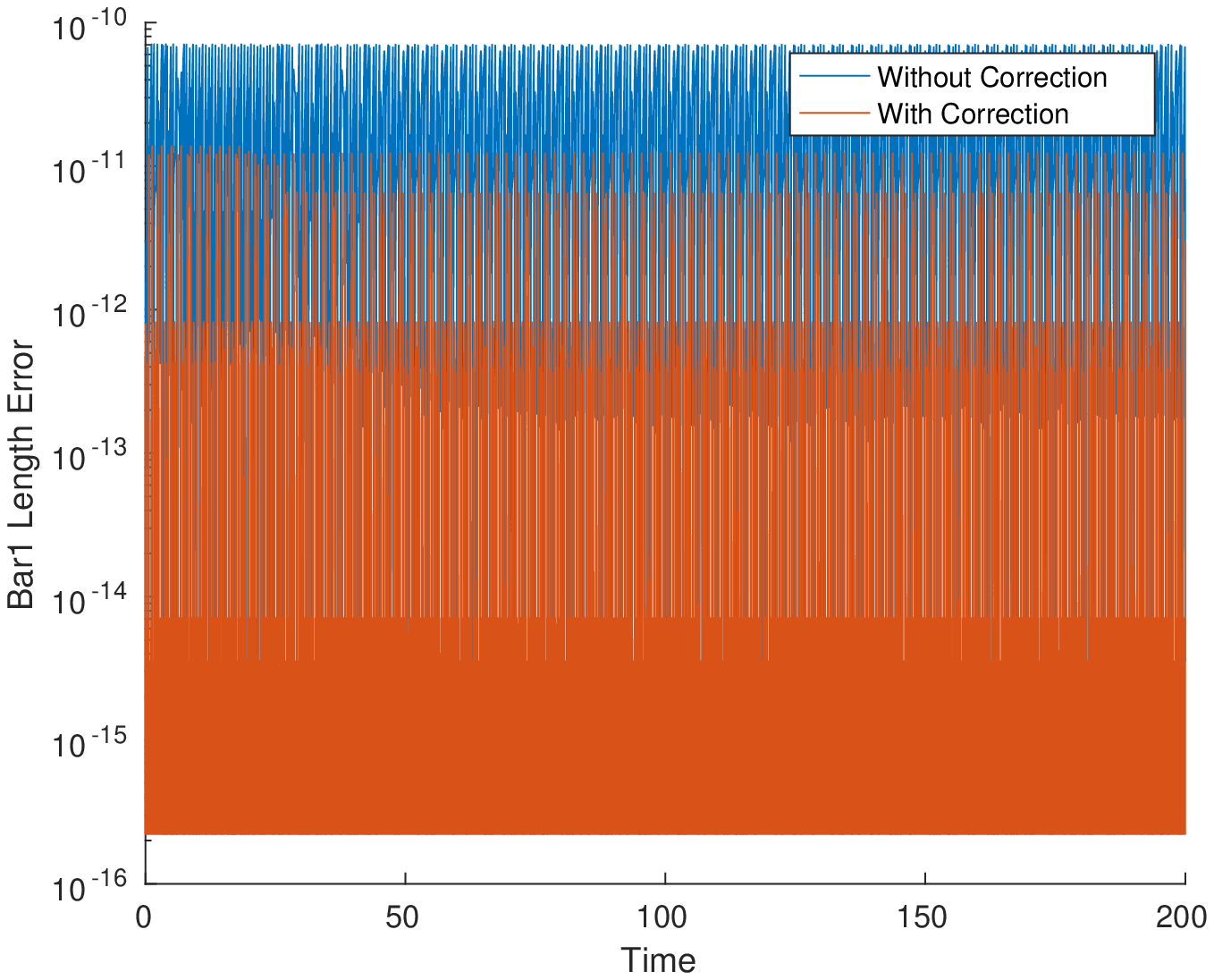}
\label{fig:Bar1LengthErrTol10EnergyCorrection_Tbar}}\hfill
% \caption{Bar Length constraint violation observed for Bar 1 in the T-bar example, observed with and without energy correction at $10^{-10}$ relative and absolute tolerance.}
% \end{figure}
%
% \begin{figure}[!ht]
% \centering
\subfigure[Bar 2 Length Error.]
{\includegraphics[width=0.48\textwidth]
{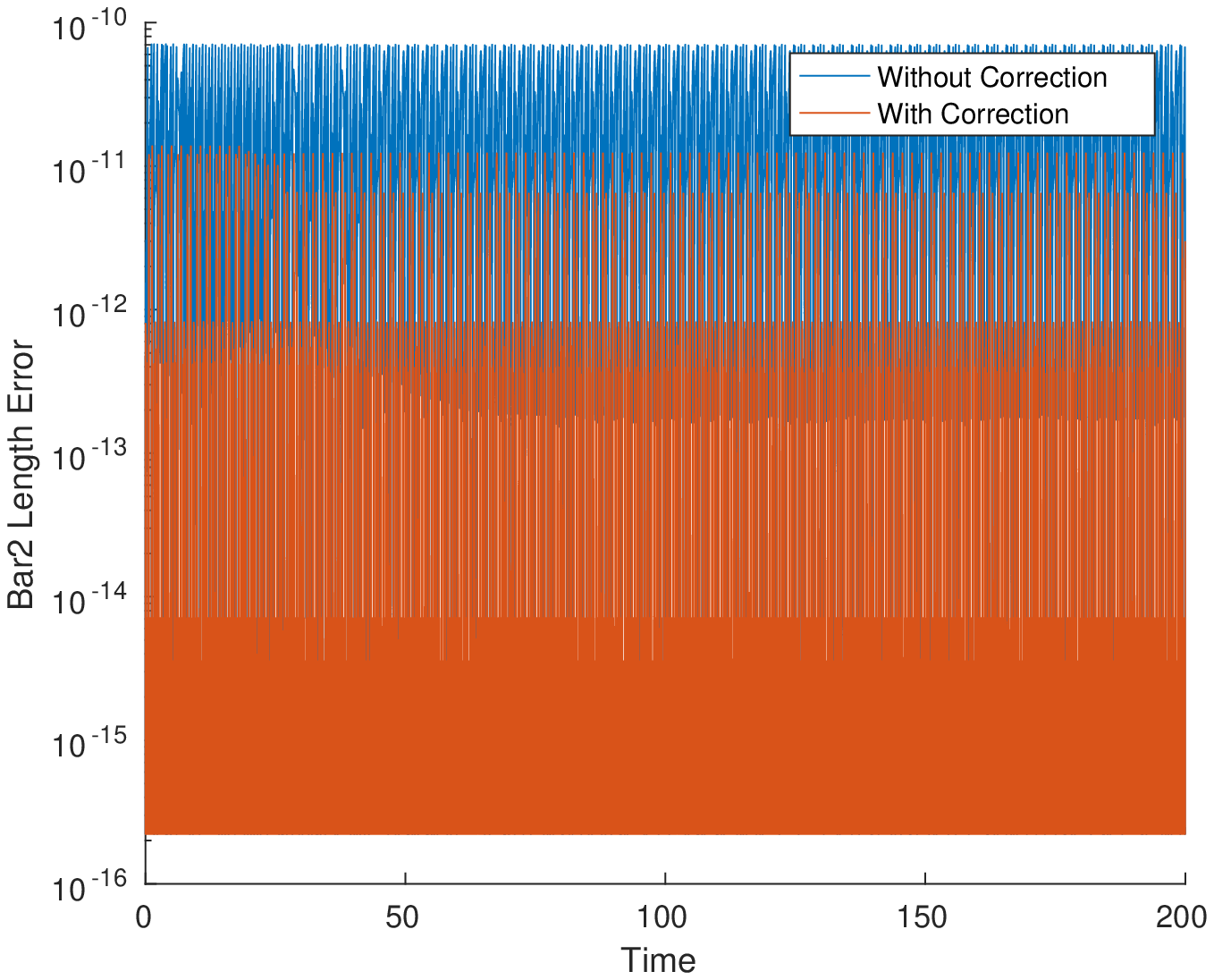}
\label{fig:Bar2LengthErrTol10EnergyCorrection_Tbar}}\hfill
% \caption{Bar Length constraint violation observed for Bar 2 in the T-bar example, observed with and without energy correction at $10^{-10}$ relative and absolute tolerance.}
\caption{Motion error, constraint and energy violations if simulated at $10^{-10}$ tolerance with and without energy correction in the example of the 2-bar structure shown in Figure \ref{fig:Tbar}.}
\end{figure*}

\begin{figure*}[!ht]
\centering
\subfigure[Motion Error of Node 3.]
{\includegraphics[width=0.48\textwidth]
{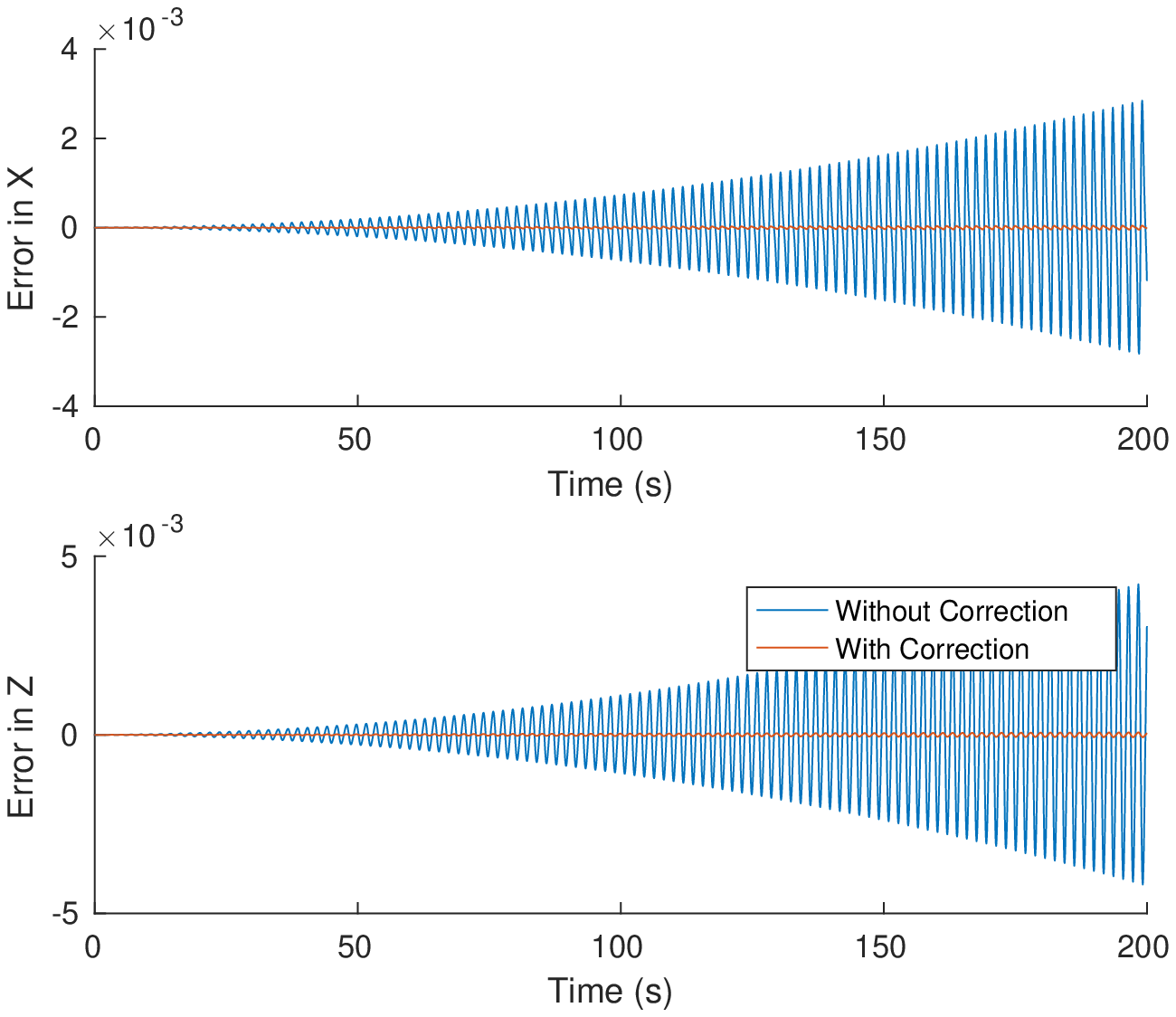}
\label{fig:MotionErrTol66EnergyCorrection_Tbar}}\hfill
% \caption{Difference in motion of node 3 in the T-bar example, observed with and without energy correction at $10^{-6}$ relative and absolute tolerance.}
% \end{figure}
%
% \begin{figure}[!ht]
% \centering
\subfigure[Energy violation of \eqref{eqn:conservation}.]
{\includegraphics[width=0.48\textwidth]
{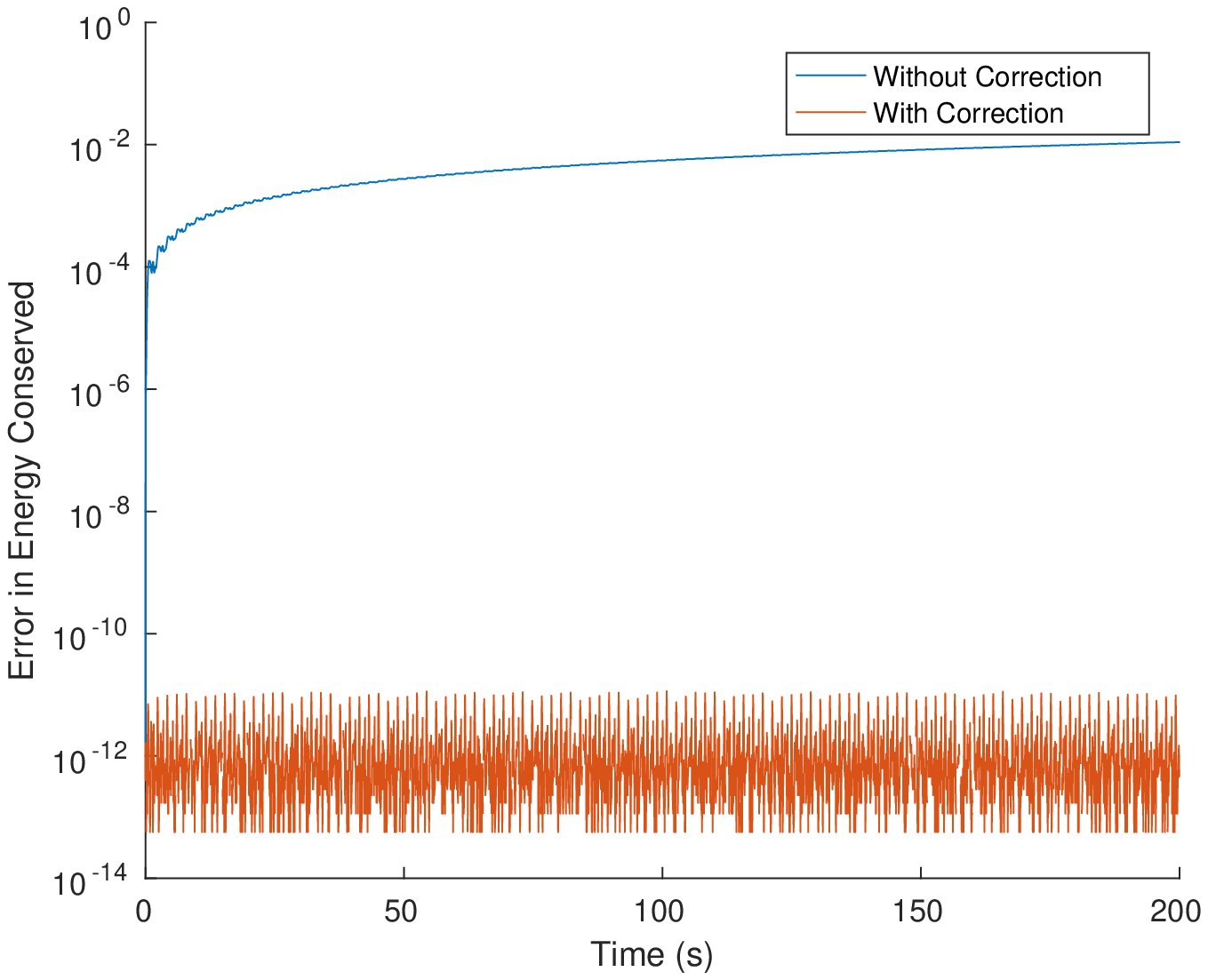}
\label{fig:EnergyErrTol66EnergyCorrection_Tbar}}\hfill
% \caption{Bar Length constraint violation observed for Bar 2 in the T-bar example, observed with and without energy correction at $10^{-10}$ relative and absolute tolerance.}
% \end{figure}
%
% \begin{figure}[!ht]
% \centering
\subfigure[Bar 1 Length Error.]
{\includegraphics[width=0.48\textwidth]
{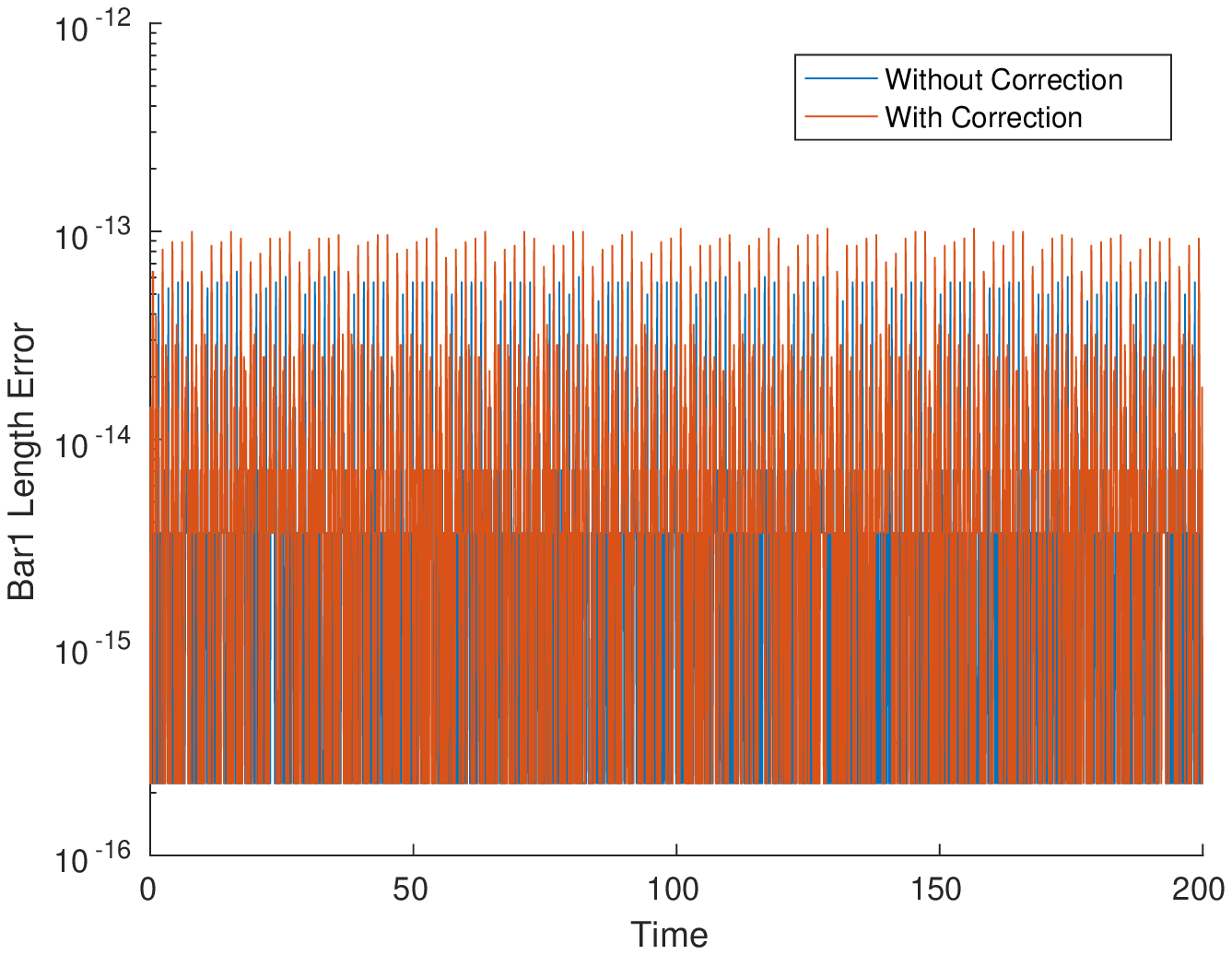}
\label{fig:Bar1LengthErrTol66EnergyCorrection_Tbar}}\hfill
% \caption{Bar Length constraint violation observed for Bar 1 in the T-bar example, observed with and without energy correction at $10^{-6}$ relative and absolute tolerance.}
% \end{figure}
%
% \begin{figure}[!ht]
% \centering
\subfigure[Bar 2 Length Error.]
{\includegraphics[width=0.48\textwidth]
{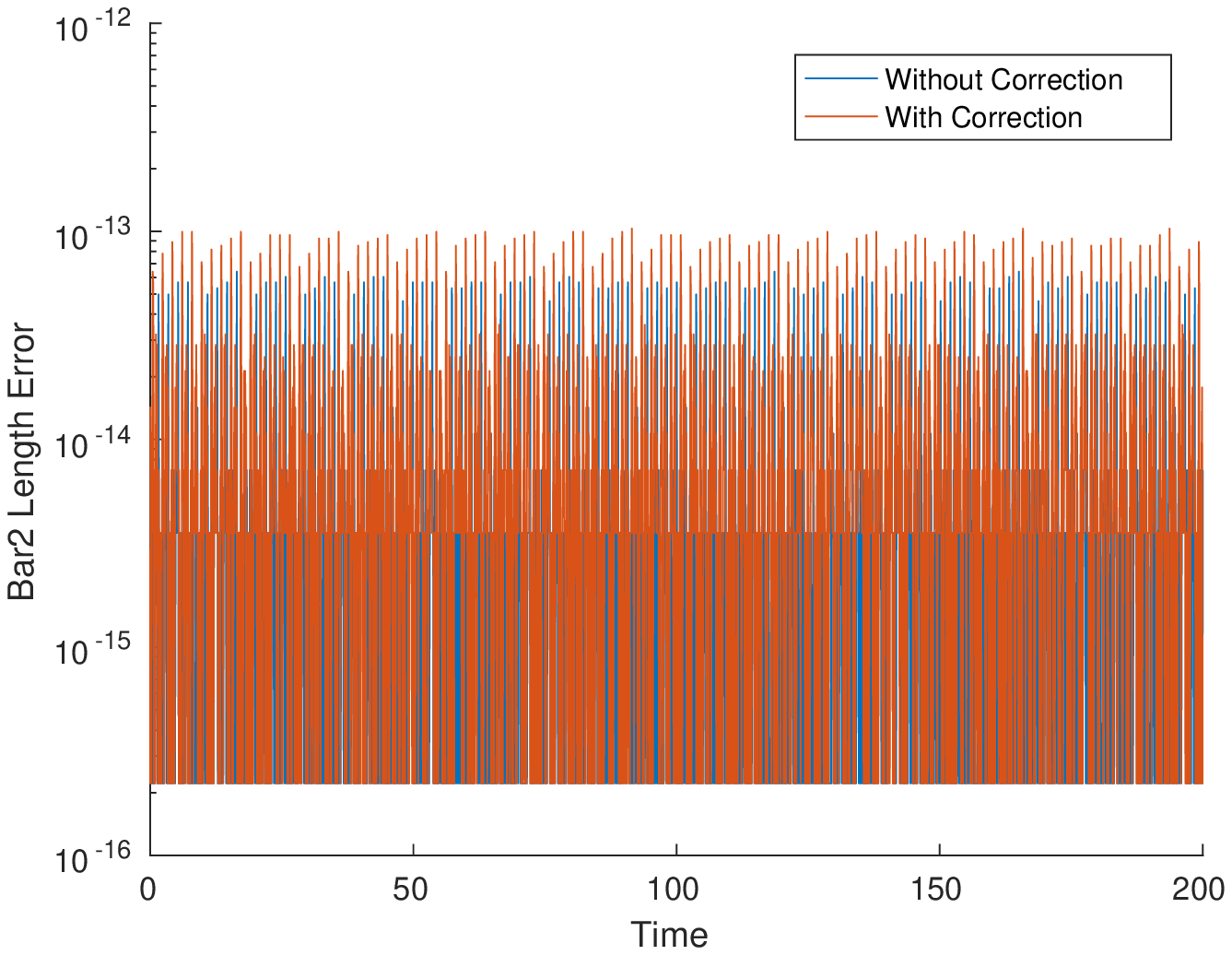}
\label{fig:Bar2LengthErrTol66EnergyCorrection_Tbar}}\hfill
% \caption{Bar Length constraint violation observed for Bar 2 in the T-bar example, observed with and without energy correction at $10^{-6}$ relative and absolute tolerance.}
\caption{Motion error, constraint and energy violations if simulated at $10^{-6}$ tolerance with and without energy correction in the example of the 2-bar structure shown in Figure \ref{fig:Tbar}.}
\end{figure*}

\begin{figure*}[!ht]
\centering
\subfigure[Motion error in X of Node 3.]
{\includegraphics[width=0.75\textwidth]
{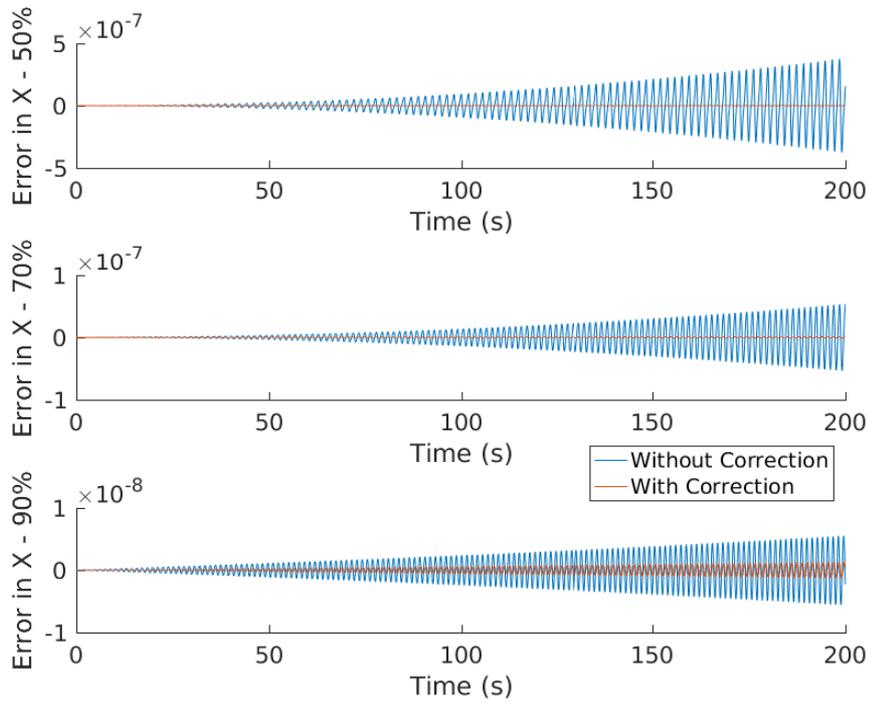}
\label{fig:MotionErrTol10EnergyCorrectionRLX_Tbar}}\hfill
% \caption{Difference in motion of node 3 in the T-bar example, observed with and without energy correction at relative and absolute tolerance set at $10^{-10}$ for different rest lengths.}
% \end{figure}
%
% \begin{figure}[!ht]
% \centering
\subfigure[Motion error in Z of Node 3.]
{\includegraphics[width=0.75\textwidth]
{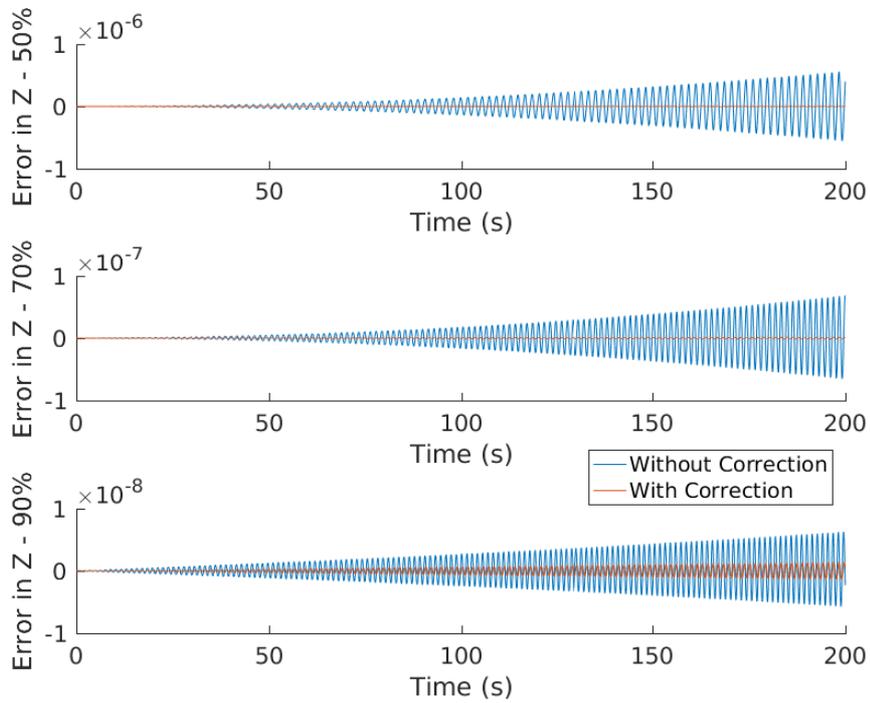}
\label{fig:MotionErrTol10EnergyCorrectionRLZ_Tbar}}\hfill
% \caption{Bar Length constraint violation observed for Bar 2 in the T-bar example, observed with and without energy correction at $10^{-6}$ relative and absolute tolerance.}
\caption{Motion error of Node 3 if simulated at $10^{-10}$ tolerance with and without energy correction at different rest lengths in the example of the 2-bar structure shown in Figure \ref{fig:Tbar}.}
\end{figure*}

To investigate the impact of the energy preservation scheme of the T-bar, we simulated the T-bar example at 2 different tolerance settings ($10^{-6}$ and $10^{-10}$), and at 3 different rest lengths($50\%$, $70\%$, and $90\%$).
We found a considerably stronger effect at a higher tolerance setting than at a lower one, for the same rest length $50\%$, as indicated by the order of magnitude of motion errors in Figures \ref{fig:MotionErrTol66EnergyCorrection_Tbar} and \ref{fig:MotionErrTol10EnergyCorrection_Tbar} respectively.
Figures \ref{fig:Bar1LengthErrTol10EnergyCorrection_Tbar}, \ref{fig:Bar2LengthErrTol10EnergyCorrection_Tbar},
\ref{fig:Bar1LengthErrTol66EnergyCorrection_Tbar} and \ref{fig:Bar2LengthErrTol66EnergyCorrection_Tbar} demonstrate the consistency of the direct constraint correction scheme in stabilizing geometric constraint violations below a specified norm bound, despite a large difference in the order of magnitude of tolerance.
In Figure \ref{fig:MotionErrTol10EnergyCorrectionRLX_Tbar} and \ref{fig:MotionErrTol10EnergyCorrectionRLZ_Tbar},
the T-bar example is simulated at a tolerance of $10^{-10}$ for rest lengths of $50\%$, $70\%$, and $90\%$. Evidently, larger deformations in the $50\%$ case bring energy correction into play more effectively.

% Comparing the motion error in Figure \ref{fig:motionEnergy_TbarL} where the proposed method \eqref{eqn:correction_final} and \eqref{eqn:correction_direct} are used, one should notice that the energy correction does not improve the accuracy of the motion significantly because of the nature of the correction scheme's approach to energy preservation.
% Specifically, $E(\q^u,\dot{\q}^u)$ in \eqref{eqn:energyCorrection} does not reveal the true variation because the errors of the kinetic and potential energy in $E(\q^u,\dot{\q}^u)$ may cancel each other out. For example, if the potential energy is violated negatively while the kinetic energy is violated positively, the variation of the total energy is canceled out so that the true error cannot be seen from the equation. The energy correction therefore does not modify the motion completely. The same observation was made in \cite{blajer2002elimination} as well.

\begin{figure*}[!ht]
\centering
\subfigure[Motion of Tip and Connection Points.]
{\includegraphics[width=0.48\textwidth]{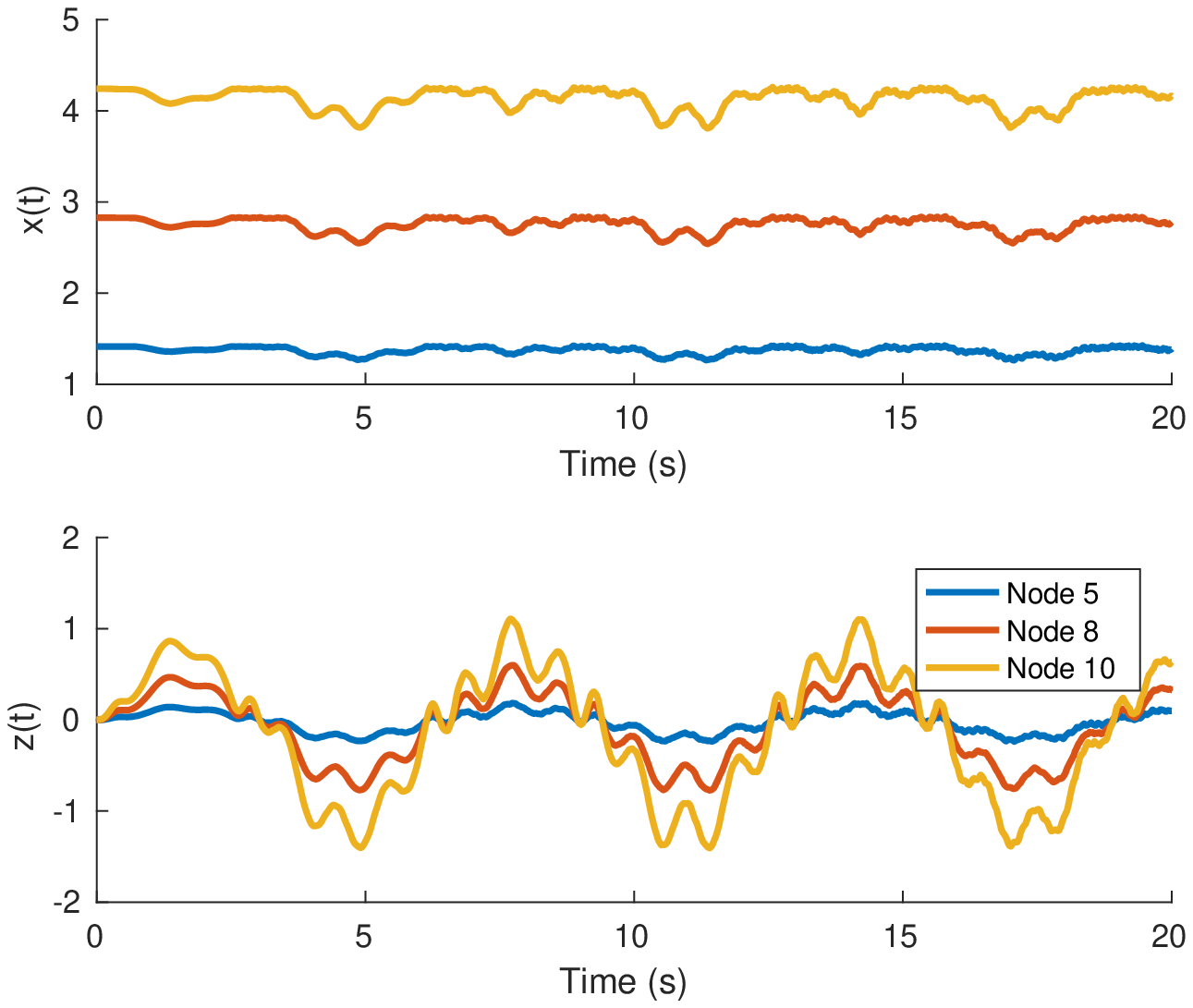}
\label{fig:Motion_arm}}\hfill
% \caption{Motion trajectories of the tip and the connection points between the stations in the example of the arm.}
% \end{figure}
%
% \begin{figure}[!ht]
% \centering
\subfigure[Motion of Nodes 4,7,12,14.]
{\includegraphics[width=0.48\textwidth]{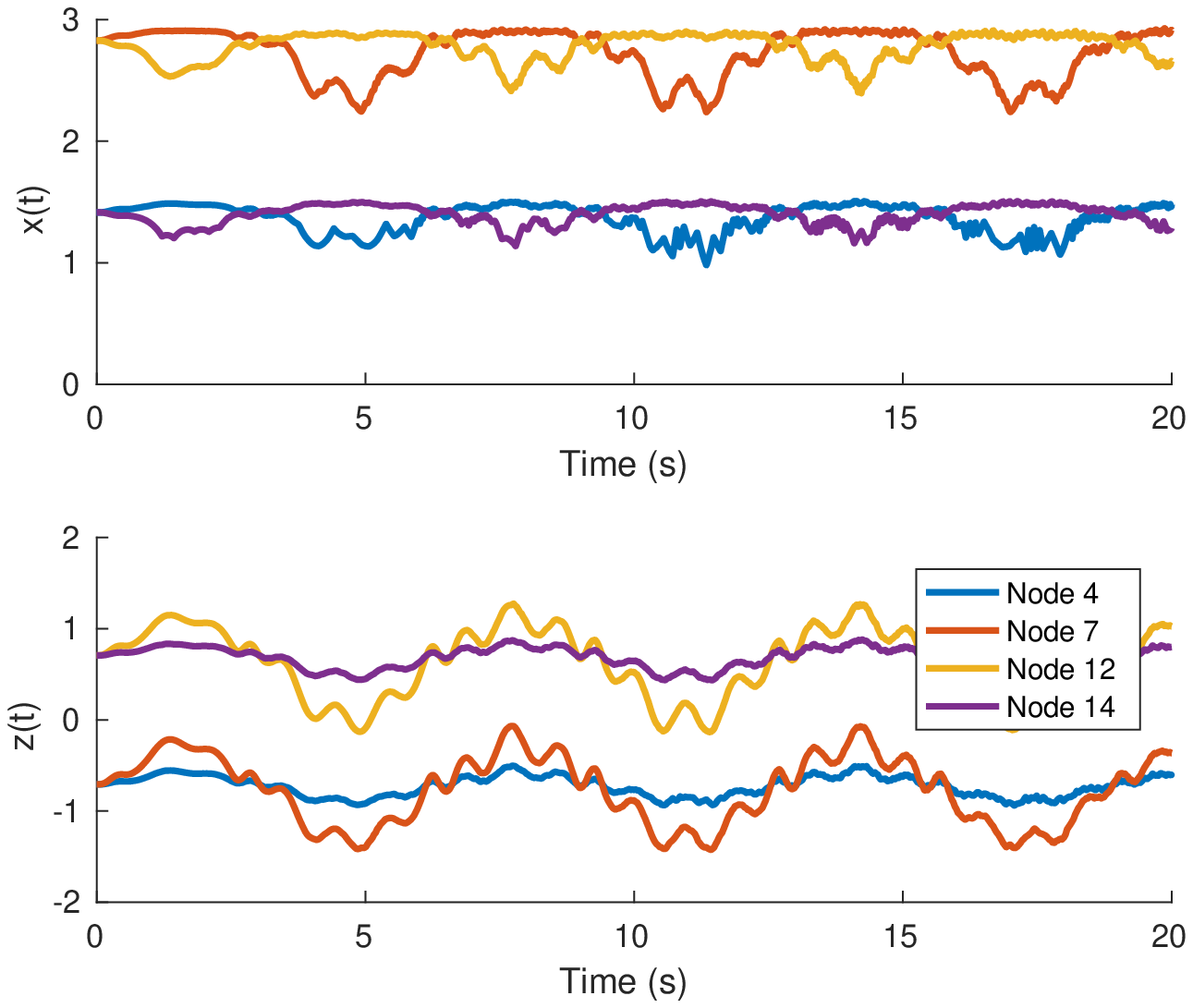}
\label{fig:Motion2_arm}}\hfill
% \caption{Motion trajectories of the node \#4,\#7,\#12,\#14 in the example of the arm.}
\caption{Motion trajectories of the particular nodes in the example of the arm shown in Figure \ref{fig:Arm}.}
\end{figure*}

\begin{figure*}[!ht]
\subfigure
{\includegraphics[width=0.45\textwidth]{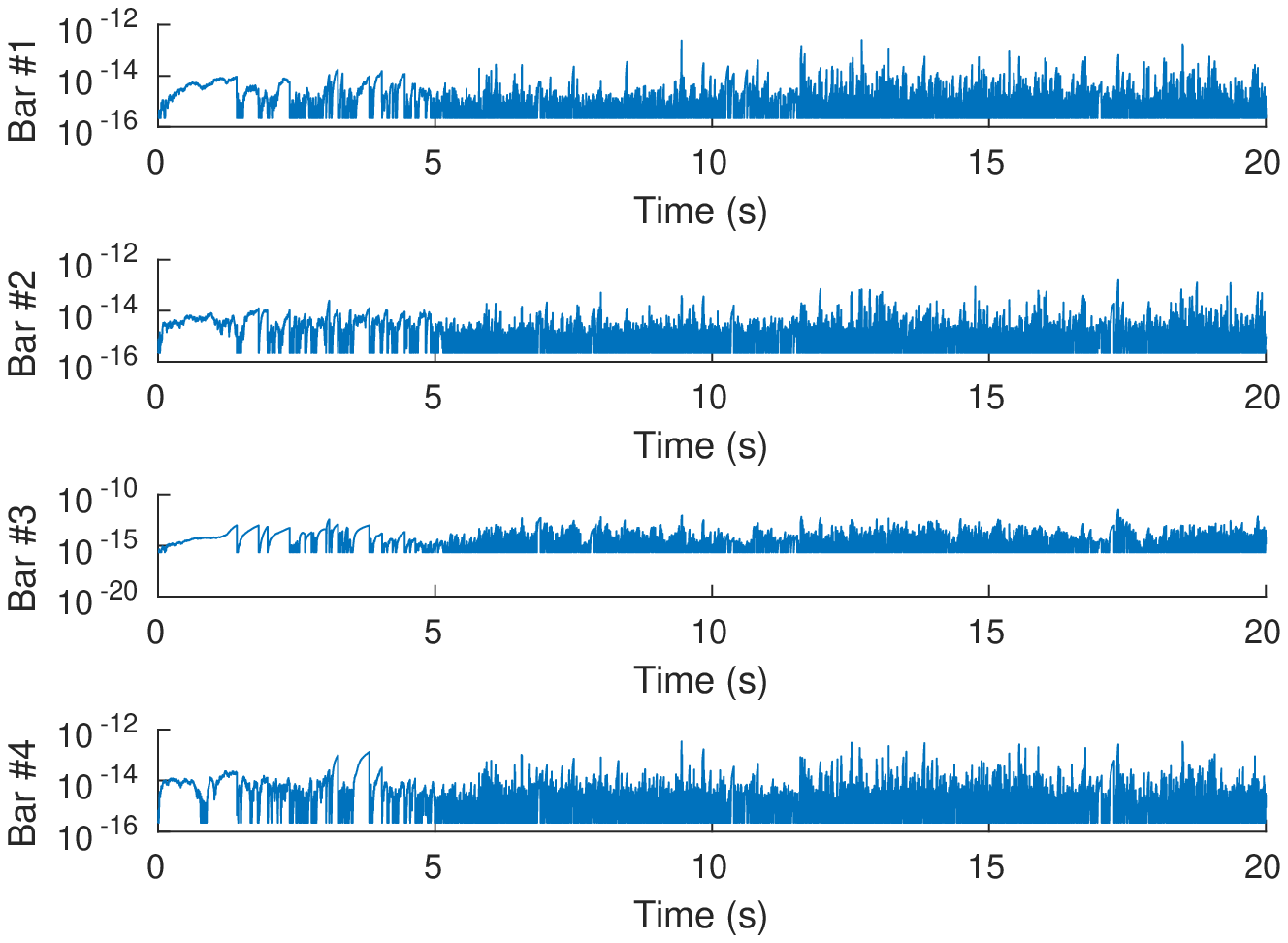}}\hfill
\subfigure
{\includegraphics[width=0.45\textwidth]{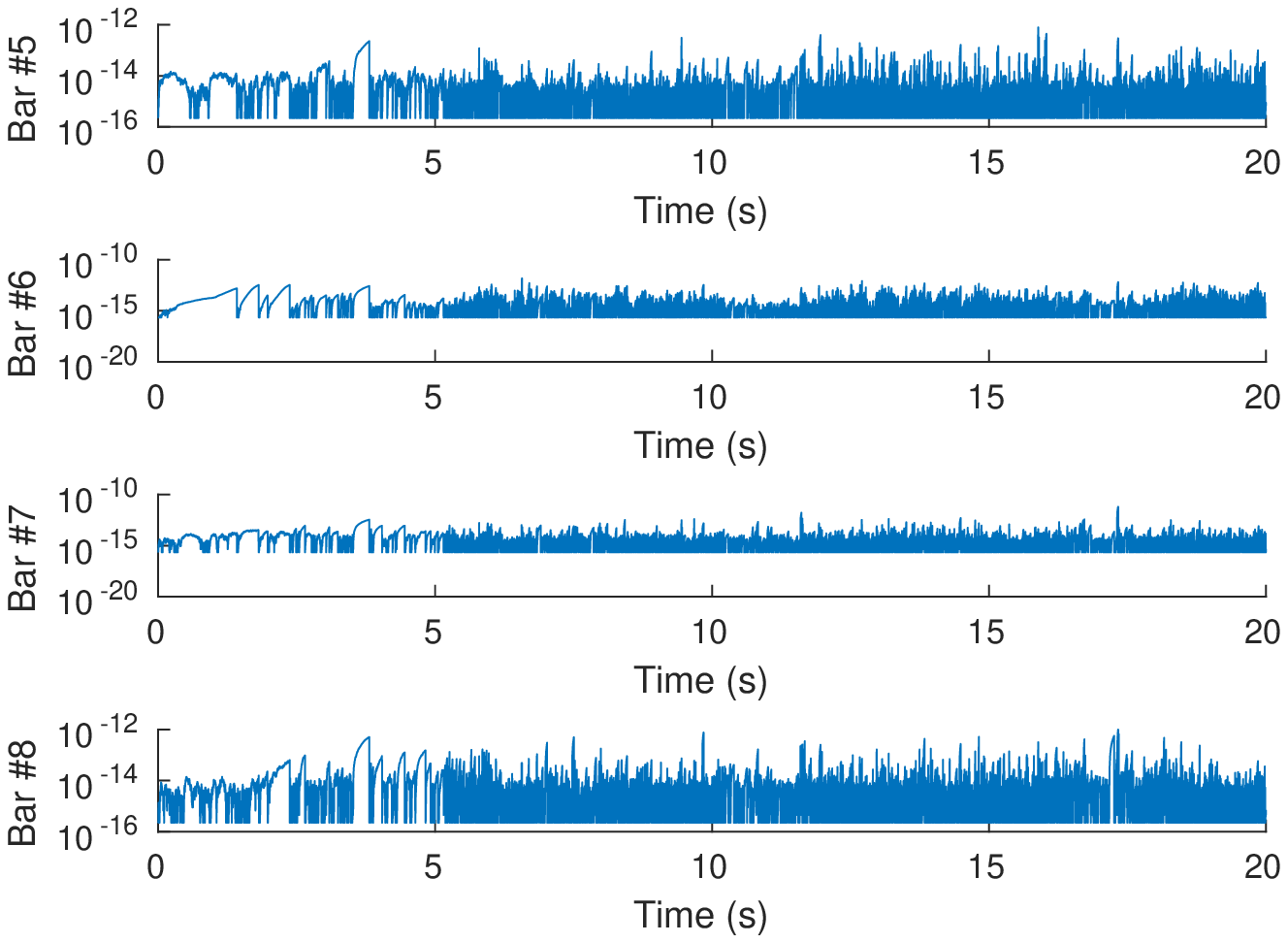}}\hfill
\subfigure
{\includegraphics[width=0.45\textwidth]{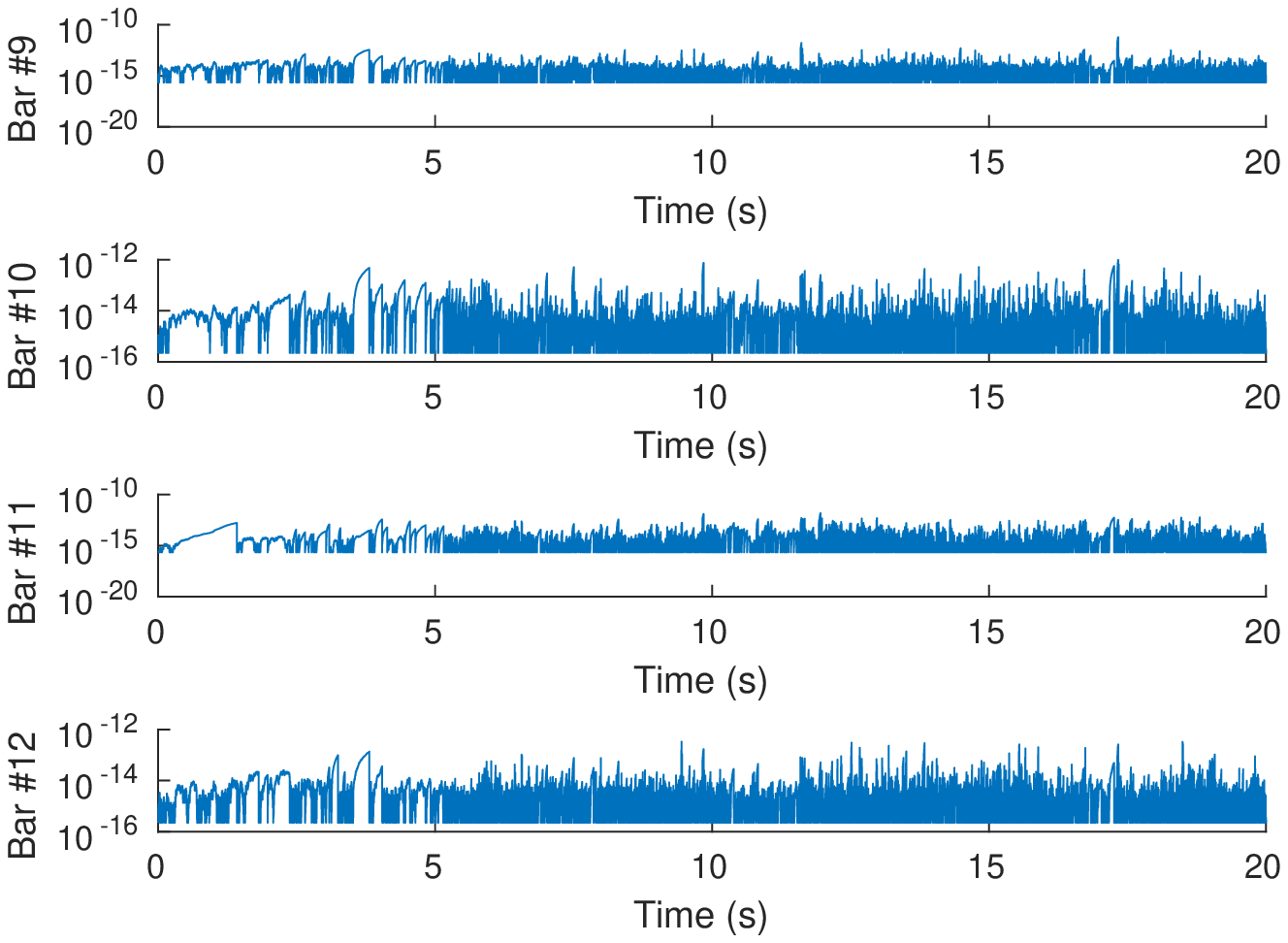}}\hfill
\subfigure
{\includegraphics[width=0.45\textwidth]{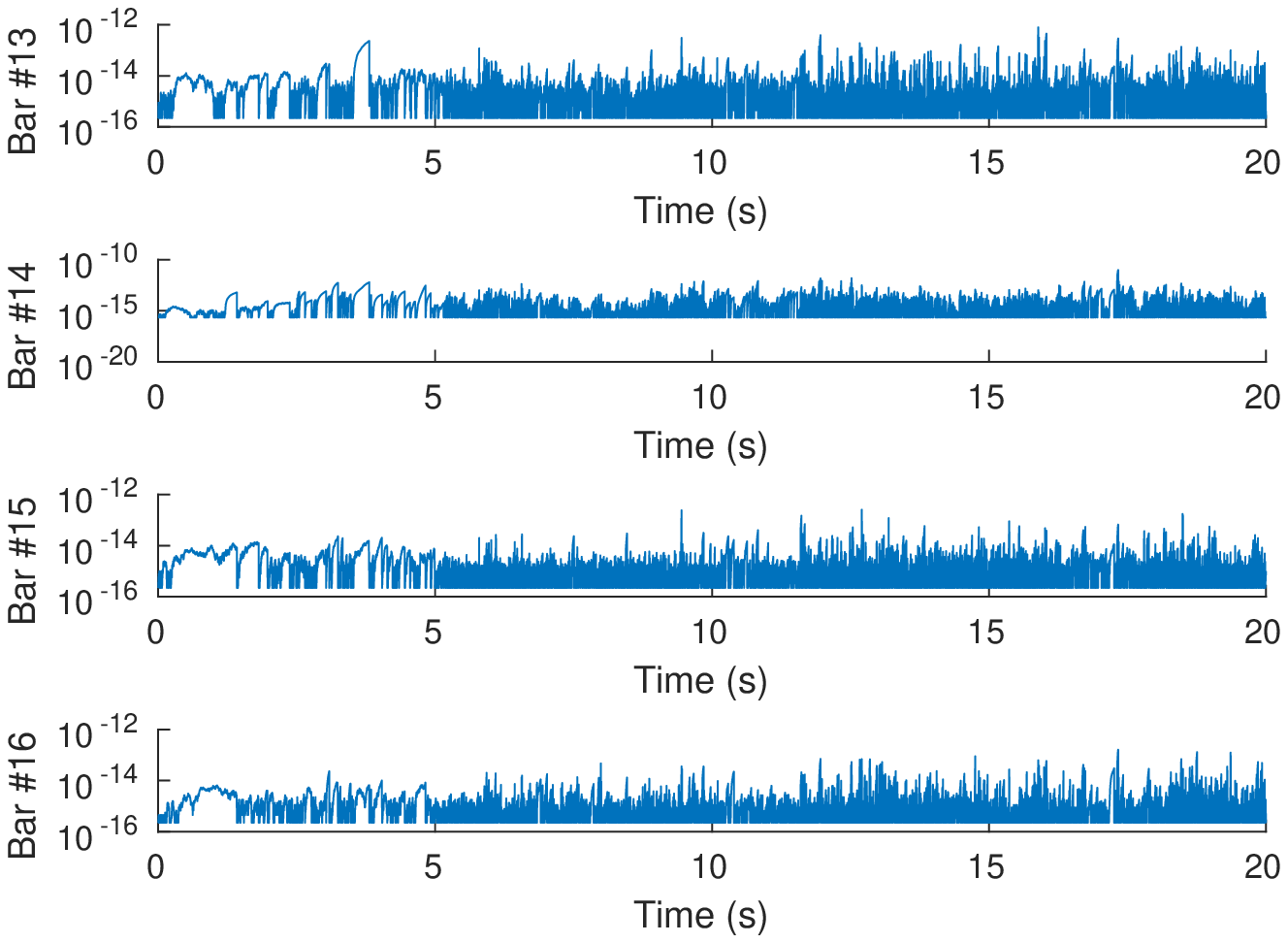}}\hfill
\caption{Bar length errors in the example of the arm shown in Figure \ref{fig:Arm}, where the value of zero is set to the minimum positive double precision number.}
\label{fig:Bar_arm}
\end{figure*}

\begin{figure*}[!ht]
\centering
% \subfigure[Energy violation of \eqref{eqn:conservation}.]
{\includegraphics[width=0.4\textwidth]{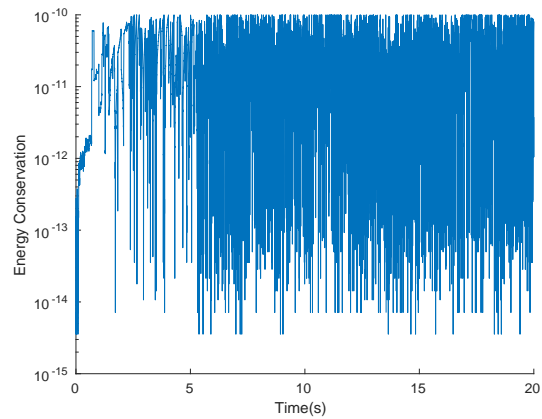}}
% \caption{Energy violation of \eqref{eqn:conservation} in the example of the arm.}
\caption{Energy violation \eqref{eqn:conservation} of the arm shown in Figure \ref{fig:Arm}.}
\label{fig:Energy_ARM}
\end{figure*}

\begin{figure*}[!ht]
\centering
\subfigure[Motion of nodes 4,7,8,13.]
{\includegraphics[width=0.48\textwidth]{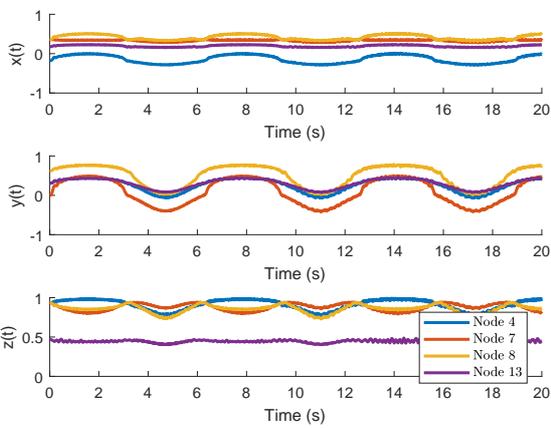}
\label{fig:motion_ball}}\hfill
% \caption{Motion trajectories of the top 3 nodes and the payload in the example of the ball.}
% \end{figure}
%
% \begin{figure}[!ht]
% \centering
\subfigure[Energy violation of \eqref{eqn:conservation}.]
{\includegraphics[width=0.48\textwidth]{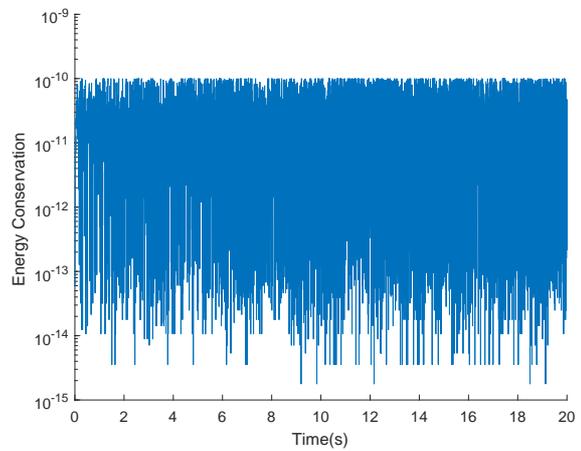}
\label{fig:energy_ball}}\hfill
% \caption{Energy violation of \eqref{eqn:conservation}.}
\caption{Motion and energy violations in the example of the ball shown in Figure \ref{fig:ball}.}
\end{figure*}

\begin{figure*}[!ht]
\subfigure[Bar length error.]
{\includegraphics[width=0.48\textwidth]{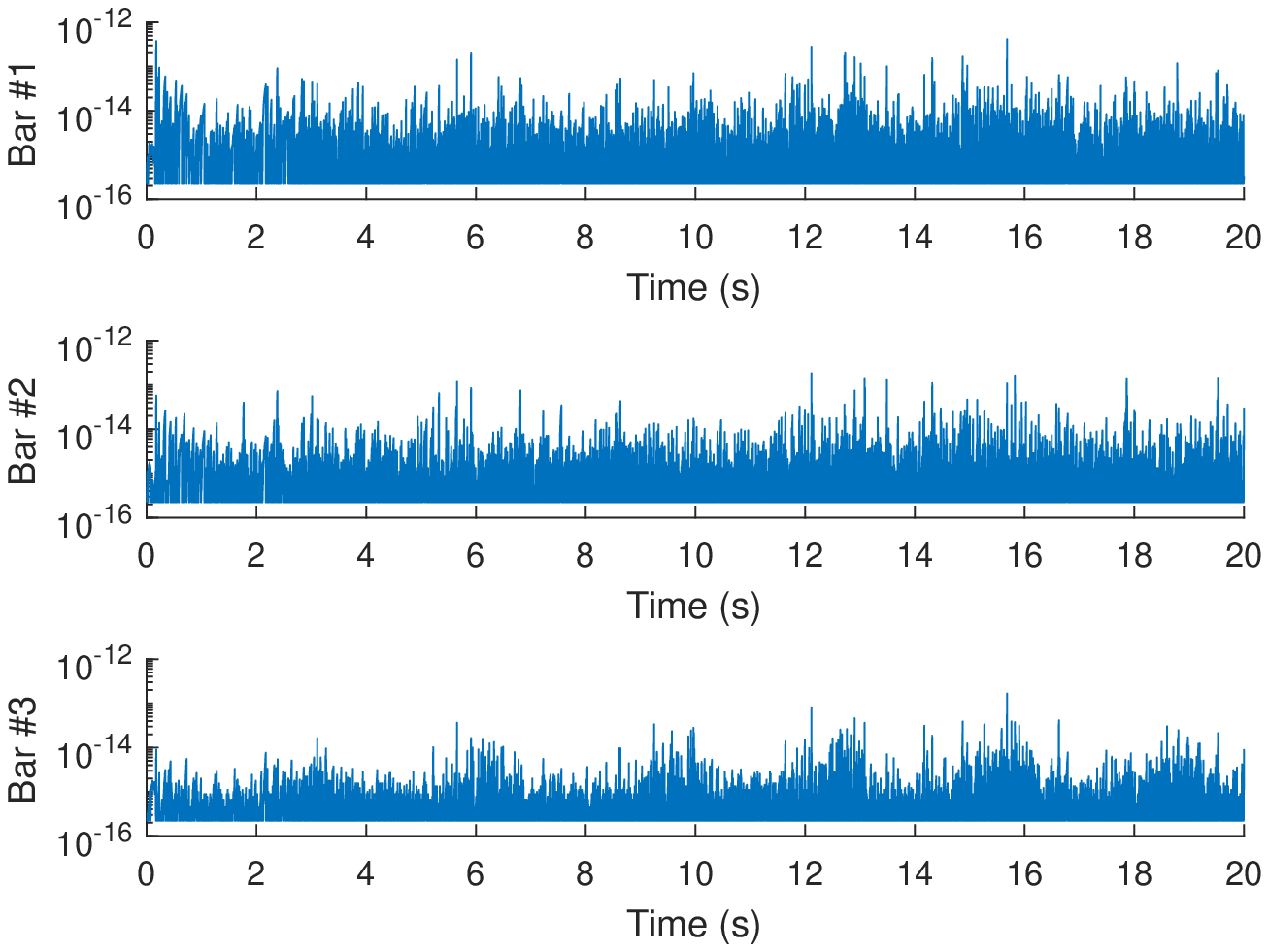}}\hfill
\subfigure[Bar length error.]
{\includegraphics[width=0.48\textwidth]{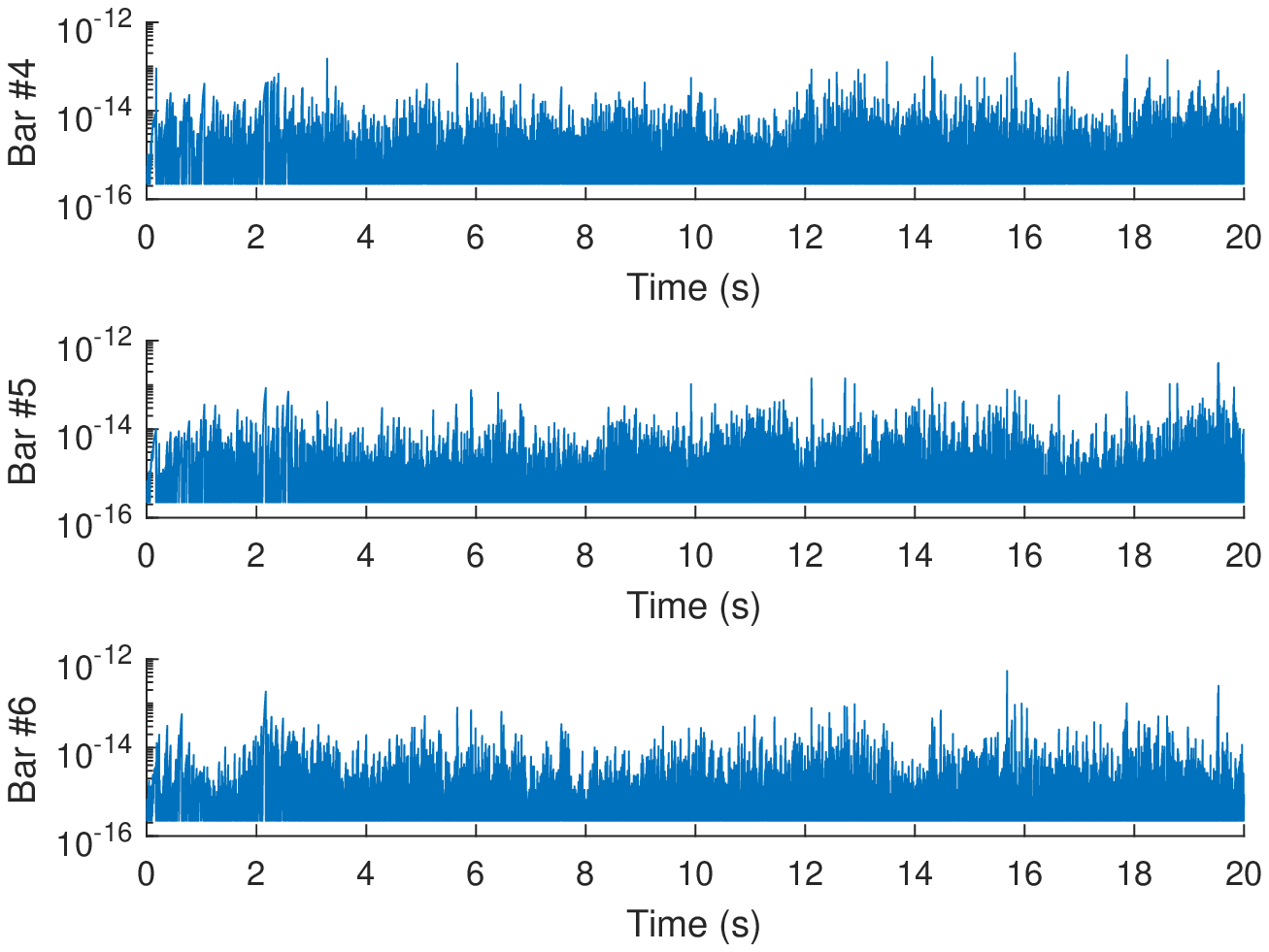}}\hfill
\caption{Bar length errors in the example of the ball shown in Figure \ref{fig:ball}, where the value of zero is set to the minimum positive double precision number.}
\label{fig:bar_ball}
\end{figure*}

For the example of the robotic arm, Figure \ref{fig:Motion_arm} shows the motion of node 5, node 8 and node 10 for 20 seconds. Since we've simulated the structure with nylon strings (Young's modulus: 2 GPa), the structure appears to be chattering intermittently.
Figures \ref{fig:Bar_arm} and \ref{fig:Energy_ARM} show the constraint violations of bar length and energy. The order of the violations testifies to the stability of the constraints for problems involving intricate geometries. One can observe that the bar length constraints of bars \#3, \#6, \#11 amd \#14 are violated more than others in the observed time period and nodes \#4,\#7,\#12,\#14 in Figure \ref{fig:Motion2_arm} are vibrating in higher frequencies, which implies a positive correlation between constraint variations and motion frequencies. Figure \ref{fig:motion_ball} depicts the motion of the 3-dimensional ball which is in accordance with the high stiffness of the strings.
Preserving the order of constraint violations as observed in the second example, Figure \ref{fig:bar_ball} and \ref{fig:energy_ball} demonstrate the ability of the implemented correction method to maintain stability of the constraints despite an increase in complexities associated with 3 dimensions. The videos capturing the real-time motion of the structures and the relevant code to generate simulations are publicly available \cite{Tadiparthi2019}.

\begin{figure*}[!ht]
\subfigure[Motion of Node 3.]
{\includegraphics[width=0.48\textwidth]{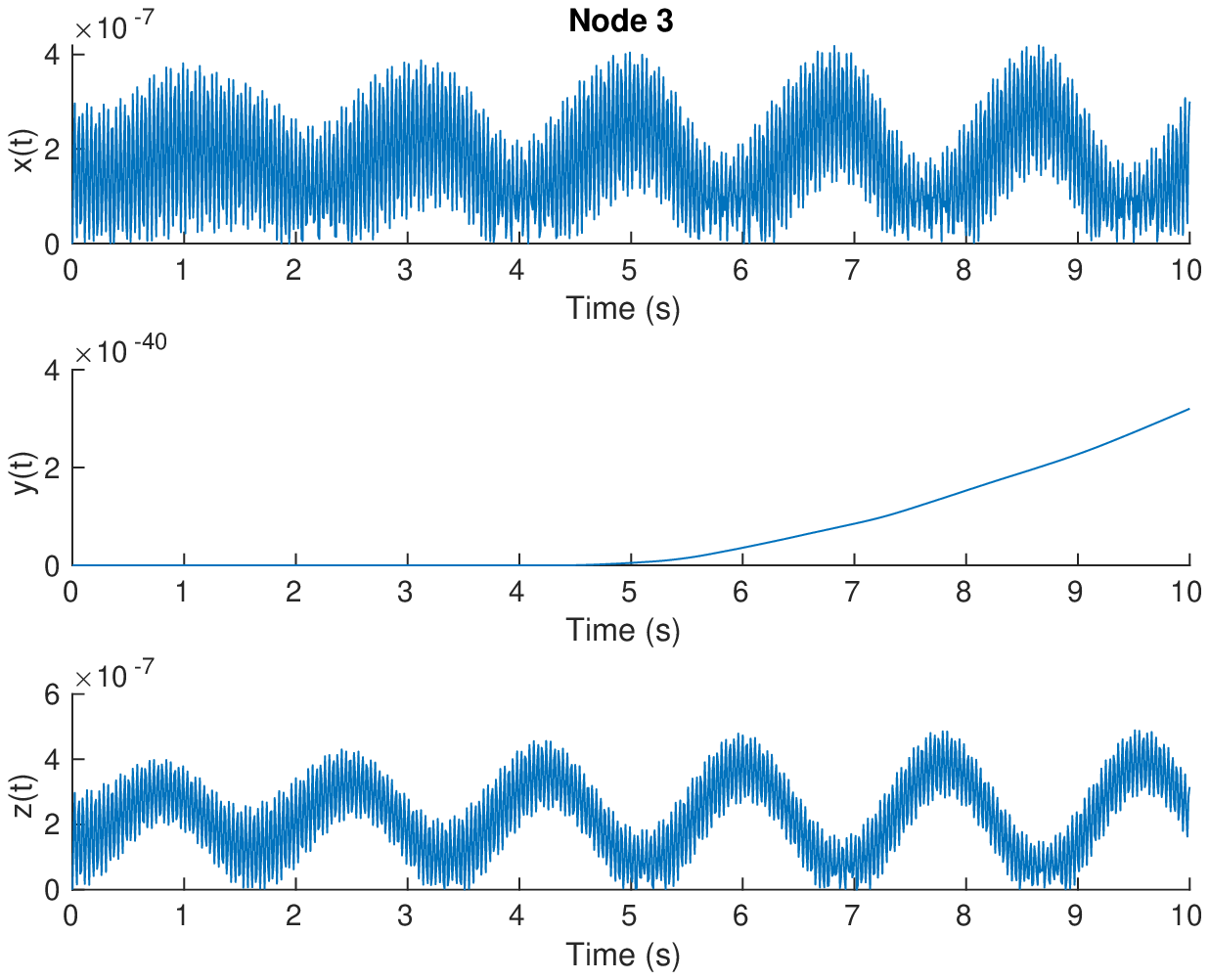}}\hfill
\subfigure[Motion of Node 4.]
{\includegraphics[width=0.48\textwidth]{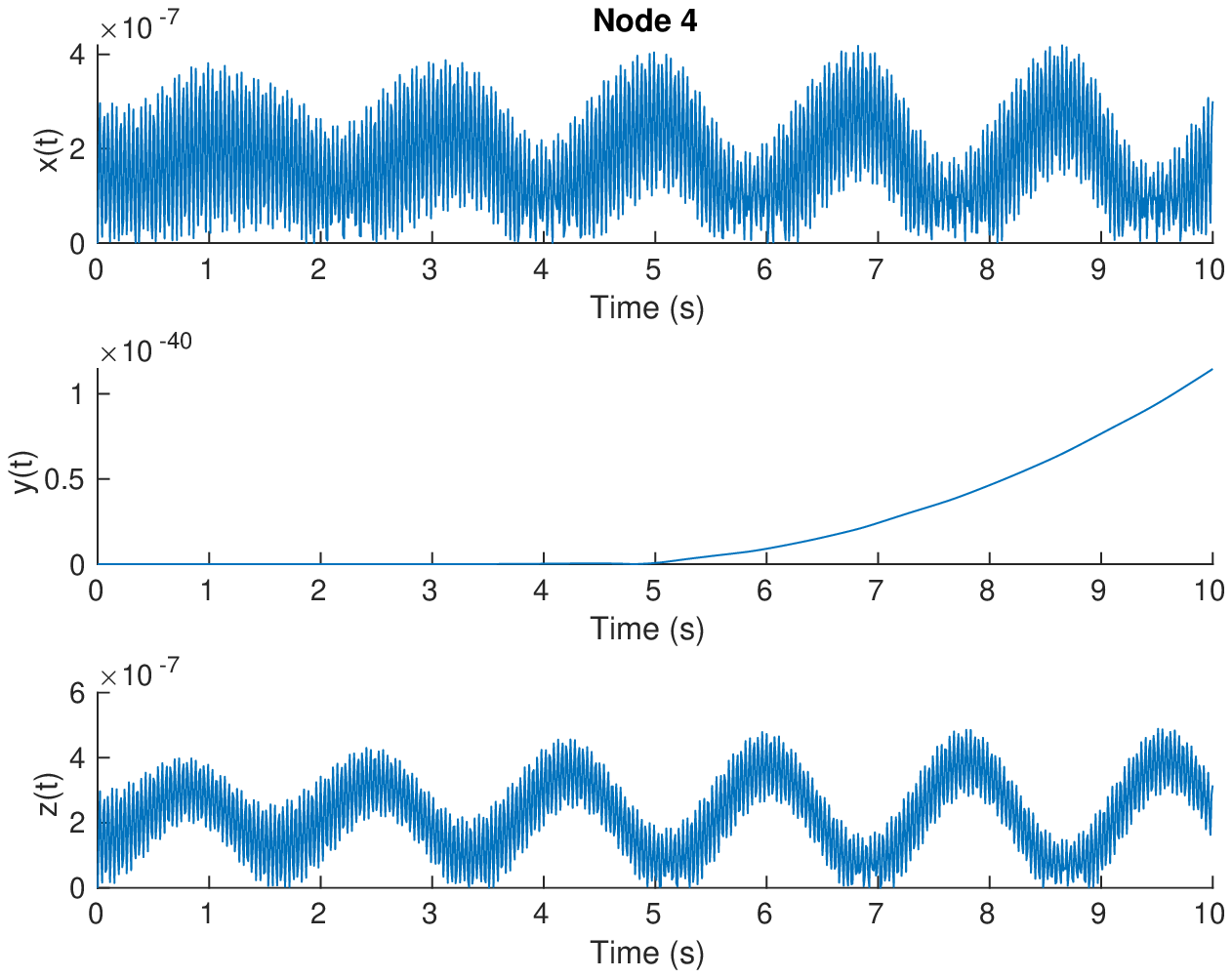}}\hfill
\caption{Difference in Motion for the Tbar shown in Figure \ref{fig:Tbar}, but with compressible bars, compared against Tbar with rigid bars but no correction applied.}
\label{fig:Tbar_flex}
\end{figure*}

Further, we redo the constrained T-bar dynamics, but substitute the rigid bars in the structure with compressible ones. While this elasticity allows us to relax the bar length constraints during simulation, the DAE becomes a stiff problem, requiring us to choose a solver from the stiff ODE suite, e.g. \cite{shampine1997matlab}. As we can observe from figure \ref{fig:Tbar_flex},  the difference in motion from simulations of a T-bar with compressible bars against a T-bar with rigid bars but without any constraint correction implemented, is of the order of $10^{-7}$.

\begin{figure*}[!ht]
\subfigure[Motion of Node 3.]
{\includegraphics[width=0.48\textwidth]{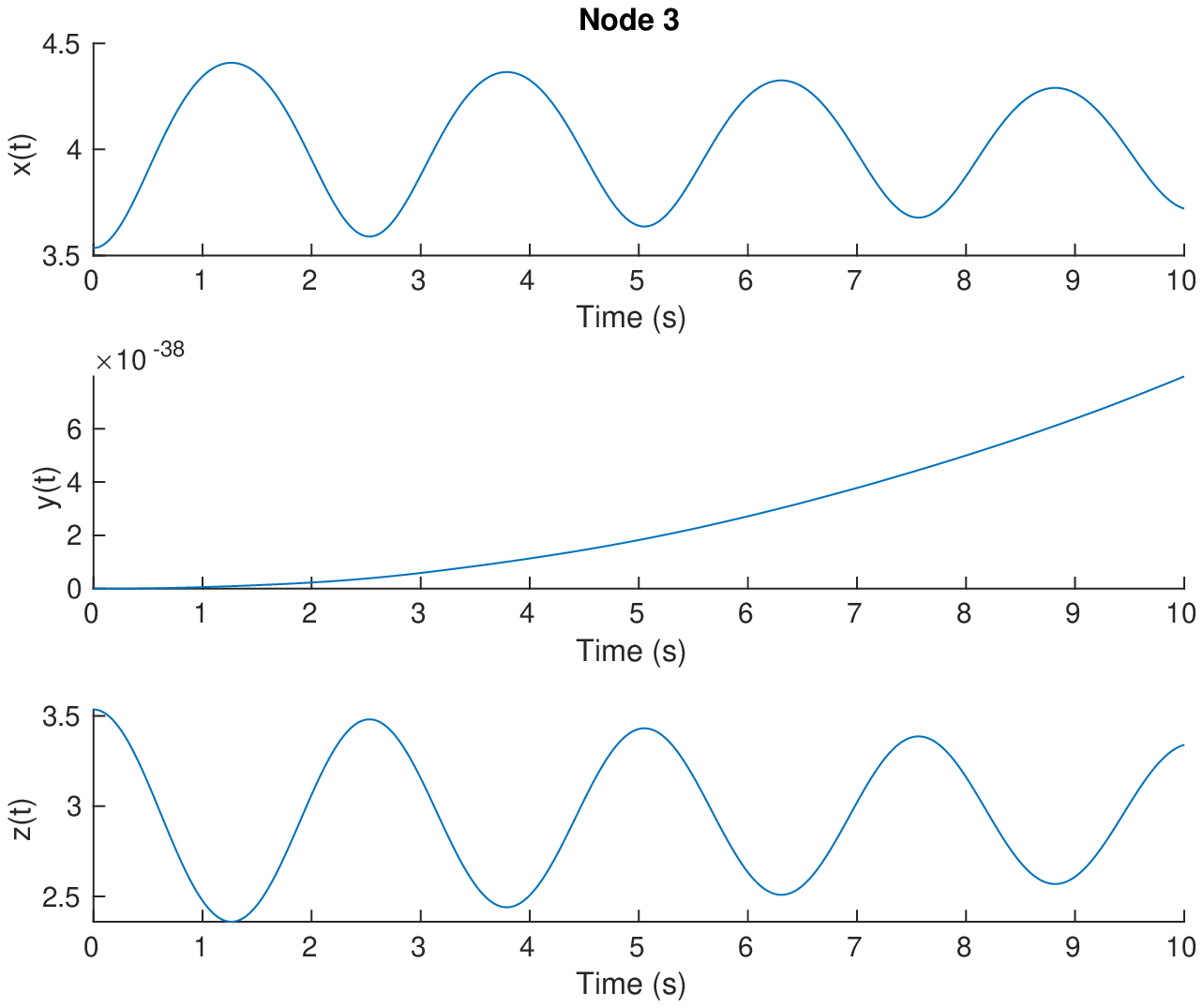}}\hfill
\subfigure[Motion of Node 4.]
{\includegraphics[width=0.48\textwidth]{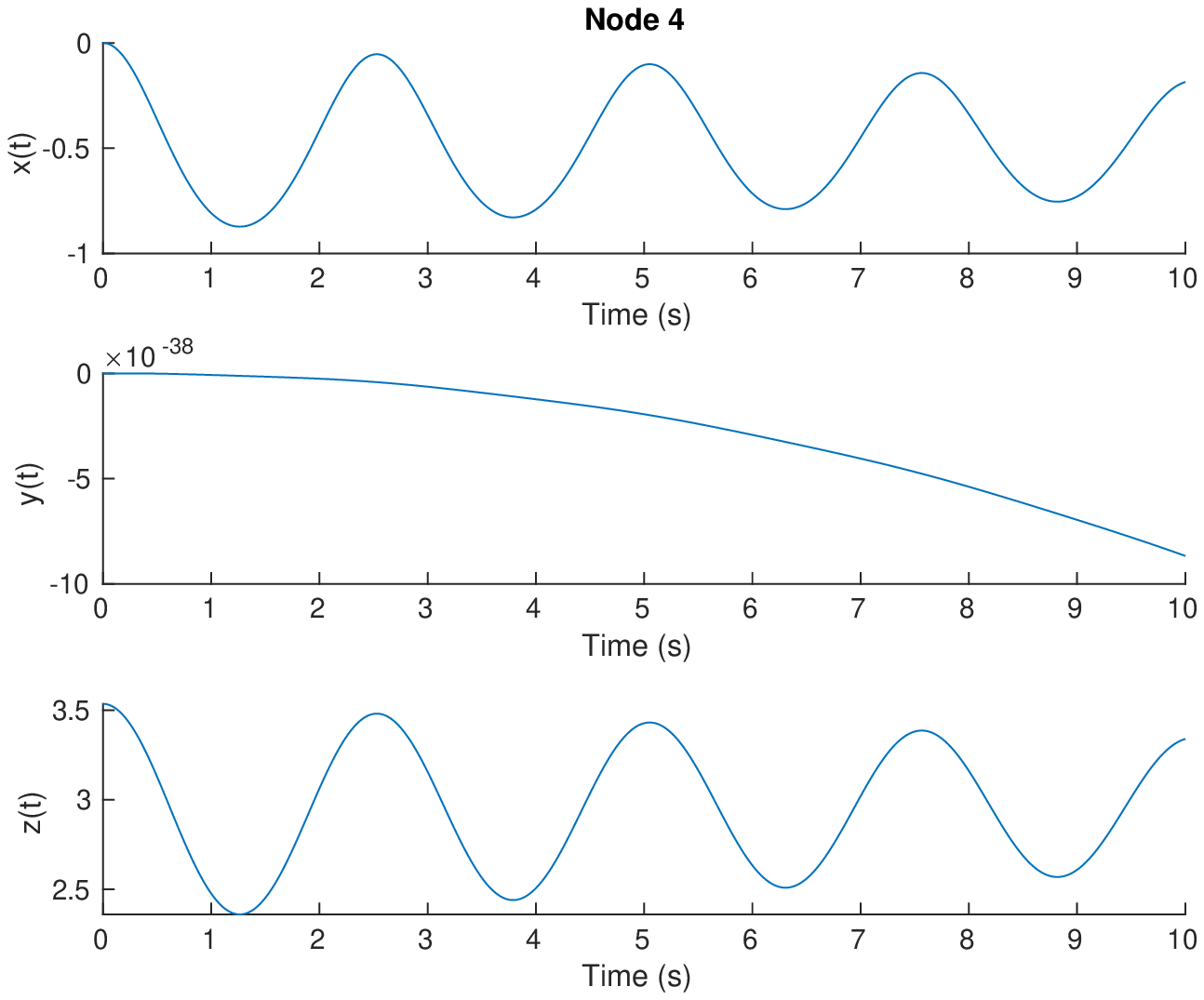}}\hfill
\caption{Motion plots for the Tbar shown in Figure \ref{fig:Tbar}, using compressible bars made of HDPE.}
\label{fig:Tbar_soft}
\end{figure*}

Figure \ref{fig:Tbar_soft} shows the motion of the unfixed nodes in the same T-bar structure being simulated with compressible bars, but this time, the bars are assumed to be of a different material. High Density Poly Ethylene (HDPE), a polymer-based material used to construct pneumatic struts in inflatable tensegrity systems \cite{dessi2019self}, replaces conventionally used metals like steel or aluminium. The mean computation time for simulating a T-bar for 10 seconds using soft HDPE bars was $\sim$ 39.18 seconds, while that for using aluminium bars was $\sim$ 502.2 seconds, a difference of nearly 1100 $\%$. Evidently, using a softer bar material speeds up the computation by a significant margin.

\begin{figure*}[!ht]
	\centering
	{\includegraphics[width=0.4\textwidth]{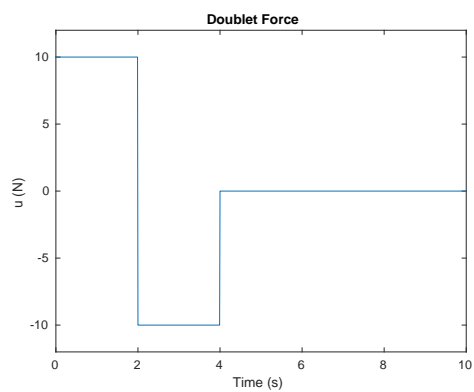}}
	% \caption{Energy violation of \eqref{eqn:conservation} in the example of the arm.}
	\caption{Compressible T-bar using soft and metallic bars subjected to doublet force of 10 N. }
	\label{fig:doublet}
	\end{figure*}

\begin{figure*}[!ht]
\subfigure[Deviation of Node 3.]
{\includegraphics[width=0.48\textwidth]{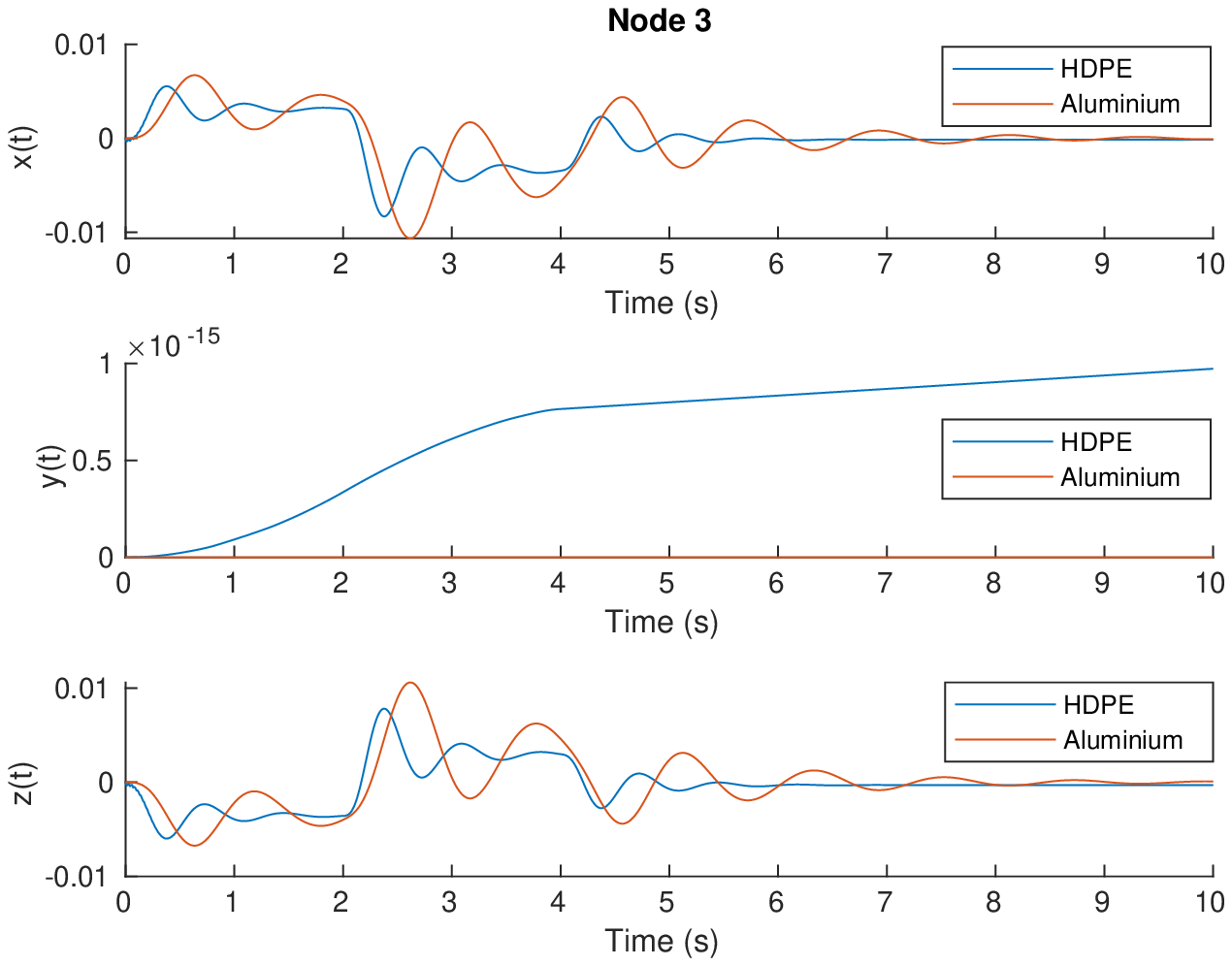}}\hfill
\subfigure[Deviation of Node 4.]
{\includegraphics[width=0.48\textwidth]{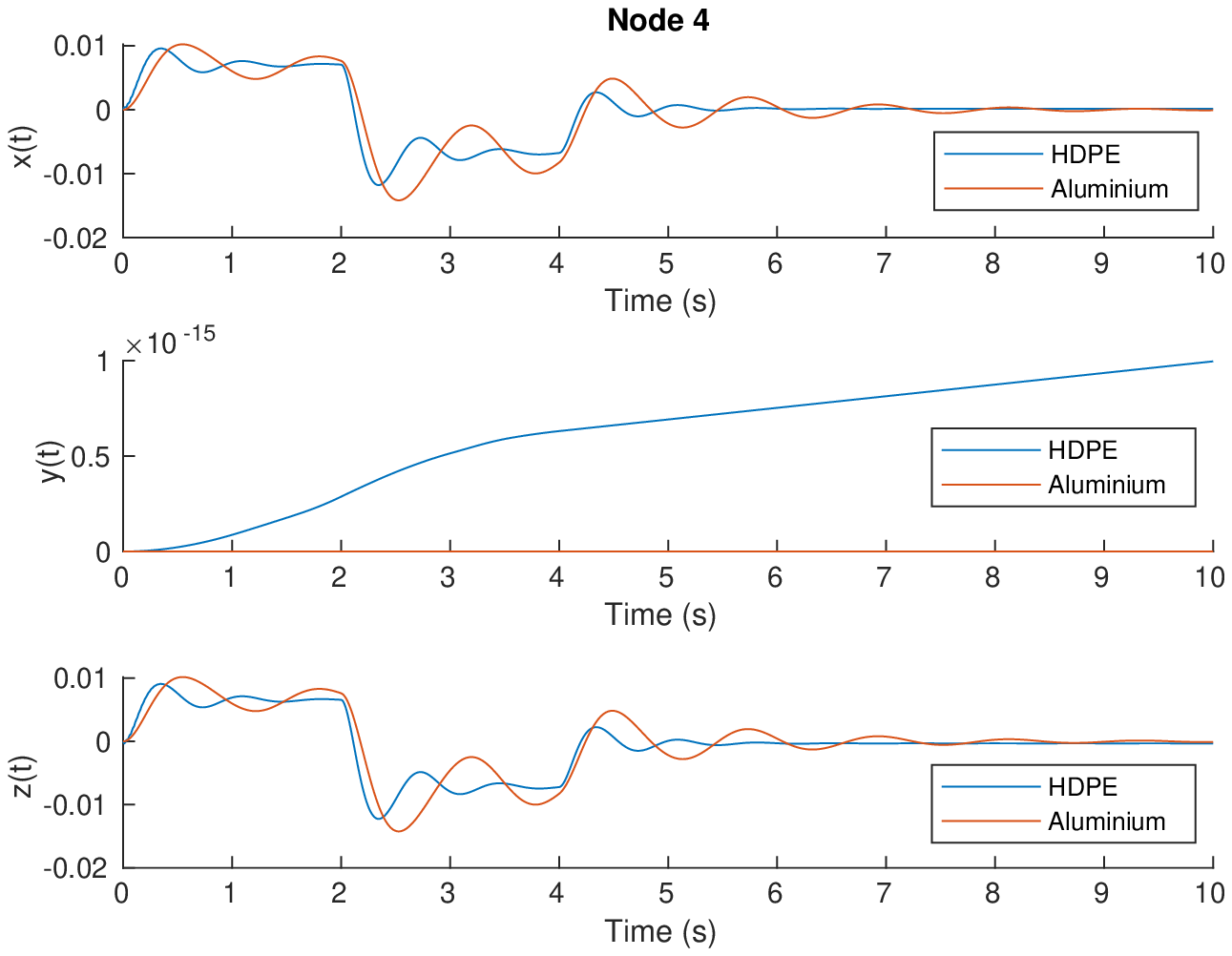}}\hfill
\caption{Deviation of free nodes from equilibrium for the Tbar shown in Figure \ref{fig:Tbar}, using compressible bars made of HDPE and Aluminium.}
\label{fig:Tbar_distSoftRigid}
\end{figure*}

\begin{figure*}[!ht]
	\centering
	{\includegraphics[width=0.4\textwidth]{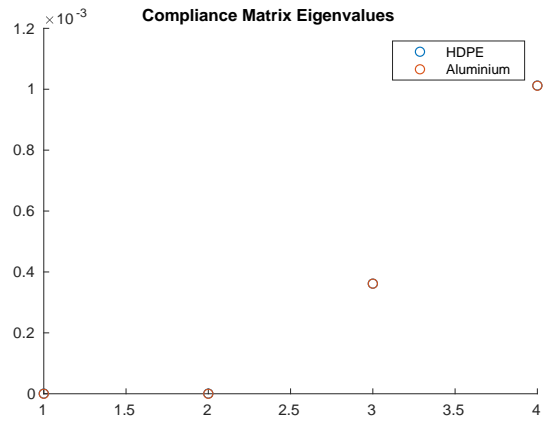}}
	% \caption{Energy violation of \eqref{eqn:conservation} in the example of the arm.}
	\caption{Eigenvalues of the compliance matrix for the T-bar in Figure \ref{fig:Tbar} using soft and metallic bars.}
	\label{fig:eig_compTbar}
	\end{figure*}

Finally, we sought to investigate the characteristics of the two systems (soft vs metallic) from a control engineer's perspective.
% Would softening the bars dramatically lower robustness of the system?
The linearized models help us analyse the \textit{norms} of the system \cite{doyle2013feedback}, which in this case, implies the relationship between the size of the output signal's energy or amplitude to that of the input signal's energy.
Even though the linearized model may differ significantly from the true nonlinear model at points far away from equilibrium, they allow us to draw preliminary inferences regarding metrics like system robustness to disturbance.
Table \ref{tab:TbarSoftRigid} summarizes the properties of the two types of bars we used for simulating the T-bar using the compressible formulation.
The resulting observations concerning the system norms and corresponding input-output relationships are also present in the table.
Figure \ref{fig:Tbar_distSoftRigid} presents the deflection of the free nodes when the structure is subjected to a doublet force of amplitude 10 N (fig \ref{fig:doublet}).
Clearly, the T-bar with rigid metallic bars exhibits larger deviations when subject to the same force, when compared to the T-bar with softer bars.
% \textcolor{red}{Discuss System norm}
Note that due to the large difference in stiffnesses of the bars and the strings in the T-bar structure, the eigenvalues of the resulting compliance matrix do not differ meaningfully when we replace the conventionally used aluminium bars with those made of HDPE, as can be seen in figure \ref{fig:eig_compTbar}.
This essentially means that given a certain force, amplitudes of static deflection would be remarkably similar for either kind of bar material.
In such scenarios, given budgetary considerations on structural weight, it would be worth investigating whether replacing stiff metallic bars with bars made of softer materials would yield comparable strength for lower mass.

\begin{table*}[!ht]
\centering
\begin{tabular}[center]{|l||*{2}{r|}}\hline
	\diagbox{Attribute}{Type of Bar}
	&\makebox[3em]{Soft}&\makebox[3em]{Metallic}\\
	\hline \hline
	Material  & HDPE & Aluminium\\
\hline
	Density ($kg/m^3$) & 960 & 2700 \\
	Young's Modulus ($GPa$)& 1 & 68 \\
	Total Mass ($kg$) & 75.39 & 212.05  \\
 \hline
	Norms of Linear System: & {} & {}\\
	$||{G}||_2$ & 0.1059 & 0.1752 \\
	$||{G}||_{\infty}$ & 0.1718 & 0.4832 \\
\hline
	Input-Output Relationships from Nonlinear Response: & {} & {}\\
	$\cfrac{||{y}||_{\infty}}{||u||_2}$ & 4.79$e^{-4}$ & 5.3$e^{-4}$ \\
	$\cfrac{||{y}||_{2}}{||u||_2}$ & 7.22$e^{-4}$ & 8.14$e^{-4}$ \\
\hline
\end{tabular} \\
\caption{Comparison of soft and metallic bars in the compressible T-bar example}
\label{tab:TbarSoftRigid}
\end{table*}

% \textcolor{blue}{
\section{Conclusions}
Informally characterized as a network of axially loaded bars and cables, tensegrity structures are an increasingly popular subset of multibody systems finding applications in several diverse domains ranging from space robotics to biological modeling.
In this paper, equations of motion for analysing the dynamics of tensegrity structures were developed in Cartesian coordinates, i.e., a non-minimum set of coordinates, using Lagrangian mechanics.
% }

% \textcolor{blue}{
The use of non-minimum coordinates necessitates an active effort to prevent constraint violations, since numerical errors creep up on the constraint space when integrating the governing equations of motion of an overparameterized system.
A direct correction approach was employed to ensure that constraints are adhered to, not only in position and velocity, but also in total mechanical energy.
Simulations comparing our approach with the commercially available Simscape Multibody illustrate the superiority of our formulation in terms of numerical accuracy.
The correction method ensures stability of constraints in 3-dimensional complex tensegrity systems as well.
The differences with and without correction grow as the structure exhibits larger motion amplitudes, or more commonly when the error of numerical integration is unchecked for long simulation time spans.
Further, a linearization of the associated equations of motion was derived to facilitate future work in control.
% }

% \textcolor{blue}{
We also considered compressible bars, wherein bars are no longer geometrically constrained.
This will allow a deeper investigation into advanced compliant tensegrity structures that might incorporate softer bars in their architecture.
We conducted one such analysis comparing a linearized compressible system with  2 different kinds of bars, one made using soft HDPE and the other with conventionally used aluminium.
It revealed the limited effect of bar rigidity on the amplitudes of static deflection, an interesting insight that raises further questions concerning bar material choices given strength and mass considerations.
% }

% Equations of motion %\eqref{eqn:finalEqn}
% for tensegrity structures were developed in Cartesian coordinates, i.e. a non-minimum set of coordinates, using Lagrangian mechanics
% % Tensegrity structures are naturally constrained since the r
% , and the linearization of the associated EOMs was introduced
% %in \eqref{eqn:linDyn}
% symbolically. Since states at the position and velocity levels may drift away from the constraint space due to numerical computations, the direct correction approach %described in
% % \eqref{eqn:correction_solution}
% was used so that the violations in both geometric constraints and total mechanical energy are eliminated. Furthermore, we considered compressible bars, where rigid bars are no longer geometrically constrained, but the high Young's modulus of the bars results in a stiff problem. A simple tensegrity structure was simulated with different parameters to show the improvement in accuracy by using the proposed approach. Besides, we showed that the differences between numerical simulation performed with and without correction become more prominent when a structure has larger motion or when the error of numerical integration is increasing. Further, we simulated tensegrity structures with significantly more complex geometries to illustrate the consistency of the proposed method towards constraint stabilization. Finally, we demonstrated the extensibility of our formulation to systems with compressible bars, thereby allowing deeper investigation into advanced tensegrity structures that might incorporate softer bars in their architecture.

\section*{Acknowledgement}
This work was supported by NSF IUSE/PFE: RED: REvolutionizing Diversity Of Engineering (REDO-E) Award Number:1730693; and NASA NIAC Phase II grant, on Tensegrity Approaches to In-Space Construction of a 1g Growable Habitat.

\bibliographystyle{spmpsci}
\bibliography{raktim}

%% Figures

%% Figures to establish that 10^-14 tolerance is as good as accurate without constraint stabilisation

%% Computation Time Plot

%% Figures to discuss effectiveness of energy correction
% Tol 10^-10

% Tol 10^-6

% Rest Length Plot

%% End all figures for T-bar\label{fig:energy_ball}

% \begin{figure}[!ht]
% \centering
% \subfigure[Node 3.]
% {\includegraphics[width=0.48\textwidth]{figs/TBAR_MotionErrorEnergy_Node3}\label{fig:motionEnergy_TbarL1}}\hfill
% \subfigure[Node 4.]
% {\includegraphics[width=0.48\textwidth]{figs/TBAR_MotionErrorEnergy_Node4}\label{fig:motionEnergy_TbarL2}}\hfill
% \caption{Motion difference between the proposed method with and without the energy correction in the example of T-bar.}
% \label{fig:motionEnergy_TbarL}
% \end{figure}

%% ARM figures

%% BALL

%% Compressible Tbar

%% Compressible Tbar

%% END all Figures

%% References







% \end{thebibliography}

\end{document}